\documentclass[11pt,twoside,nohyper]{pallis}
\usepackage{epsfig}
\usepackage{latexsym}
\usepackage{euscript}
\usepackage{amssymb}
\usepackage{pstricks}
\usepackage{fancyhed}

\newcommand{\beq}{\begin{equation}}
\newcommand{\eeq}{\end{equation}}
\newcommand{\ben}{\begin{equation*}}
\newcommand{\een}{\end{equation*}}
\newcommand{\brr}{\begin{array}}
\newcommand{\err}{\end{array}}
\newcommand{\bc}{\begin{center}}
\newcommand{\ec}{\end{center}}

\newcommand{\bea}{\begin{eqnarray}}
\newcommand{\eea}{\end{eqnarray}}
\newcommand{\bean}{\begin{eqnarray*}}
\newcommand{\eean}{\end{eqnarray*}}
\newcommand{\ftn}{\footnotesize}
\newcommand{\nsz}{\normalsize}
\newcommand{\ssz}{\scriptsize}

\newcommand{\vH}{{\mbox{$\bar H$}}}

\newcommand{\vrho}{{\mbox{$\bar\rho$}}}
\newcommand{\vq}{{\mbox{$\bar q$}}}
\newcommand{\vQ}{{\mbox{$\bar Q$}}}
\newcommand{\vV}{{\mbox{$\bar V$}}}
\newcommand{\vGamma}{{\mbox{$\bar \Gamma$}}}
\newcommand{\vVo}{{\mbox{$\bar V_0$}}}

\newcommand{\vtf}{{\mbox{$\vtau_{_{\rm F}}$}}}
\newcommand{\vtns}{{\mbox{$\vtau_{_{\rm NS}}$}}}

\newcommand{\vtft}{{\mbox{$\vtauf_{_{\rm T}}$}}}
\newcommand{\vtt}{{\mbox{$\vtau_{_{\rm T}}$}}}

\newcommand{\sv}{{\mbox{$\langle \sigma v \rangle$}}}

\newcommand{\brhofi}{{\mbox{$\bar\rho_{\phi_{\rm I}}$}}}
\newcommand{\brhoqi}{{\mbox{$\bar\rho_{q_{\rm I}}$}}}

\newcommand{\brhof}{{\mbox{$\bar\rho_{\phi}$}}}
\newcommand{\brhoq}{{\mbox{$\bar\rho_{q}$}}}

\newcommand{\rhofi}{{\mbox{$\rho_{\phi_{\rm I}}$}}}
\newcommand{\rhoqi}{{\mbox{$\rho_{q_{\rm I}}$}}}
\newcommand{\brhoRi}{{\mbox{$\vrho_{_{\rm RI}}$}}}

\newcommand{\btkr}{{\mbox{$\btau_{_{\rm KR}}$}}}
\newcommand{\btkp}{{\mbox{$\btau_{_{\rm K \phi}}$}}}
\newcommand{\btrhp}{{\mbox{$\btau_{_{\rm RH}}$}}}
\newcommand{\btrh}{{\mbox{$\btau_{_{\rm RD}}$}}}
\newcommand{\btpr}{{\mbox{$\btau_{_{\phi \rm R}}$}}}
\newcommand{\btast}{{\mbox{$\btau_{\ast}$}}}
\newcommand{\btf}{{\mbox{$\btau_{_{\rm F}}$}}}
\newcommand{\btt}{{\mbox{$\btau_{_{\rm T}}$}}}
\newcommand{\btfn}{{\mbox{$\btau_{_{\rm f}}$}}}

\newcommand{\vqkp}{{\mbox{$\vq_{_{\rm K\phi}}$}}}

\newcommand{\Tkr}{{\mbox{$T_{{\rm KR}}$}}}
\newcommand{\Tkp}{{\mbox{$T_{{\rm K\phi}}$}}}

\newcommand{\Tpr}{{\mbox{$T_{\phi{\rm R}}$}}}

\newcommand{\btfkp}{{\mbox{$\btauf_{_{\rm K\phi}}$}}}

\def\openep{\leavevmode\hbox{\normalsize$\iota$\kern-3.8pt$^$-}}
\def\vtau{\leavevmode\hbox{\normalsize$\tau$\kern-5.pt$\iota$}}
\def\vtauf{\leavevmode\hbox{\ftn$\tau$\kern-4.pt$\iota$}}
\def\btau{\leavevmode\hbox{\normalsize$\tilde\tau$\kern-5.pt$\iota$}}
\def\btauf{\leavevmode\hbox{\ftn$\tilde\tau$\kern-4.pt$\iota$}}

\title{\huge K{\Large INATION}-D{\Large OMINATED} R{\Large EHEATING
AND \\} C{\Large OLD} D\kern-1.pt{\Large ARK} M\kern-1.pt{\Large
ATTER} A\Large BUNDANCE}

\author{\Large C. P\kern-1.5pt\nsz ALLIS\\
Physics Division, School of Technology,\\
Aristotle University of Thessaloniki, \\
541 24 Thessaloniki, GREECE \\ \vspace{5pt}
\email{kpallis@auth.gr}}

\abstract{\vspace{5pt}\\ $~~~~~$ We consider the decay of a
massive particle under the complete or partial domination of the
kinetic energy density generated by a quintessential exponential
model and we impose a number of observational constraints,
originating from nucleosynthesis, the present acceleration of the
universe and the dark-energy-density parameter. We show that the
presence of kination causes a prolonged period during which the
temperature is frozen to a plateau value, much lower than the
maximal temperature achieved during the process of reheating in
the absence of kination. The decoupling of a cold dark matter
particle during this period is analyzed, its relic density is
calculated both numerically and semi-analytically and the results
are compared with each other. Using plausible values (from the
viewpoint of particle models) for the mass and the thermal
averaged cross section times the velocity of the cold relic, we
investigate scenaria of equilibrium or non-equilibrium production.
In both cases, acceptable results for the cold dark matter
abundance can be obtained, by constraining the initial energy
density of the decaying particle, its decay width, its mass and
the averaged number of the produced cold relics. The required
plateau value of the temperature is, in most cases, lower than
about $40~{\rm GeV}$.

{\sl\bfseries Published in} {\sl  Nucl.
Phys.} {\bf B751}, 129 (2006)}

\begin{document}

\setcounter{page}{1} \pagestyle{fancyplain}

\addtolength{\headheight}{.5cm}

\rhead[\fancyplain{}{ \bf \thepage}]{\fancyplain{}{K{\ftn
INATION}-D{\ftn OMINATED} R{\ftn EHEATING AND} CDM A{\ftn
BUNDANCE}}} \lhead[\fancyplain{}{ \leftmark}]{\fancyplain{}{\bf
\thepage}} \cfoot{}

\section{I{\ftn NTRODUCTION}}\label{intro}

\hspace{.67cm} A plethora of recent data \cite{wmap, snae}
indicates that the energy-density content of the universe is
comprised of Cold (mainly \cite{wmapl}) Dark Matter (CDM) and Dark
Energy (DE) with density parameters \cite{wmap}:
\beq {\sf (a)}~~\Omega_{\rm CDM}=0.24\pm0.1~~\mbox{and}~~{\sf
(b)}~~\Omega_{\rm DE}=0.73\pm0.12, \label{cdmba}\eeq
respectively, at $95\%$ confidence level (c.l.). Several
candidates and scenarios have been proposed so far for the
explanation of these two unknown substances.

As regards CDM, the most natural candidates \cite{candidates} are
the weekly interacting massive particles, $\chi$'s. The most
popular of these is the lightest supersymmetric (SUSY) particle
(LSP) \cite{goldberg}. However, other candidates \cite{lkk} arisen
in the context of the extra dimensional theories should not be
disregarded. In light of eq.~(\ref{cdmba}), the $\chi$-relic
density, $\Omega_{\chi}h^2$, is to satisfy the following range of
values:
\beq {\sf (a)}~~0.09\lesssim \Omega_{\chi}h^2~~\mbox{and}~~ {\sf
(b)}~~\Omega_{\chi}h^2\lesssim0.13. \label{cdmb}\eeq

Obviously, the $\Omega_{\chi}h^2$ calculation crucially depends on
the adopted assumptions. According to the standard cosmological
scenario (SC) \cite{kolb}, $\chi$'s {\sf (i)} are produced through
thermal scatterings in the plasma, {\sf (ii)} reach chemical
equilibrium with plasma and {\sf (iii)} decouple from the cosmic
fluid at a temperature $T_{\rm F}\sim (10-20) ~{\rm GeV}$ during
the radiation-dominated (RD) era (we do not consider in our
analysis, modification to the Friedmann equations due to a low
brane-tension as in ref.~\cite{seto} and some other CDM candidates
(see e.g. ref.~\cite{gravitino}) which require a somehow different
cosmological set-up). The assumptions above fix the form of the
relevant Boltzmann equation, the required strength of the $\chi$
interactions for equilibrium production (EP) and lead to an
isentropic cosmological evolution during the $\chi$ decoupling:
The Hubble parameter is $H\propto T^2$ with temperature $T\propto
R^{-1}$ ($R$: the scale factor of the universe). In this context,
the $\Omega_{\chi}h^2$ calculation depends only on two parameters:
The $\chi$ mass, $m_\chi$ and the thermal-averaged cross section
of $\chi$ times velocity, $\langle \sigma v \rangle$. If, in
addition, a specific particle model is adopted, $\langle \sigma v
\rangle$ can be derived from $m_\chi$ and the residual particle
spectrum of the theory (see e.g. refs.~\cite{lkk, spanos}).
Consequently, by imposing the CDM constraint -- eq.~(\ref{cdmb})
-- some particle models (such as the Constrained Minimal SUSY
Model (CMSSM) \cite{Cmssm, spanos}) can be severely restricted
whereas some others (e.g. the models of refs.~\cite{su5} and
\cite{wells} which produce higgsino  or wino LSP respectively) can
be characterized as cosmologically uninteresting due to the very
low obtained $\Omega_{\chi}h^2$.

In a couple of recent papers \cite{scnr, jcapa}, we investigated
model-independently (from the viewpoint of particle physics) how
the calculation of $\Omega_{\chi}h^2$ is modified, when one or
more of the assumptions of the SC are lifted (for similar
explorations, see refs.~\cite{Kam, McDonald, riotto, salati,
masiero}). Namely, in ref.~\cite{scnr} we assumed that $\chi$'s
{\sf (i$^\prime$)} decouple during a decaying-massive-particle,
$\phi$, dominated era (and mainly before reheating) {\sf
(ii$^\prime$)} do or do not reach chemical equilibrium with the
thermal bath {\sf (iii$^\prime$)} are produced by thermal
scatterings and directly by the $\phi$ decay (which, naturally
arises even without direct coupling \cite{drees}).  The key point
in our investigation is that the reheating process is not
instantaneous \cite{turner}. During its realization, the maximal
temperature, $T_{\rm max}$, is much larger than the so-called
reheat temperature, $T_{\rm RH}$, which can be taken to be lower
than $T_{\rm F}$. Also, for $T>T_{\rm RH}$, $H\propto T^4$ with $T
\propto R^{-3/8}$ and an entropy production occurs (in contrast
with the SC). As a consequence, in the context of this (let name
it, Low Reheating) scenario (LRS) the $\Omega_{\chi}h^2$
calculation depends also on $T_{\rm RH}$, the mass of the decaying
particle, $m_\phi$, and the averaged number of the produced
$\chi$'s, $N_\chi$. We found \cite{scnr} that, for fixed $m_\chi$
and $\sv$, comfortable satisfaction of eq.~(\ref{cdmb}) can be
achieved by constraining $T_{\rm RH}$ (to values lower than
$20~{\rm GeV}$), $m_\phi$ and $N_\chi$. E.g., $\Omega_{\chi}h^2$
decreases with respect to (w.r.t) its value in the SC
\cite{fornengo} for low $N_\chi$'s and increases for larger
$N_\chi$'s \cite{moroi, fujii} (with fixed $m_\chi$ and $\sv$).
Both EP and non-EP are possible for commonly obtainable $\sv$'s
\cite{gondolonn}. \vspace{-.8pt}

Another role that a scalar field could play when it does not
couple to matter (in contrast with the former case) is this of
quintessence \cite{early}. This scalar field $q$ (not to be
confused with the deceleration parameter \cite{wmapl}) is supposed
to roll down its potential undergoing three phases during its
cosmological evolution: Initially its kinetic energy, which
decreases as $T^6$, dominates and gives rise to a possible novel
period in the universal history termed ``kination''
\cite{kination}. Then, $q$ freezes to a value close to the Planck
scale and by now its potential energy, adjusted so that
eq.~(\ref{cdmba}{\sf b}) is met, becomes dominant. In
ref.~\cite{jcapa}, we focused on a range of the exponential
potential \cite{wet} parameters, which can lead to a simultaneous
satisfaction of several observational data (arising from
nucleosynthesis, acceleration of the universe and the DE density
parameter) in conjunction with the existence of an early totally
kination-dominated (KD) era, for a reasonable region of initial
conditions \cite{brazil, german}. As a consequence, during the KD
era we have: $H\propto T^3$ with $T\propto R^{-1}$. Under the
assumption that the $\chi$-decoupling occurs after the
commencement of the totally KD phase -- the assumptions {\sf (i)}
and {\sf (ii)} were maintained -- we found that in this scenario
(let name it QKS) $\Omega_{\chi}h^2$ increases w.r.t its value in
the SC \cite{salati} (with fixed $m_\chi$ and $\sv$) and we showed
that this enhancement can be expressed as a function of the
quintessential density parameter at the eve of nucleosynthesis,
$\Omega_q(\vtns)$. Moreover, values of $\Omega_q(\vtns)$ close to
its upper bound require $\sv$ to be almost three orders of
magnitude larger than this needed in the SC so as eq.~(\ref{cdmb})
is fulfilled. It is obvious that the QKS although beneficial
\cite{prof} for some particle-models \cite{su5, wells} can lead to
an utter exclusion of some other simple, elegant and predictive
particle models such as the CMSSM \cite{salati, prof, jcapa}.

\begin{table}[!t]
\begin{center}
\begin{tabular}{|c|c|c|c|} \hline
{\bf  SC} & {\bf  LRS} & {\bf QKS} & {\bf KRS}
\\ \hline \hline
$\brhoq=\brhof=0$&$\brhofi\gg\brhoRi,~\brhoq=0$&
$\brhoqi\gg\brhoRi,~\brhof=0$&$\brhoqi\gg\brhofi\gg\brhoRi$\\
$H\propto T^2$& $H\propto T^4$&$H\propto T^3$&$H\propto T^3$\\
$T\propto R^{-1}$&$T\propto R^{-3/8}$&$T\propto R^{-1}$&$T={\rm
cst}$\\
$sR^3={\rm cst}$&$sR^3\neq{\rm cst}$&$sR^3={\rm
cst}$&$sR^3\neq{\rm cst}$\\
$N_\chi=0$&$N_\chi\neq0$&$N_\chi=0$&$N_\chi\neq0$\\ \hline
\end{tabular}
\end{center}\vspace*{-.155in}
\caption{\sl\ftn Differences and similarities of the KRS with the
SC, LRS and the QKS (the various symbols are explained in
sec.~\ref{sec:num}, the subscript I is referred to the onset of
each scenario and cst stands for ``constant'').\label{tab1}}
\vspace*{-.2in}
\end{table}

It would be certainly interesting to examine if the latter
negative result could be evaded, invoking the coexistence of the
two scenaria above (LRS and QKS) i.e. if a low $T_{\rm RH}$, which
can assist us to the reduction of $\Omega_{\chi}h^2$, can be
compatible with a quintessential KD phase. To this aim we first
investigate the dynamics of an oscillating field ($\phi$) under
the complete or partial domination of the kinetic energy density
of another field, $q$. We name this novel cosmological set-up KD
Reheating (KRS). A similar situation has been just approximately
explored  in refs.~\cite{liddle, feng}, under the name ``curvaton
reheating'' for considerably higher scales (in their case $q$ is
restricted to drive quintessential \cite{qinf} or steep
\cite{steep} inflation and $\phi$ is constrained so as it acts as
curvaton \cite{curv}). The numerical integration of the relevant
equations reveals that a prominent period of constant maximal
temperature, $T_{\rm PL}$, arises (surprisingly, similar findings
have been reported in ref.~\cite{notari} for another cosmological
set-up). $T_{\rm PL}$ turns out to be much lower than $T_{\rm
max}$ obtained in the LRS with the same initial $\phi$ energy
density (the similarities and the differences between the various
scenarios can easily emerge from table~\ref{tab1}). On the other
hand, the evolution of $q$ is just slightly affected and so, it
can successfully play the role of quintessence, similarly to the
QKS. The resultant $\Omega_{\chi}h^2$ reaches the range of
eq.~(\ref{cdmb}) with (i$^{\prime\prime}$)
$N_\chi\sim(10^{-7}-10^{-5})$ when $T_{\rm PL}\ll T_{\rm F}$ (type
I non-EP), (ii$^{\prime\prime}$) $N_\chi\sim0$ when $T_{\rm
PL}\sim T_{\rm F}$ (type II non-EP). On the other hand, EP which
is activated for $T_{\rm PL}>T_{\rm F}$ requires a tuning of $\sv$
in order that interesting $\Omega_{\chi}h^2$'s are obtained. The
required $T_{\rm PL}$'s are mostly lower than 40 GeV
\cite{referee}.

The framework of the KRS is described in sec.~\ref{sec:num}, while
the analysis of the $\Omega_{\chi}h^2$ calculation is displayed in
sec.~\ref{sec:neut}. Some numerical particle-model-independent
applications of our findings are realized in sec.~\ref{ap}.
Finally, sec.~\ref{con} summarizes our results. Throughout the
text and the formulas, brackets are used by applying disjunctive
correspondence, natural units ($\hbar=c=k_{\rm B}=1$) are assumed,
the subscript or superscript $0$ is referred to present-day values
(except for the coefficient $V_0$) and $\ln~[\log]$ stands for
logarithm with basis $e~[10]$.

%\newpage

\section{D{\ftn YNAMICS OF} K{\ftn INATION}-D{\ftn OMINATED}
R{\ftn EHEATING}} \label{sec:num}

\hspace{.562cm} According to the KRS, we consider the coexistence
of two spatially homogeneous, scalar fields $q$ and $\phi$. The
field $q$ represents quintessence. The particle $\phi$ with mass
$m_\phi$ can decay with a rate $\Gamma_\phi$ into radiation,
producing an average number $N_{\chi}$ of $\chi$'s with mass
$m_{\chi}$, rapidly thermalized. We, also, let open the
possibility  that ${\chi}$'s are produced through thermal
scatterings in the bath. The system of equations which governs the
cosmological evolution is presented in sec.~\ref{Beqs} and a
numerically robust form of this system is extracted in
sec.~\ref{Neqs}. In sec.~\ref{reqq}  we present the various
observational restrictions and in sec. \ref{Seqs} we derive useful
expressions, which fairly approximate our numerical results.

\subsection{R{\ssz ELEVANT} E{\ssz QUATIONS}}
\label{Beqs}

\hspace{.562cm} The cosmological evolution of $q$ obeys the
homogeneous Klein-Gordon equation:
\beq \mbox{\sf (a)}~~\ddot q+3H\dot
q+V_{,q}=0,~~\mbox{where}~~\mbox{\sf (b)}~~V=V_0 e^{-\lambda
q/m_{_{\rm P}}}\label{qeq} \eeq
is the adopted potential for the $q$-field, $,q$ [dot] stands for
derivative w.r.t $q$ [the cosmic time, $t$] and the Hubble
parameter, $H$, is written as:
\beq \label{Hb} \mbox{\sf (a)}~~H=\frac{1}{\sqrt{3}m_{\rm
P}}\sqrt{\rho_{\chi}+\rho_q +\rho_\phi+\rho_{_{\rm
R}}}~~\mbox{with}~~\mbox{\sf (b)}~~\rho_{\chi}= m_{\chi}
n_{\chi}~~\mbox{and}~~\mbox{\sf (c)}~~\rho_q=\frac{1}{2}\dot
q^2+V,\eeq
%
%\begin{eqnarray} \label{Hini}
%&& H=\sqrt{\left(\rho_{\chi}+\rho_q +\rho_\phi+\rho_{_{\rm R}}
%\right)/3m_{_{\rm P}}^2}~~\mbox{with}\\
%
%&& \rho_{\chi}= m_{\chi}
%n_{\chi}~~\mbox{and}~~\rho_q=\frac{1}{2}\dot q^2+V,
%\end{eqnarray}
%
the energy densities of $\chi$ and $q$ correspondingly and
$m_{_{\rm P}}=M_{\rm P}/\sqrt{8\pi}$ (where $M_{\rm
P}=1.22\times10^{19}~{\rm GeV}$ is the Planck scale). The energy
density of radiation [$\phi$], $\rho_{_{\rm R}}~[\rho_\phi]$, and
the number density of $\chi$, $n_{\chi}$, satisfy the following
equations \cite{chung, moroi}
($\Delta_\phi=(m_\phi-N_{\chi}m_{\chi})/m_\phi$):
\numparts
\begin{eqnarray}
&& \dot \rho_\phi+3H\rho_\phi+\Gamma_\phi \rho_\phi=0,\label{nf}
\\
&& \dot \rho_{_{\rm R}}+4H\rho_{_{\rm R}}-\Gamma_\phi
\rho_\phi-2m_{\chi}\langle \sigma v \rangle \left( n_{\chi}^2 -
n_{\chi}^{\rm eq2}\right)=0, \label{rR}\\
&& \dot n_{\chi}+3Hn_{\chi}+\langle \sigma v \rangle \left(
n_{\chi}^2 - n_{\chi}^{\rm eq2}\right)-\Gamma_\phi N_{\chi}
\rho_\phi/\Delta_\phi m_\phi=0,\label{nx}
\end{eqnarray}
\endnumparts
\hspace{-.14cm}where the equilibrium number density of $\chi$'s,
$n_{\chi}^{\rm eq}$, obeys the Maxwell-Boltzmann statistics:
\begin{equation} \label{neq}
n_{\chi}^{\rm eq}(x)=\frac{g}{(2\pi)^{3/2}}
m_{\chi}^3\>x^{3/2}\>e^{-1/x}P_2(1/x),~~\mbox{where}~~
x=T/m_{\chi},
\end{equation}
$g=2$ is the number of degrees of freedom of ${\chi}$ and
$P_n(z)=1+(4n^2-1)/8z$ is obtained by asymptotically expanding the
modified Bessel function of the second kind of order $n$.

The temperature, $T$, and the entropy density, $s$, can be found
using the relations:
\beq \mbox{\sf (a)}~~\rho_{_{\rm R}}=\frac{\pi^2}{30}g_{\rho*}\
T^4~~\mbox{and}~~\mbox{\sf (b)}~~s=\frac{2\pi^2}{45}g_{s*}\ T^3,
\label{rs}\eeq
where $g_{\rho*}(T)~[g_{s*}(T)]$ is the energy [entropy] effective
number of degrees of freedom at temperature $T$. Their numerical
values are evaluated by using the tables included in {\tt
micrOMEGAs} \cite{micro}, originated from the {\sf DarkSUSY}
package \cite{dark}.

Finally, to keep contact with the LRS, we express the decay width
of $\phi$, $\Gamma_\phi$, in terms of a temperature $T_\phi$
through the relation:
\begin{equation}
\Gamma_\phi =5\sqrt{\frac{\pi^3 g_{\rho*}(T_\phi)}{45}}
\frac{T_\phi^2}{M_{\rm P}}=\sqrt{\frac{5\pi^3
g_{\rho*}(T_\phi)}{72}} \frac{T_\phi^2}{m_{_{\rm P}}} \label{GTrh}
\cdot
\end{equation}
Note that the prefactor 5 is different from our choice in
ref.~\cite{scnr} (4) with $T_\phi=T_{\rm RH}$ and several others
(e.g. 6 \cite{ham} and 2 \cite{riotto, gondolon}). Our final
present choice is justified in sec.~\ref{fdom}.

\subsection{N{\ssz UMERICAL} I{\ssz NTEGRATION}}
\label{Neqs}

\hspace{.562cm} The integration of the equations above can be
realized successively in two steps: The first step concerns the
completion of the KD reheating process, while the second one
regards the residual running of $q$ from the onset of RD era until
today.

\subsubsection{First step of integration.}
The numerical integration of eqs.~(\ref{qeq}) and
(\ref{nf})--(\ref{nx}) is facilitated by absorbing the dilution
terms. To this end, we find it convenient to define the following
dimensionless variables \cite{chung, riotto, scnr} (recall that
$R$ is the scale factor):
\begin{equation} \label{fdef}
f_\phi=\rho_\phi R^3,~f_{\rm R}=\rho_{_{\rm R}}R^4,~
f_{\chi}^{[\rm eq]}=n^{[\rm eq]}_{\chi} R^3~~\mbox{and}~~ f_q=\dot
qR^3.
\end{equation}
Converting the time derivatives to derivatives  w.r.t the
logarithmic time \cite{brazil, german} (the value of $R_{\rm I}$
in this definition turns out to be numerically irrelevant):
\beq \btau=\ln\left(R/R_{\rm
I}\right)~\Rightarrow~R^\prime=R~~\mbox{and}~~R=R_{\rm I
}e^{\btauf}\label{dtau} \eeq
eqs.~(\ref{qeq}) and (\ref{nf})--(\ref{nx}) become (prime denotes
derivation w.r.t $\btau$):
\numparts
\begin{eqnarray}
Hf^\prime_q&=&-V_{,q} R^3,\label{fq}\\
Hf^\prime_\phi&=&-\Gamma_\phi f_\phi,\label{ff}\\
H R^2 f^\prime_{\rm R}&=&\Gamma_\phi f_\phi R^3+2 m_{\chi} \langle
\sigma v \rangle \left(f_{\chi}^2 - f_{\chi}^{\rm eq2}\right),
\label{fR}
\\
HR^3f^\prime_{\chi}&=& -\langle\sigma v\rangle \left(f_{\chi}^2 -
f_{\chi}^{\rm eq2}\right)+\Gamma_\phi N_{\chi} f_\phi R^3/
\Delta_\phi m_\phi,\label{fn}
\end{eqnarray}
\endnumparts
\hspace{-.14cm}where $H$ and $T$ can be expressed correspondingly,
in terms of the variables in eq.~(\ref{fdef}), as:
\begin{equation} \label{H2exp}
H=\frac{1}{\sqrt{3R^3}m_{_{\rm P}}}\sqrt{m_{\chi}
f_{\chi}+f^2_q/2R^3+VR^3 + f_\phi +f_{\rm R}/R
}~~\mbox{and}~~T=\left(\frac{30\ f_{\rm R}}{\pi^2 g_{\rho\ast}
R^4}\right)^{1/4}\cdot\end{equation}
The system of eqs.~(\ref{fq})--(\ref{fn}) can be solved from 0 to
$\btau_{\rm f}\sim 25$, imposing the following initial conditions
(recall that the subscript I is referred to quantities defined at
$\btau=0$):
\begin{equation}
f_\phi(0)R_{\rm I}^3 =\brhofi\rho^0_{\rm c},~f_{\rm
R}(0)=f_{\chi}(0)=0,~ q(0)=0 ~~\mbox{and}~~\dot
q(0)=\sqrt{2\rho^0_{\rm c}}(m_\phi/H_0). \label{init}
\end{equation}
Note that, unlike in the LRS \cite{scnr}, the numerical choice of
$f_\phi(0)$ and $\dot q(0)$ is crucial for the result of
$\Omega_{\chi} h^2$. We let the first one as a free parameter,
while we determine the second one via $m_\phi$, assuming that the
$\phi$ oscillations commence at $H_{\rm I}\simeq m_\phi$
\cite{chung, feng,liddle}. Since $H_{\rm I}^2\simeq\dot
q^2(0)/6m^2_{_{\rm P}}$, we obtain the last condition in
eq.~(\ref{init}).

\subsubsection{Second step of integration.} When
$\rho_{_{\rm R}}\gg\rho_\phi$ the transition into the pure RD era
has been terminated and the evolution of $q$ can be continued by
employing the formalism of ref.~\cite{jcapa}. More precisely, we
can define a transition point, $\btt$, through the relation
$\rho_{_{\rm R}}(\btt)/\rho_\phi(\btt)\sim 100$ and then, find the
corresponding $\vtt$, via the formula \cite{jcapa}:
\beq \label{rhotau}\mbox{\sf (a)}~~\rho_{_{\rm
R}}(\btt)=\rho^0_{_{\rm R}}\frac{g_{\rho*}}{g^0_{\rho*}}
\left(\frac{g^0_{s*}}{g_{s*}}\right)^{4/3}e^{-4\vtft}~~\mbox{where}~~\mbox{\sf
(b)}~~\vtau=\ln\left(R/R_0\right).\eeq
The running of $q$ can be realized from $\vtt$ to 0, by following
the procedure described in sec. 2.1.3 of ref.~\cite{jcapa} with
initial conditions:
\begin{equation}
q(\vtt)=q(\btt) ~~\mbox{and}~~\dot q(\vtt)=(f_q/R^3)(\btt).
\label{initb}
\end{equation}

%\newpage
\subsection{I{\ssz MPOSED} R{\ssz EQUIREMENTS}} \label{reqq}

\hspace{.565cm}  We briefly describe the various criteria that we
impose on our model.
\subsubsection{KD ``Constraint''.} We focus our attention on the range of parameters which
ensure an absolute domination of the $q$-kinetic energy at
$\btau=0$. This can be achieved, when:
\beq\mbox{\sf (a)}~~\Omega^{\rm
I}_q=\Omega_q(0)=1~~\mbox{with}~~\mbox{\sf
(b)}~~\Omega_q=\rho_q/(\rho_q+\rho_{_{\rm
R}}+\rho_\chi+\rho_\phi)\label{domk}\eeq
the quintessential energy density parameter.
\subsubsection{Nucleosynthesis (NS) Constraint.} The
presence of $\rho_q$ and $\rho_\phi$ have not to spoil the
successful predictions of Big Bang NS which commences at about
$\vtns=-22.5$ corresponding to $T_{\rm NS}=1~{\rm MeV}$
\cite{oliven}. Taking into account the most up-to-date analysis of
ref.~\cite{oliven}, we adopt a rather conservative upper bound on
$\Omega_q(\vtns)$, less restrictive than that of ref.~\cite{nsb}.
Namely, we require:
\beq\mbox{\sf (a)}~~\Omega_q^{\rm
NS}=\Omega_q(\vtns)\leq0.21~~\mbox{($95\%$
c.l.)}~~\mbox{and}~~\mbox{\sf (b)}~~T_{\rm RD}\geq1~{\rm MeV},
\label{nuc}\eeq
where $T_{\rm RD}$ is defined as the largest temperature of the RD
era and 0.21 corresponds to additional effective neutrinos species
$\delta N_\nu<1.6$ \cite{oliven}. We do not consider extra
contribution (potentially large \cite{liddle}) in the left hand
side (l.h.s) of eq.~(\ref{nuc}) due to the energy density of the
gravitational waves \cite{giova} generated during a possible
former transition from inflation to KD epoch \cite{qinf}. The
reason is that inflation could be driven by another field
different to $q$ and so, any additional constraint arisen from
that period would be highly model dependent.

\subsubsection{Coincidence Constraint.} The present value of
$\rho_q$, $\rho^0_q$, must be compatible with the preferred range
of eq.~(\ref{cdmba}{\sf b}). This can be achieved by adjusting the
value of $\vVo$. Since, this value does not affect crucially our
results (especially on the CDM abundance), we decide to fix
$\vrho^0_q=\rho^0_q/\rho^0_{\rm c}$ to its central experimental
value, demanding:
\beq \Omega^0_q=\vrho^0_q=0.73.\label{rhoq0}\eeq

\subsubsection{Acceleration Constraint.} A successful
quintessential scenario has to account for the present-day
acceleration of the universe, i.e. \cite{wmap},
\beq\mbox{\sf (a)}~~-1\leq w_q(0)\leq-0.78~~\mbox{($95\%$ c.l.)}
~~\mbox{with}~~\mbox{\sf (b)}~~w_q=(\dot q^2/2-V)/(\dot q^2/2+V)
\label{wq}\eeq
the barotropic index of the $q$-field. In our case, we do not
succeed to avoid \cite{german} the eternal acceleration ($w^{\rm
fp}_q>-1/3$) which is disfavored by the string theory.

\subsubsection{Residual Constraints.\label{resc}} In our scanning, we take
into account the following less restrictive but also not so
rigorous bounds, which, however, do not affect crucially our
results:

\beq {\sf (a)}~~10^3~{\rm GeV}\leq m_\phi\lesssim10^{14}~{\rm GeV}
~~\mbox{and}~~{\sf (b)}~~N_{\chi}\leq 1. \label{para}\eeq
The lower bound of eq. (\ref{para}{\sf a}) is imposed so as the
decay of $\phi$ to a pair of $\chi$'s with mass at most $500~{\rm
GeV}$ is kinematically allowed. The upper bound of
eq.~(\ref{para}{\sf a}) comes from the COBE constraints
\cite{cobe} on the spectrum of gravitational waves produced at the
end of inflation \cite{qinf}. In particular, the later constraint
impose an upper bound on $H_{\rm I}\lesssim10^{14}$ \cite{jcapa}
which is translated to an upper bound on $m_\phi$, due to our
initial condition $H_{\rm I}=m_\phi$. Note that this constraint is
roughly more restrictive that this which arises from the
requirement for the thermalization of the $\phi$-decay products.
Indeed, in order the decay products of the $\phi$-field are
thermalized within a Hubble time, through $2\to3$ processes we
have to demand $m_\phi\lesssim8\times10^{14}~{\rm GeV}$
\cite{sarkar}. The later is crucial so that
eqs.~(\ref{nf})--(\ref{nx}) are applicable. Finally, the bound of
eq.~(\ref{para}{\sf b}) comes from the arguments of the appendix
of ref.~\cite{moroi}.

\subsection{S{\ssz EMI}-A{\ssz NALYTICAL} A{\ssz PPROACH}}
\label{Seqs}

\hspace{.562cm} We can obtain a comprehensive and rather accurate
approach of the KRS dynamics, following the strategy of
refs.~\cite{scnr, jcapa}. Despite the fact that our main interest
is focused on the $q$-domination ($\rhofi<\rhoqi$) analyzed in
sec.~\ref{qdom}, we briefly review in sec.~\ref{fdom} the dynamics
of the decaying-$\phi$-domination ($\rhofi>\rhoqi$) for
completeness, clarity and better comparison. We first (see
sec.~\ref{norm}) introduce a set of normalized quantities which
simplify significantly the relevant formulas.

\subsubsection{Normalized quantities.\label{norm}} In terms of
the following dimensionless quantities:
\beq \label{vrhos}\mbox{\sf (a)}~~\vrho_{_{\rm
\phi[R]}}=\rho_{_{\rm \phi[R]}}/\rho^0_{\rm c },~~\mbox{\sf
(b)}~~\vVo=V_0/\rho^0_{\rm c}~~\mbox{and}~~\mbox{\sf
(c)}~~\vq=q/\sqrt{3}m_{_{\rm P}}.\eeq
eqs.~(\ref{qeq}) and (\ref{Hb}) take the form (we use $\rho^0_{\rm
c}=8.099\times10^{-47}h^2~{\rm GeV^4}$ with $h=0.72$):
%
%\beq \mbox{\sf (a)}~~\vQ=\vH\vq^\prime ~~\mbox{and}~~\mbox{\sf
%(b)}~~\vH\vQ^\prime+3\vH\vQ+\bar V_{,\bar
%q}=0~~\mbox{with}~~\mbox{\sf (c)}~~\vH^2\simeq\vrho_q+\vrho_{_{\rm
%R}}+\vrho_{\phi}, \label{vH} \eeq
%
%where the following quantities have been also defined:
%
%\beq \mbox{\sf (a)}~~\vV=\vVo e^{-\sqrt{3}\lambda
%\vq},~~\vH=H/H_0,~~\vQ=Q/\sqrt{\rho^0_{\rm
%c}}~~\mbox{and}~~\mbox{\sf
%(b)}~~\vrho_q=\vQ^2/2+\vV.\label{vrhoq}\eeq
%
\bea  \mbox{\sf (a)}~~\vQ=\vH\vq^\prime ~~\mbox{and}~~\mbox{\sf
(b)}~~\vH\vQ^\prime+3\vH\vQ+\bar V_{,\bar
q}=0~~\mbox{with}~~\mbox{\sf (c)}~~\vH^2\simeq\vrho_q+\vrho_{_{\rm
R}}+\vrho_{\phi},~~~~ \label{vH}\\
\mbox{where}~~\mbox{\sf (a)}~~\vV=\vVo e^{-\sqrt{3}\lambda
\vq},~~\vH=H/H_0,~~\vQ=Q/\sqrt{\rho^0_{\rm
c}}~~\mbox{and}~~\mbox{\sf
(b)}~~\vrho_q=\vQ^2/2+\vV.~~~~\label{vrhoq}\eea
We do not present the normalized forms of the residual
eqs.~(\ref{ff})-(\ref{fn}), since we do not use them in our
analysis below.

\subsubsection{Decaying-$\phi$-Domination.\label{fdom}} When
$\rhofi>\rhoqi$, the evolution of the universe at the epoch before
the completion of reheating, $T\gg T_{\rm RH}$, is dominated by
$\rho_{\phi}$ (transition to a $q$-dominated phase is not possible
since $\rho_q$ decreases steeper than $\rho_{\phi}$).
Consequently,
\beq \label{Hfdom}\mbox{\sf (a)}~~\vH\simeq\brhof^{1/2}~~\mbox{and
so, from eq.~(\ref{ff}) we obtain: {\sf (b)}}~~\vrho_\phi=\brhofi
e^{-3\btauf}.\eeq
Substituting eqs.~(\ref{Hfdom}) in eq.~(\ref{fR}) and ignoring the
last term in its right hand side (r.h.s), we can easily solve it,
with result:
\beq \label{rRfdom}\vrho_{_{\rm
R}}=\frac{2}{5}\vGamma_\phi\brhofi^{1/2}
\left(e^{-3\btauf/2}-e^{-4\btauf}\right)~~\mbox{with}~~
\vGamma_\phi=\Gamma_\phi/H_0. \eeq
The function $\vrho_{_{\rm R}}(\btau)$ (in accordance with
refs.~\cite{riotto,scnr}) reaches at
\beq \btau_{_{\rm max}}\simeq\ln(1.48)=0.39,~~\mbox{a maximum
value}~~\vrho_{_{\rm Rmax}}\simeq0.14\
\vGamma_\phi\brhofi^{1/2}.\label{rhomax}\eeq
with corresponding $T_{\rm max}$ derived through eq.~(\ref{rs}{\sf
a}). The completion of the reheating is realized at
$\btau=\btrhp$, such that:
\beq \label{trh} \rho_{_{\rm
R}}(\btrhp)=\rho_\phi(\btrhp)~\Rightarrow~\btrhp\simeq-
\frac{2}{3}\ln\frac{2}{5}\vGamma_\phi\brhofi^{-1/2}.\eeq
Equating the r.h.s of eqs.~(\ref{Hfdom}{\sf b}) and (\ref{rs}{\sf
a}) for $\btau=\btrhp$ and solving the resultant equation w.r.t
$\Gamma_\phi$ we get eq.~(\ref{GTrh}) with $T_\phi=T_{\rm RH}$. We
checked that the resulting prefactor 5 assists us to approach more
satisfactorily than in ref.~\cite{scnr} the numerical solution of
$\rho_\phi=\rho_{_{\rm R}}$.

\subsubsection{$q$-Domination.\label{qdom}} When
$\rhoqi>\rhofi$, an intersection of $\rho_q$ with $\rho_\phi$ at
$\btau=\btkp$ is possible, since $\rho_q$ decreases faster than
$\rho_\phi$. If this is realized, we obtain a KRS with partial
$q$-domination ($q$-PD) whereas if it is not, we obtain a KRS with
total (or complete) $q$-domination ($q$-TD). In either case, the
KRS terminates at $\btau=\btrh$ which is defined as the
commencement of the RD era (to avoid confusion with the case of
sec.~\ref{fdom} we do not use the symbol $\btrhp$, any more). In
particular,
\beq \label{btrh} \btrh=\btkr~~\mbox{for $q$-TD,
or}~~\btrh=\btpr~~\mbox{for $q$-PD,}\eeq
where at $\btkr~[\btpr]$ the transition from the KD
[$\phi$-dominated] to RD phase occurs.

For $q$-TD and $\btau<\btrh$ or $q$-PD and $\btau<\btkp$, the
cosmological evolution is dominated by $\rho_{q}$ in
eq.~(\ref{vrhoq}{\sf b}), where the term $\vV$ is negligible --
see eq.~(\ref{init}). Therefore,
\beq \label{Hqdom}\mbox{\sf (a)}~~\vH\simeq\brhoq^{1/2}~~\mbox{and
so, from eq.~(\ref{vH}{\sf b}) we obtain: {\sf
(b)}}~~\vrho_q=\brhoqi e^{-6\btauf}.\eeq
Assuming for a while that eq.~(\ref{Hfdom}{\sf b}) gives reliable
results for the $\rho_\phi$ evolution even with $q$-domination
(see sec.~\ref{sec:fdom}.b), we can achieve a first estimation for
the value of $\btkp$, solving the equation:
\beq \label{tkp}
\rho_{\phi}(\btkp)=\rho_q(\btkp)~\Rightarrow~\btkp\simeq\ln\left(\rho_{q_{\rm
I}}/\rho_{\phi_{\rm I}}\right)/3.\eeq

Obviously when $\btkp<\btrh$ or, equivalently,
$\rho_q(\btrh)<\rho_\phi(\btrh)$ we obtain a KRS with $q$-PD. Let
us assume that eq.~(\ref{trh}) gives reliable results for $\btrh$
in the case of $q$-PD (see sec.~\ref{sec:fdom}.b). If we insert
eqs.~(\ref{tkp}) and (\ref{trh}) in the first of the inequalities
above or eqs.~(\ref{Hqdom}{\sf b}) and (\ref{Hfdom}{\sf b}) in the
second of the inequalities above and take into account that
$\vH_{\rm I}=\brhoqi^{1/2}=m_\phi/H_0$, we can extract a condition
which discriminates the $q$-TD from the $q$-PD:
\beq \label{cond} \bar\Gamma_\phi>2.5\ \brhofi H_0/
m_\phi~~\mbox{for $q$-TD, or}~~\bar\Gamma_\phi\leq2.5\ \brhofi
H_0/ m_\phi~~\mbox{for $q$-PD}.\eeq
Apart from the numerical prefactor, the condition above agrees
with this of ref.~\cite{liddle}. Using the terminology of that
reference, during the $q$-TD [$q$-PD] the $\phi$-field decays
before [after] it becomes the dominant component of the universe.
In the following, we present the main features of the cosmological
evolution in each case, separately.

\paragraph{2.4.3.a Total $q$-Domination.\label{sec:qdom}} Substituting eqs.~(\ref{Hqdom})
into eq.~(\ref{ff}) we obtain:
\beq \label{rfqdom} \vrho_\phi\simeq \brhofi\
\exp\left(-3\btau-\frac{1}{3}c_{_{q\phi}}\
e^{3\btauf}\right)~~\mbox{with}~~c_{_{q\phi}}=\vGamma_\phi\
\brhoqi^{-1/2}=\Gamma_\phi/m_\phi. \eeq
The difference of the expression above from eq.~(\ref{Hfdom}{\sf
b}) is the presence of the second term in the exponent, which
(although does not cause dramatic changes in the early
$\rho_\phi$-evolution) is crucial for obtaining a reliable
semi-analytical expression for the $\vrho_{_{\rm R}}$ evolution
for $\btau\leq\btrh$. Indeed, inserting eqs.~(\ref{rfqdom}) in
eq.~(\ref{fR}) and ignoring the last term in its r.h.s, we end up
with the following:
\beq \label{rRqdom}\vrho_{_{\rm R}}\simeq c_{_{q\phi}}\ \brhofi\
I_{\rm R}\ e^{-4\btauf}~~\mbox{with}~~I_{\rm
R}(\btau)=\int_0^{\btauf} d\btau_{\rm i}\ \exp\left(4\btau_{\rm
i}-\frac{1}{3}c_{_{q\phi}} e^{3\btauf_{\rm i}}\right).\eeq

For relatively small $\btau$ (so as $4\btau\gg c_{_{q\phi}}
e^{3\btauf}/3$) the integration of $I_{\rm R}$ can be realized
analytically and we can derive the simplified formula:
\beq \label{rRqdoma}\vrho_{_{\rm R}}\simeq\frac{1}{4}c_{_{q\phi}}\
\brhofi\ \left(1-e^{-4\btauf}\right)~~\mbox{for}~~\btau\lesssim
\btau_{_{\rm PLf}},\eeq
where $\btau_{_{\rm PLf}}$ can be found by solving numerically the
equation $4\btau_{_{\rm PLf}}\simeq10 c_{_{q\phi}}
e^{3\btauf_{_{\rm PLf}}}/3$. From eq.~(\ref{rRqdoma}) we can
easily induce that the function $\rho_{_{\rm R}}(\btau)$ takes
rapidly (for $\btau\gtrsim\btau_{_{\rm RPL}}=1.5$) a maximal
plateau value:
\beq \rho_{_{\rm RPL}}\simeq\frac{1}{4}c_{_{q\phi}}\rho_{\phi_{\rm
I}} ~~\mbox{for}~~1.5\lesssim\btau\lesssim \btau_{_{\rm PLf
}},\label{rhomaxq}\eeq
Combining the previous expression with eq.~(\ref{rs}{\sf a}), we
can estimate quite accurately the constant, plateau value of
temperature:
\beq\label{Tpl} T_{\rm PL}\simeq \left(7.5\
c_{_{q\phi}}\rho_{\phi_{\rm I}}/\pi^2g_{\rho*}(T_{\rm
PL})\right)^{1/4}.\eeq

When $4\btau<c_{_{q\phi}} e^{3\btauf}/3$, the first term in the
exponent of eq.~(\ref{rRqdom}) can be neglected and the
$\vrho_{_{\rm R}}$ evolution can be approximated by the
expression:
\bea\nonumber\vrho_{_{\rm R}}&\simeq & \frac{1}{4}c_{_{q\phi}}
\brhofi\ e^{-4\btauf}\Big[ \left(e^{-4\btauf_{_{\rm
PLf}}}-1\right)\\ &+&\frac{1}{3}\left({\rm
Ei}(-\frac{c_{_{q\phi}}}{3}e^{3\btauf})-{\rm
Ei}(-\frac{c_{_{q\phi}}}{3}e^{3\btauf_{_{\rm
PLf}}})\right)\Big]~~\mbox{for}~~\btau\gtrsim \btau_{_{\rm
PLf}}\label{rRqdomb}\eea
where ${\rm Ei}(z)$ is the second exponential integral function,
defined as ${\rm Ei}(z)=-\int_{-z}^{\infty}e^{-t}/tdt$. However,
the contribution to $\vrho_{_{\rm R}}$ of the terms in the second
line of eq.~(\ref{rRqdomb}) turns out to be numerically
suppressed. Therefore, we can deduce that $\vrho_{_{\rm R}}$ for
$\btau\gtrsim \btau_{_{\rm PLf}}$ decreases as in the RD phase.

Using eqs.~(\ref{rRqdom}), (\ref{rfqdom}) and (\ref{Hqdom}{\sf b})
for solving numerically the equations:
\beq \label{eqs}\mbox{\sf (a)}~~\vrho_{_{\rm
R}}(\btpr)=\vrho_\phi(\btpr)~~\mbox{and}~~\mbox{\sf
(b)}~~\vrho_{_{\rm R}}(\btkr)=\vrho_q(\btkr),\eeq
we can determine the points $\btpr~[\btkr]$ where $\vrho_{_{\rm
R}}$ commences to dominates over $\vrho_\phi~[\vrho_q]$ (see
fig.~\ref{fig1} and table~\ref{t1}). In the present case, as we
anticipated in eq.~(\ref{btrh}), the hierarchy is $\btpr<\btkr$
and so, $\btkr$ can be identified as $\btrh$.

The $q$-evolution can be easily derived, inserting
eq.~(\ref{Hqdom}{\sf a}) into eq.~(\ref{vH}{\sf b}) and ignoring
the negligible third term in its l.h.s. Namely,
\beq\label{qk} \vq\simeq\sqrt{2}\ \btau~~(\Rightarrow
~\vq^\prime=\sqrt{2})~~\mbox{for}~~\btau\leq\btkr.\eeq
For $\btau>\btkr$, we have $\vH\simeq\vrho_{_{\rm R}}^{1/2}$.
Inserting the latter into eq.~(\ref{vH}{\sf b}), we can similarly
extract:
\beq\label{qfq} \vq=\vq_{_{\rm
KR}}+\sqrt{2}\left(1-e^{-(\btauf-\btauf_{_{\rm
KR}})}\right)~~\mbox{for}~~\btkr<\btau\eeq
where $\vq_{_{\rm KR}}=\vq(\btau_{_{\rm KR}})$. It is obvious from
eq.~(\ref{qfq}) that $q$ freezes at about $\btau_{_{\rm
KF}}\simeq\btau_{_{\rm KR}}+6$ to the following value:
\beq\label{qfqa} \vq_{_{\rm F}}\simeq\vq_{_{\rm
KR}}+\sqrt{2}~~(\Rightarrow
~\vq^\prime=0)~~\mbox{for}~~\btau_{_{\rm KF}}\leq\btau. \eeq
Comparing these results with those derived in the context of the
QKS (see ref.~\cite{jcapa}), we conclude that the presence of
$\phi$ modifies just the value of $\btau_{_{\rm KR}}$ -- due to
the presence of eq.~(\ref{rRqdom}) -- and so, it does not affect
crucially the $q$-evolution.

\paragraph{2.4.3.b Partial $q$-Domination. \label{sec:fdom}} When $\btau<\btkp$,
the $\rho_\phi~[\rho_q]$ evolution is given by
eq.~(\ref{Hqdom}{\sf b}) [eq.~(\ref{rfqdom})]. Equating the r.h.s
of these equations for $\btau=\btkp$, a more accurate result for
$\btkp$ can be achieved than this obtained from eq.~(\ref{tkp}).
However, the correction becomes more and more negligible as
$\brhoqi/\brhofi$ decreases (note that eq. (\ref{domk}) remains
always valid).

For $\btau>\btkp$, the $\vrho_\phi~[\vrho_{_{\rm R}}]$ evolution
takes the form that it has in the decaying-$\phi$-dominated era --
see eq.~(\ref{Hfdom}{\sf b}) [eq.~(\ref{rRfdom})]. Namely,
inserting eq.~(\ref{Hfdom}{\sf a}) into  eqs.~(\ref{ff}) and
(\ref{fR}) and integrating from $\btkp$ to $\btau>\btkp$, we
arrive at the following results:
%, ignoring the last term of the r.h.s of eq.~(\ref{fR})
\numparts \bea \label{rffdom} && \vrho_\phi=\vrho_\phi(\btkp)\
e^{-3(\btauf-\btfkp)}\\\mbox{and}~~&& \vrho_{_{\rm
R}}=\vrho_{_{\rm R }}(\btkp)\ e^{-4(\btauf-\btfkp)}
+\frac{2}{5}\vGamma_\phi\brhofi^{1/2}
e^{-3\btauf/2}\left(1-e^{-5(\btauf-\btfkp)/2}\right),\label{rRqpdom}\eea
\endnumparts
\hspace{-.14cm}where $\vrho_\phi(\btkp)$ [$\vrho_{_{\rm R
}}(\btkp)$] can be derived from eq.~(\ref{Hqdom}{\sf b})
[eq.~(\ref{rfqdom})].

Employing the expressions above and eq.~(\ref{Hqdom}{\sf b}) for
solving numerically eq.~(\ref{eqs}{\sf b}) [eq.~(\ref{eqs}{\sf
a})], we can determine the points $\btkr~[\btpr]$ where
$\vrho_{_{\rm R}}$ commences to dominate over
$\vrho_q~[\vrho_\phi]$ (see fig.~\ref{fig2} and table~\ref{t1}).
In the present case, as we anticipated in eq.~(\ref{btrh}), the
hierarchy is $\btkr<\btpr$ and so, $\btpr$ can be identified as
$\btrh$. The resultant $\btpr$ approaches $\btrhp$ obtained by
eq.~(\ref{trh}) as $\brhoqi/\brhofi$ decreases.

As regards the $q$-evolution, this obeys eq.~(\ref{qk}) for
$\btau\leq\btkp$. For $\btau>\btkp$, inserting
eq.~(\ref{Hfdom}{\sf a}) into eq.~(\ref{vH}{\sf b}) and ignoring
the negligible third term in its l.h.s, we can extract:
\beq\label{qff} \vq=\vq_{_{\rm
K\phi}}+\frac{2}{3}\sqrt{2}\left(1-e^{-3(\btauf-\btauf_{_{\rm
K\phi}})/2}\right)~~\mbox{for}~~\btkp<\btau\eeq
where $\vqkp=\vq(\btkp)\simeq\sqrt{2}\ln\left(\rho_{q_{\rm
I}}/\rho_{_{\rm \phi I}}\right)/3$. It is obvious from
eq.~(\ref{qff}) that $q$ freezes at about $\btau_{_{\rm
KF}}\simeq\btau_{_{\rm K\phi}}+3$ to the following value:
%(possible correction due to the use of eq.~(\ref{rffdom})
%is numerically negligible)
\beq\label{qfff} \vq_{_{\rm
F}}\simeq\vqkp+2\sqrt{2}/3~~(\Rightarrow
~\vq^\prime=0)~~\mbox{for}~~\btkp\leq\btau. \eeq
Comparing these results with those derived in the previous case,
we conclude that for $q$-PD, $q$ takes its constant value during
the $\phi$-dominated epoch and that $\vq_{_{\rm F}}-\vqkp$ is less
than $\vq_{_{\rm F}}-\vq_{_{\rm KR}}$ in the $q$-TD. On the other
hand, $\rho_q$ continues its evolution according to
eq.~(\ref{Hqdom}{\sf b}) until $\btau_{_{\rm PL}}\gg\vtau_{_{\rm
KF}}$ (note that $\vq^\prime(\btau_{_{\rm KF}})=0$ but
$\vQ(\btau_{_{\rm KF}})\neq0$) where $\vrho_q$ reaches its
constant value, $\vrho_{q_{\rm F}}=\vV(q_{_{\rm F}})$. The point
$\btau_{_{\rm PL}}$ can be easily found (compare with
ref.~\cite{jcapa}):
\beq \vQ^2(\btau_{_{\rm PL}})/2=\vV(\vq_{_{\rm
F}})~\Rightarrow~\btau_{_{\rm PL}}=\lambda\vq_{_{\rm
F}}/2\sqrt{3}-\ln(\vVo/\vrho_{q_{\rm I}})/6.\eeq

\section{C{\ftn OLD} D{\ftn ARK} M{\ftn ATTER} A{\ftn BUNDANCE}}
\label{sec:neut}
\hspace{.562cm} Our final aim is the $\Omega_{\chi}h^2$
calculation, which is based on the well known formula
\cite{gelmini}:
\begin{equation}
\label{om1} \Omega_{\chi}=\rho_{\chi}^0/\rho_{\rm c}^0=
(s_0/\rho_{\rm c}^0)(n_\chi/s)(\btau_{_{\rm
f}})m_{\chi}~\Rightarrow~\Omega_{\chi}h^2= 2.741 \times 10^8\
(f_\chi/sR^3)(\btau_{_{\rm f}})\ m_{\chi}/\mbox{GeV},
\end{equation}
where a background radiation temperature of $T_0=2.726~^0$K is
taken for the computation of $s_0$ and $\rho_{\rm c}^0$ and
$\btau_{_{\rm f}}\sim 25$ is chosen large enough so as $f_\chi$ is
stabilized to a constant value (with $g$'s fixed to their values
at $T_\phi$). Based on the semi-analytical expressions of
sec.~\ref{Seqs}, we can proceed to an approximate computation of
$f_\chi$, which facilitates the understanding of the problem and
gives, in most cases, accurate results.

We assume that $\chi$'s are thermalized with plasma (see
sec.~\ref{resc}) and non-relativistic ($m_\chi>T$) around the
`critical' points $\btau_\ast$ or $\btau_{_{\rm F}}$ (see below).
Our semi-analytical treatment relies on the reformulated Boltzmann
eq.~(\ref{fn}). Note that, due to the prominent
constant-temperature phase during the KRS, we are not able to
absorb the dilution term of eq.~(\ref{nx}) by defining a variable
$Y=n_\chi/s^{\nu_s}$ with $\nu_s$ dependent on the form of $T-R$
relation as, e.g., in refs.~\cite{gelmini, scnr, jcapa}. However,
we checked that our present analysis based on eq.~(\ref{fn}) --
see also ref.~\cite{riotto} -- is quite generic and applicable to
any other case.

Since $f_\chi$ is stabilized to a constant value at large enough
$\btau_{_{\rm f}}\sim 25$, we find it convenient to split the
semi-analytical integration of eq.~(\ref{fn}) into two distinct,
successive regimes: One for $0<\btau\leq\btrh$ (see
sec.~\ref{before}) and one for $\btrh<\btau\leq\btau_{_{\rm f}}$
(see sec.~\ref{after} \cite{dreesn}). In both regimes we single
out two fundamental subcases: $\chi$'s do (sec.~\ref{sec:eq} and
\ref{sec:eqa}) or do not maintain (secs.~\ref{sec:noneq} and
\ref{sec:noneqa}) chemical equilibrium with plasma. In the latter
case for $0<\btau\leq\btrh$, two extra subcases can be
distinguished: The type I and II non-EP. The conditions which
discriminates the various possibilities are specified in
sec.~\ref{sec:noneq}. Note that in our investigation we let open
the possibility that $\chi$'s remain in chemical equilibrium even
after the onset of RD era.

\subsection{T{\ssz HE} E{\ssz VOLUTION}  B{\ssz EFORE THE}
O{\ssz NSET OF THE} RD E{\ssz RA}} \label{before}

\hspace{.562cm} During this regime, $H$ can be sufficiently
approximated by eq.~(\ref{Hqdom}{\sf a}) [eq.~(\ref{Hfdom}{\sf a})
and eq.~(\ref{rffdom})] for $q$-TD or $q$-PD and $\btau\leq\btkp$
[for $q$-PD and $\btau\geq\btkp$]. Also, $\rho_\phi$, involved in
eq.~(\ref{fn}), can be found from eq.~(\ref{rfqdom})
[eq.~(\ref{rffdom})] for $q$-TD or $q$-PD and $\btau\leq\btkp$
[for $q$-PD and $\btau\geq\btkp$]. Finally, $T$ (involved in the
computation of $f_\chi^{\rm eq}$) can be found by plugging
eq.~(\ref{rRqdom}) [eq.~(\ref{rRqpdom})] for $q$-TD or $q$-PD and
$\btau\leq\btkp$ [for $q$-PD and $\btau\geq\btkp$] in
eq.~(\ref{rs}{\sf a}). As regards the $\Omega_\chi h^2$
calculation, we can distinguish the following cases:

\subsubsection{Non-Equilibrium Production. \label{sec:noneq}}
In this case, $f_\chi\gg f_\chi^{\rm eq}$ for any $\btau<\btrh$
(type I non-EP) or $f_\chi\ll f_\chi^{\rm eq}$ for any
$\btau<\btau_\ast$ (type II non-EP). Let us consider each subcase
separately:

\paragraph{3.1.1.a Type I.} Obviously the realization of this
situation ($f_\chi\gg f_\chi^{\rm eq}$ for any $\btau<\btrh$)
requires $N_\chi\neq0$, (since if $N_\chi=0$, the maximal possible
value of $f_\chi$ is $f_\chi^{\rm eq}$). Such a suppression of
$f_\chi^{\rm eq}$ can be caused if $T_{\rm PL}\ll m_\chi/20$, as
we can deduce from eqs.~(\ref{fdef}) and (\ref{neq}). Since
$f_\chi^2-f_\chi^{\rm eq2}\simeq f_\chi^{2}$, eq.~(\ref{fn}) takes
the form:
\beq\label{Ifn} HR^3f^\prime_{\chi}= -\langle\sigma v\rangle\
f_{\chi}^2 +\Gamma_\phi N_{\chi} \rho_\phi R^6/ \Delta_\phi
m_\phi, \eeq
which can be solved numerically from $\btau=0$ to $\btrh$. In most
cases (see fig.~\ref{om}) the second term in the r.h.s of
eq.~(\ref{Ifn}) dominates over the first one and so, we can
analytically derive:
\beq \label{Ifnsol} f^{\rm
RD}_\chi=f^{N}_\chi(\btrh)~~\mbox{with}~~f^{N}_\chi(\btau)=\frac{\Gamma_\phi
N_\chi}{\Delta_\phi m_\phi}\int_0^{\btauf}d\btau_{\rm i}\
\frac{\rho_\phi R^3}{H}\cdot\eeq
The integration above can be realized numerically.

\paragraph{3.1.1.b Type II.} When
$T_{\rm PL}\sim m_\chi/20$, $f_\chi^{\rm eq}$ is not strongly
suppressed and so, the condition $f_\chi\ll f_\chi^{\rm eq}$ can
be achieved. Since  $f_\chi^2-f_\chi^{\rm eq2}\simeq -f_\chi^{\rm
eq2}$, eq.~(\ref{fn}) takes the form:
\beq\label{IIfn} HR^3f^\prime_{\chi}= \langle\sigma v\rangle\
f_{\chi}^{\rm eq2}+\Gamma_\phi N_{\chi} \rho_\phi R^6/ \Delta_\phi
m_\phi. \eeq
Integrating the latter from $\btau=0$ to $\btrh$ we arrive at
$f^{\rm RD}_\chi=f^{nN}_\chi(\btrh)$ with:
\beq\label{IIfnsol} \mbox{\sf (a)}~~
f^{nN}_\chi(\btau)=f^{n}_\chi(\btau)+f^{N}_\chi(\btau),~~\mbox{where}~~
\mbox{\sf (b)}~~f^{n}_\chi(\btau)=\int_0^{\btauf}d\btau_{\rm i}\
\sv\ f_{\chi}^{\rm eq2}/ HR^3 \eeq
and $f^{N}_\chi$ is found from eq.~(\ref{Ifnsol}). We observe that
the integrand of $f^{n}_\chi$ reaches its maximum at
$\btau=\btast$, where the maximal $\chi$-particles production
takes place. Therefore, let us summarize the conditions which
discriminates the EP from the non-EP of $\chi$'s:
\beq \label{condf} \left(f_{\chi}^{\rm
eq}/f^{nN}_{\chi}\right)(\btau_\ast)\left\{\matrix{
%\begin{array}{rl}
> 3\hfill ,& \mbox{type II non-EP (non-EPII)} \hfill \cr
< 3~~\mbox{and}~~\left(f^{N}_{\chi}/f_{\chi}^{\rm
eq}\right)(\btau_\ast)>10\hfill ,& \mbox{type I non-EP (non-EPI)}
\hfill \cr
\leq3~~\mbox{and}~~\left(f^{N}_{\chi}/f_{\chi}^{\rm
eq}\right)(\btau_\ast)\leq10 \hfill,& \mbox{EP} \hfill \cr}
%\end{array}
\right. \eeq
where the numerical values are just empirical, derived by
comparing the results of the numerical solution of
eqs.~(\ref{fq})-(\ref{fn}) with those obtained by the solution of
eq.~(\ref{Ifn}) or eq.~(\ref{IIfn}).

\subsubsection{Equilibrium Production. \label{sec:eq}} In this
case, we introduce the notion of the freeze-out temperature,
$T_{\rm F}=T(\btf)=x_{_{\rm F}}m_{\chi}$ \cite{kolb, gelmini},
which assists us to study eq.~(\ref{fn}) in the two extreme
regimes:

$\bullet$ At very early times, when $\btau\ll \btf$, $\chi$'s are
very close to equilibrium. So, it is more convenient to rewrite
eq.~(\ref{fn}) in terms of the variable
$\Delta(\btau)=f_\chi(\btau)-f^{\rm eq}_\chi(\btau)$ as follows:
\beq \label{deltaBE} \Delta^{\prime}=-{f_\chi^{\rm
eq}}^{\prime}-\sv \Delta\left(\Delta+2f_\chi^{\rm
eq}\right)/HR^3+\Gamma_\phi N_{\chi} \rho_\phi R^3/ H\Delta_\phi
m_\phi. \eeq
The freeze-out point $\btf$ can be defined by
\beq \Delta(\btf)=\delta_{\rm F}\>f_\chi^{\rm eq}(\btf)
\Rightarrow \Delta(\btf)\Big(\Delta(\btf)+2f^{\rm
eq}(\btf)\Big)=\delta_{\rm F}(\delta_{\rm F}+2)\ f_\chi^{\rm
eq2}(\btf), \label{Tf} \eeq
where $\delta_{\rm F}$ is a constant of order one, determined by
comparing the exact numerical solution of eq.~(\ref{fn}) with the
approximate under consideration one. Inserting eqs.~(\ref{Tf})
into eq.~(\ref{deltaBE}), we obtain the following equation, which
can be solved w.r.t $\btf$ iteratively:
\bea \nonumber \Big(\ln f_\chi^{\rm eq}\Big)^\prime(\btf)
&=&-\sv\delta_{\rm F} (\delta_{\rm F}+2) f_\chi^{\rm
eq}(\vtf)/(\delta_{\rm F}+1)HR^3\\&+&\Gamma_\phi N_{\chi}
\rho_\phi R^3/(\delta_{\rm F}+1) H f_\chi^{\rm
eq}(\vtf)\Delta_\phi m_\phi \label{xf}
\\ \label{xfa} \mbox{with}~~\Big(\ln f_\chi^{\rm
eq}\Big)^{\prime}(\btau)&=&3+x^\prime\frac{(16+3x)(18+25x)}{2x^2(8+15x)}\cdot
\eea
Normally, the correction to $\btf$ due to the second term in the
r.h.s of eq.~(\ref{xf}) is negligible.

$\bullet$ At late times, when $\btau\gg \btf$, $f_\chi\gg
f_\chi^{\rm eq}$ and so, $f_\chi^2-f_\chi^{\rm eq2}\simeq
f_\chi^2$. Substituting this into eq.~(\ref{fn}), the value of
$f_\chi$ at $\btrh$, $f_\chi^{\rm RH}=f_\chi(\btrh)$ can be found
by solving eq.~(\ref{Ifn}) from $\btau=\btf$ until $\btau=\btrh$
with initial condition $f_\chi(\btf)=(\delta_{\rm F}+1)f^{\rm eq
}_\chi(\btf)$. However, when $N_{\chi}=0$ or the first term in the
r.h.s of eq.~(\ref{Ifn}) dominates over the second, an analytical
solution of eq.~(\ref{Ifn}) can be easily derived. Namely
$f_\chi^{\rm RD} =f_\chi^{\rm F}(\btrh)$, where:
\beq \label{BEsol} \mbox{\sf (a)}~~f_\chi^{\rm F}(\btau) =
\left(f_\chi(\btf)^{-1}+J_{\rm F}(\btau)
\right)^{-1}~~\mbox{with}~~\mbox{\sf (b)}~~J_{\rm F}(\btau)=
\int_{\btauf_{_{\rm F }}}^{\btauf} d\btau_{\rm i}\
\frac{\sv}{HR^3}\cdot\eeq
The choice $\delta_{\rm F}=1.0\mp0.2$ provides the best agreement
with the precise numerical solution of eq.~(\ref{fn}), without to
cause dramatic instabilities.

\subsection{T{\ssz HE} E{\ssz VOLUTION} A{\ssz FTER THE}
O{\ssz NSET OF THE} RD E{\ssz RA}} \label{after}

\hspace{.562cm} During this regime, the cosmological evolution is
assumed to be RD, and so, $\vH=\sqrt{\vrho_{_{\rm R}}}$. The
evolution of $\rho_\phi$ and $\rho_{_{\rm R}}$ is sufficiently
approximated by the following expressions:
\numparts \begin{eqnarray}&& \vrho_\phi=\vrho_\phi(\btrh)\
\exp\left(-3(\btau-\btrh)-\frac{5}{4}\left(\frac{T_\phi}{T_{\rm
RD}}\right)^2\left(e^{2(\btauf-\btauf_{_{\rm RD
}})}-1\right)\right) \label{rfrdom}
\\\mbox{and}~~&& \vrho_{_{\rm R}}=\vrho_{_{\rm R}}(\btrh)\
e^{-4(\btauf-\btauf_{_{\rm RD }})},\label{rRrdom}
\end{eqnarray}
\endnumparts
\hspace{-.14cm}where $T_{\rm RD}$ corresponds to $\btrh$ defined
in eq.~(\ref{btrh}) and $\vrho_\phi(\btrh)$ is evaluated from
eq.~(\ref{rfqdom}) [eq.~(\ref{rffdom})] for $q$-TD [$q$-PD] while
$\vrho_{_{\rm R }}(\btrh)$ is found from eq.~(\ref{rRqdom})
[eq.~(\ref{rRqpdom})] for $q$-TD [$q$-PD]. In order to prove
eq.~(\ref{rfrdom}), we start from the exact solution of
eq.~(\ref{nf}) \cite{turner, kolb} which includes besides the
terms of eq.~(\ref{Hfdom}{\sf b}) an extra exponential term. We
replace the involved temporal difference, in the latter term, by
the corresponding temperature one (using the time-temperature
relation \cite{kolb} in the RD era) and $\Gamma_\phi$ by
eq.~(\ref{GTrh}), as follows:
\beq \Gamma_\phi(t-t_{_{\rm RD}})=\left(\frac{T_\phi}{T_{\rm RD
}}\right)^2\left(\frac{T_{\rm
RD}^2}{T^2}-1\right)=\left(\frac{T_\phi}{T_{\rm RD
}}\right)^2\left(e^{2(\btauf-\btauf_{_{\rm RD }})}-1\right)\eeq
where in the last step we have used the entropy conservation law,
eq.~(\ref{dtau}) and the fact that we do not expect change of
$g_{s*}$ between $T_{\rm RD}$ and $T_\phi$. Finally, $T$ (involved
in the computation of $f_\chi^{\rm eq}$) can be found by plugging
eq.~(\ref{rRrdom}) in eq.~(\ref{rs}{\sf a}).

As regards the $\Omega_\chi h^2$ calculation, we can distinguish
the following cases:

\subsubsection{Non-Equilibrium Production. \label{sec:noneqa}}
In this case, $f_\chi\gg f^{\rm eq}_\chi$ for $\btau>\btrh$. This
is the usual case we meet, when the $f_\chi$ evolution for
$\btau\leq\btrh$ has been classified in one of the cases
elaborated in secs.~\ref{sec:noneq} and ~\ref{sec:eq}. For
$\btau>\btrh$, the $f_\chi$ evolution obeys eq.~(\ref{Ifn}) with
$\rho_\phi~[\rho_{_{\rm R}}]$ given by eq.~(\ref{rfrdom})
[eq.~(\ref{rRrdom})]. This equation can be solved numerically from
$\btrh$ until $\btfn$ with initial condition $f_\chi(\btrh)=f^{\rm
RD}_\chi$ derived as we described in sec.~\ref{before}. Under some
circumstances an analytical solution can be, also, presented.
Namely,
\bea \label{BEsola} && ~~~~~~~~~~~~~~~~~~f_\chi^0 =f^{\rm f
}_\chi(\btfn)~~\mbox{with}~~f^{\rm f }_\chi(\btau)=
\left(f_\chi(\btrh)^{-1}+J_{\rm RD}(\btau)
\right)^{-1},~~\mbox{where}~~\\ && \mbox{\sf (a)}~~J_{\rm
RD}(\btau)= \int_{\btauf_{_{\rm RD }}}^{\btauf} d\btau\
\frac{\sv}{HR^3}~~\mbox{or}~~\mbox{\sf (b)}~~J_{\rm RD}(\btau)=
\frac{\Gamma_\phi N_\chi}{\Delta_\phi m_\phi}\int_{\btauf_{_{\rm
RD }}}^{\btauf} d\btau_{\rm i}\ \frac{\rho_\phi R^3}{H}\cdot
\label{Jfa}\eea
Eq.~(\ref{Jfa}{\sf a}) is applicable for $N_\chi=0$ or for $\sv\
f^{\rm RD}_\chi\gg (\Gamma_\phi N_\chi/\Delta_\phi m_\phi)
(\rho_\phi R^6)(\btrh)$ whereas eq. ~(\ref{Jfa}{\sf b}) is valid
when $\sv\ f^{\rm RD}_\chi\ll (\Gamma_\phi N_\chi/\Delta_\phi
m_\phi) (\rho_\phi R^6)(\btrh)$.

\subsubsection{Equilibrium Production. \label{sec:eqa}} In
this case, $f_\chi\sim f^{\rm eq}_\chi$ for some $\btau>\btrh$.
This can be considered as an exceptional case , since it can not
be classified in any of the cases investigated in
sec.~\ref{before} and is met for $q$-PD with very low
$\Omega_q(\vtns)$. The $\Omega_\chi h^2$ calculation is based on
the procedure described in sec.~\ref{sec:eq}. In particular,
eq.~(\ref{xf}) and (\ref{xfa}) are applicable inserting into them
eqs.~(\ref{rfrdom}) and (\ref{rRrdom}). The limits of the
integration of eq.~(\ref{Ifn}) are from $\btf$ to $\btfn$ in this
case. Under the circumstances mentioned in sec.~\ref{sec:eq}, an
analytical solution can be obtained too. Namely, $f^0_\chi=f^{\rm
F}(\btfn)$, with $f^{\rm F}(\btau)$ given by eq.~(\ref{BEsol}).

%\newpage

\section{N{\ssz UMERICAL} A{\ssz PPLICATIONS}} \label{ap}

\begin{figure}[!h]\vspace*{-.19in}
\hspace*{-.71in}
\begin{minipage}{8in}
\epsfig{file=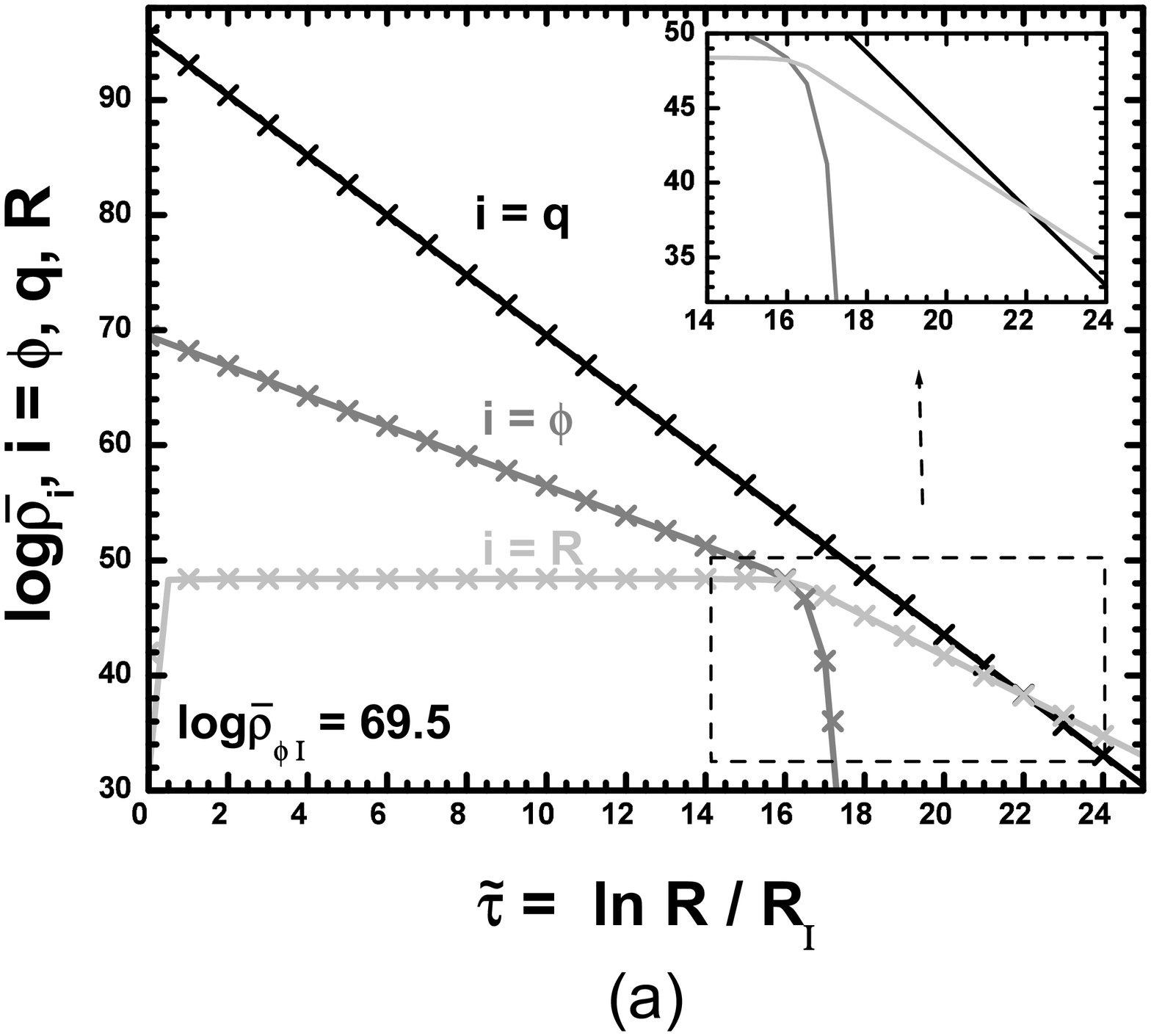,height=3.8in,angle=-90} \hspace*{-1.37 cm}
\epsfig{file=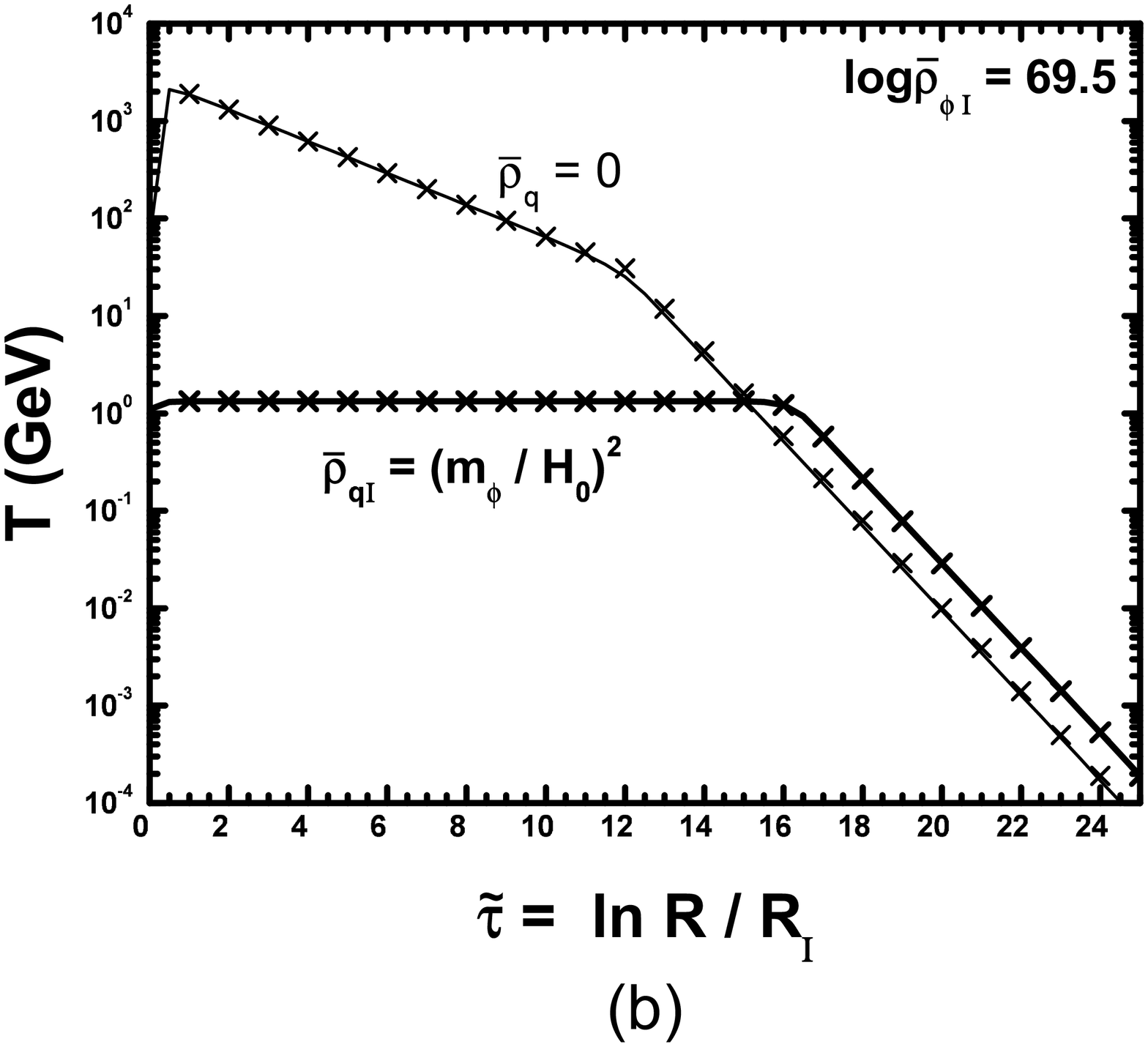,height=3.8in,angle=-90} \hfill
\end{minipage}\vspace*{-.01in}
\hfill\hspace*{-.71in}
\begin{minipage}{8in}
\epsfig{file=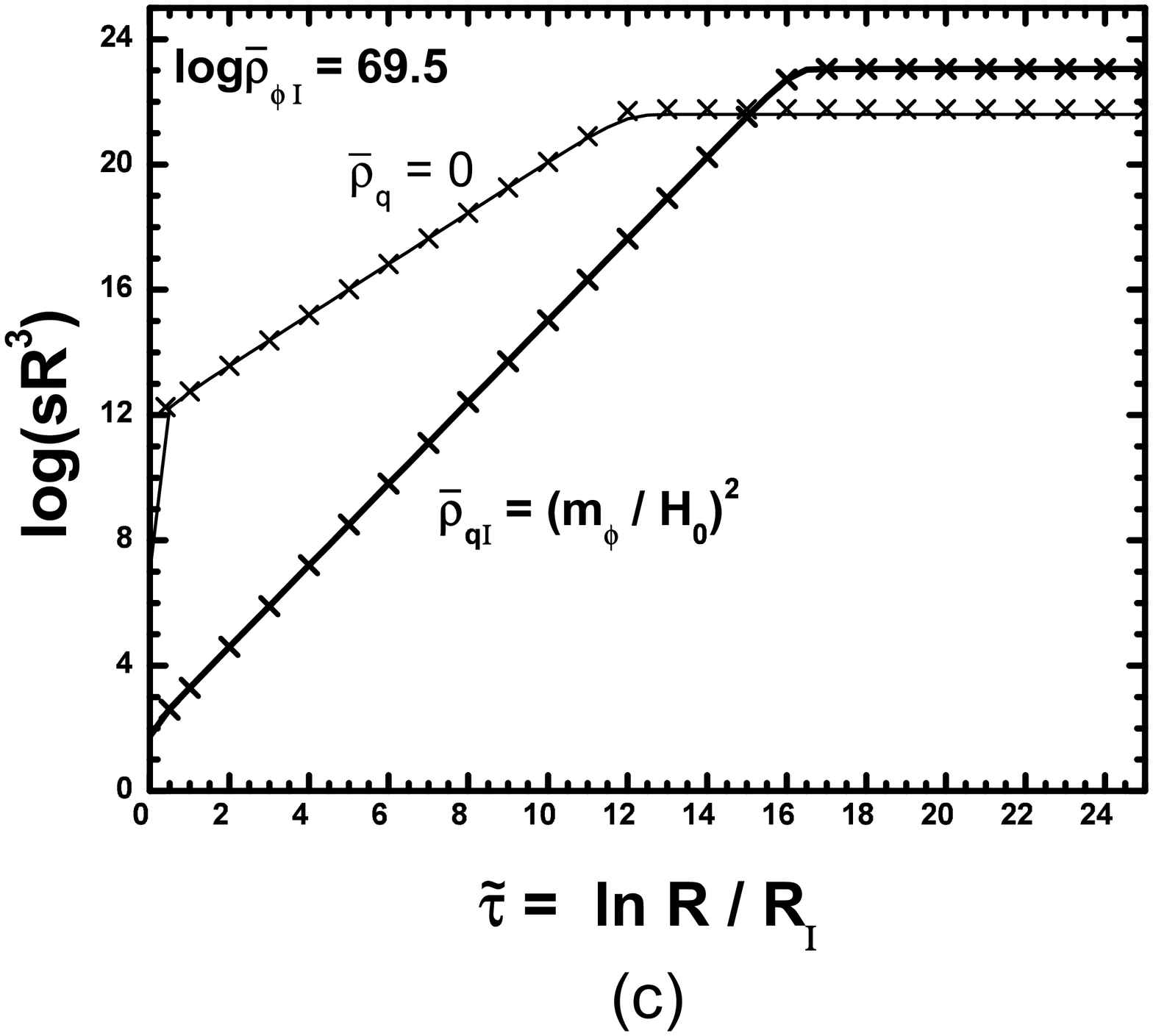,height=3.8in,angle=-90} \hspace*{-1.37 cm}
\epsfig{file=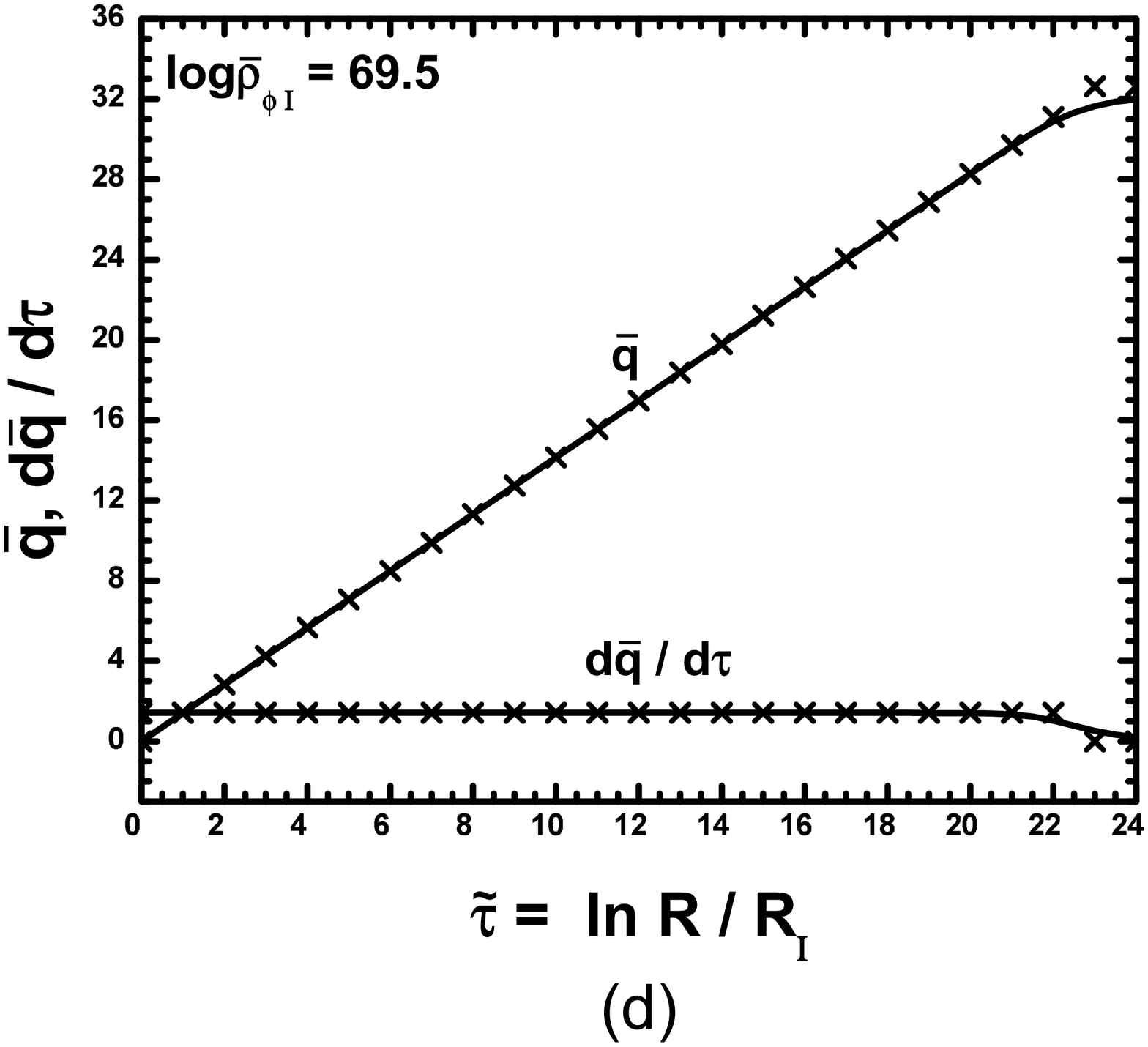,height=3.8in,angle=-90} \hfill
\end{minipage}
\hfill \caption[]{\sl\ftn The evolution as a function of $\btauf$
for $m_\phi=10^6~{\rm GeV}$, $T_\phi=30~{\rm GeV}$ and
$\log\brhofi=69.5$ of the quantities: $\log\bar\rho_i$ with $i=q$
(black line and crosses), $i=\phi$ (gray line and crosses), $i=R$
(light gray line and crosses) {\sf (a)}, $T$ for
$\brhoqi=(m_\phi/H_0)^2~[\brhoq=0]$ (bold [thin] line and crosses)
{\sf (b)}, $\log(sR^3)$ for $\brhoqi=(m_\phi/H_0)^2~[\brhoq=0]$
(bold [thin] line and crosses) {\sf (c)},  $\vq$ and $\vq^\prime$
{\sf (d)}. The solid lines [crosses] are obtained by our numerical
code [semi-analytical expressions].} \label{fig1}\vspace*{-.17in}
\end{figure}

\begin{figure}[!ht]\vspace*{-.19in}
\hspace*{-.72in}
\begin{minipage}{8in}
\epsfig{file=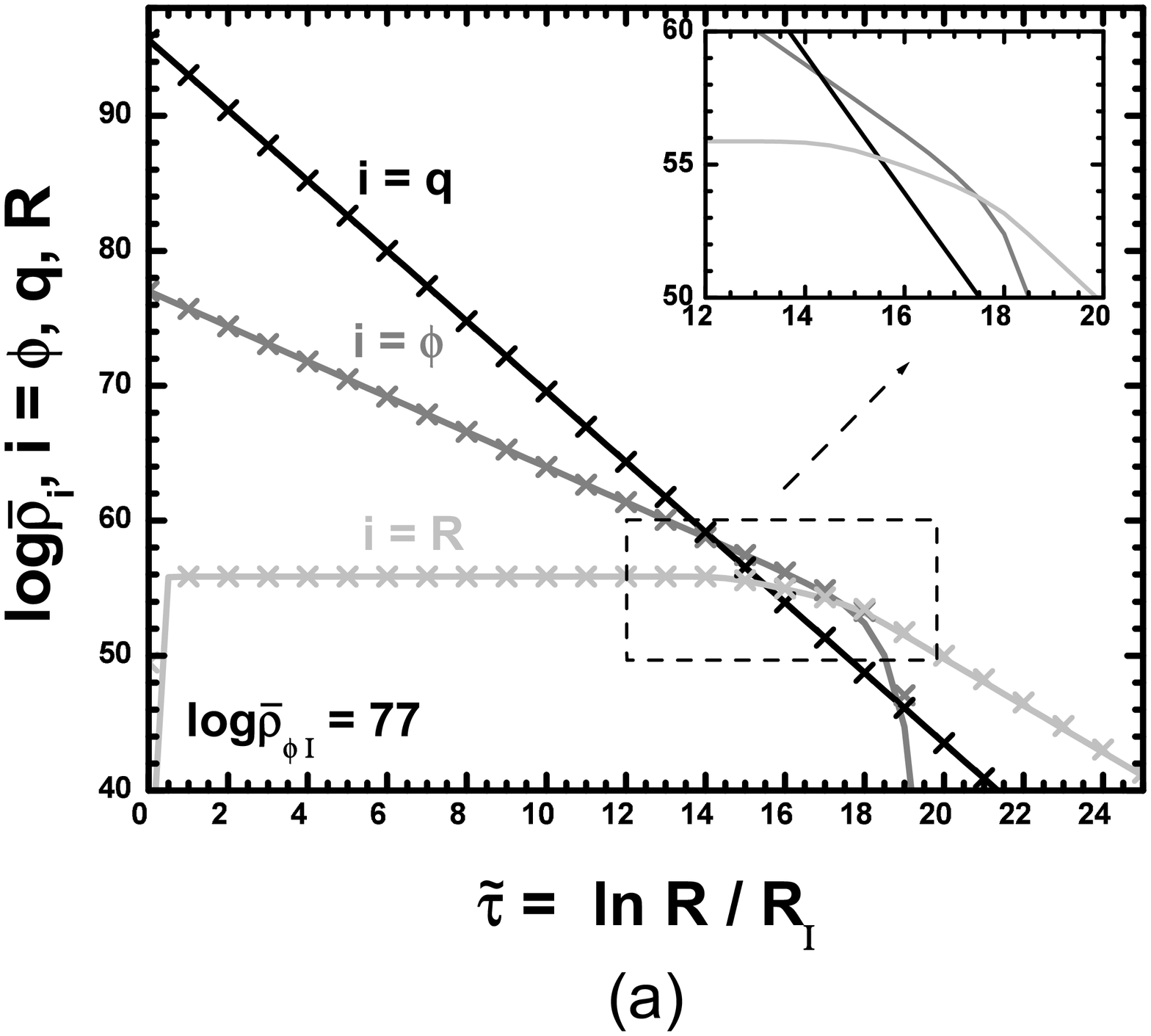,height=3.8in,angle=-90}\hspace*{-1.36 cm}
\epsfig{file=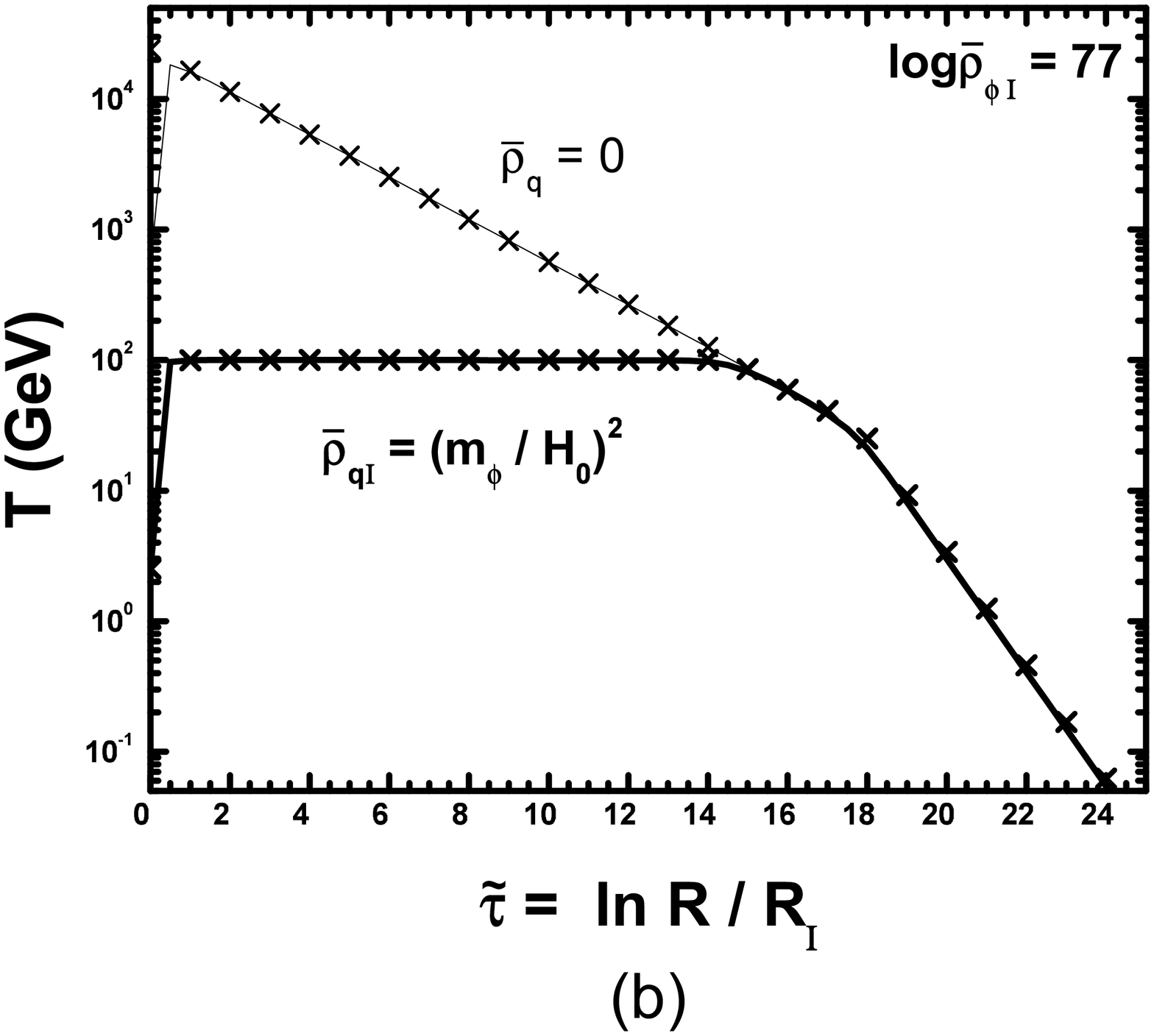,height=3.8in,angle=-90} \hfill
\end{minipage}\vspace*{-.01in}
\hfill\hspace*{-.71in}
\begin{minipage}{8in}
\epsfig{file=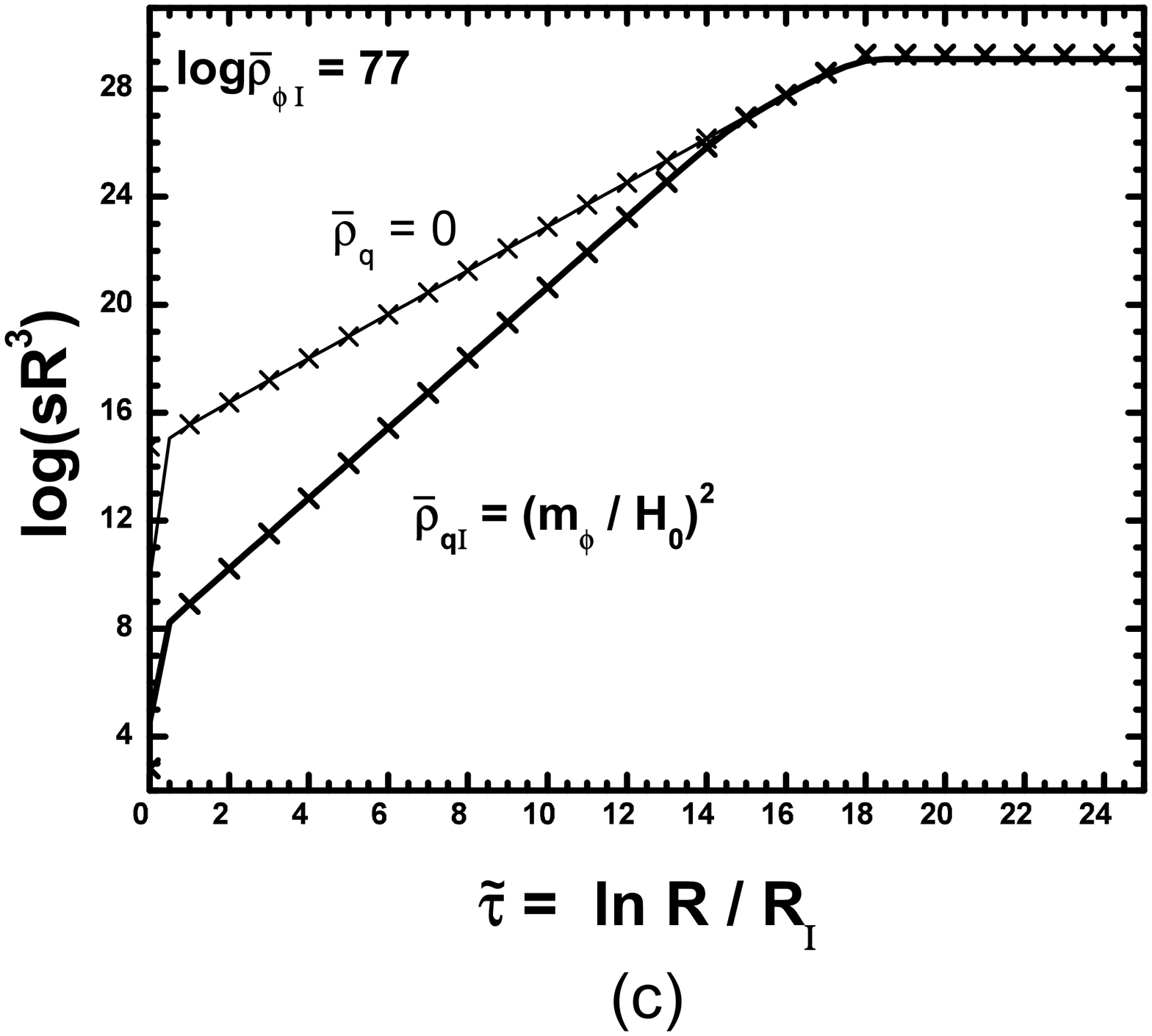,height=3.8in,angle=-90} \hspace*{-1.37 cm}
\epsfig{file=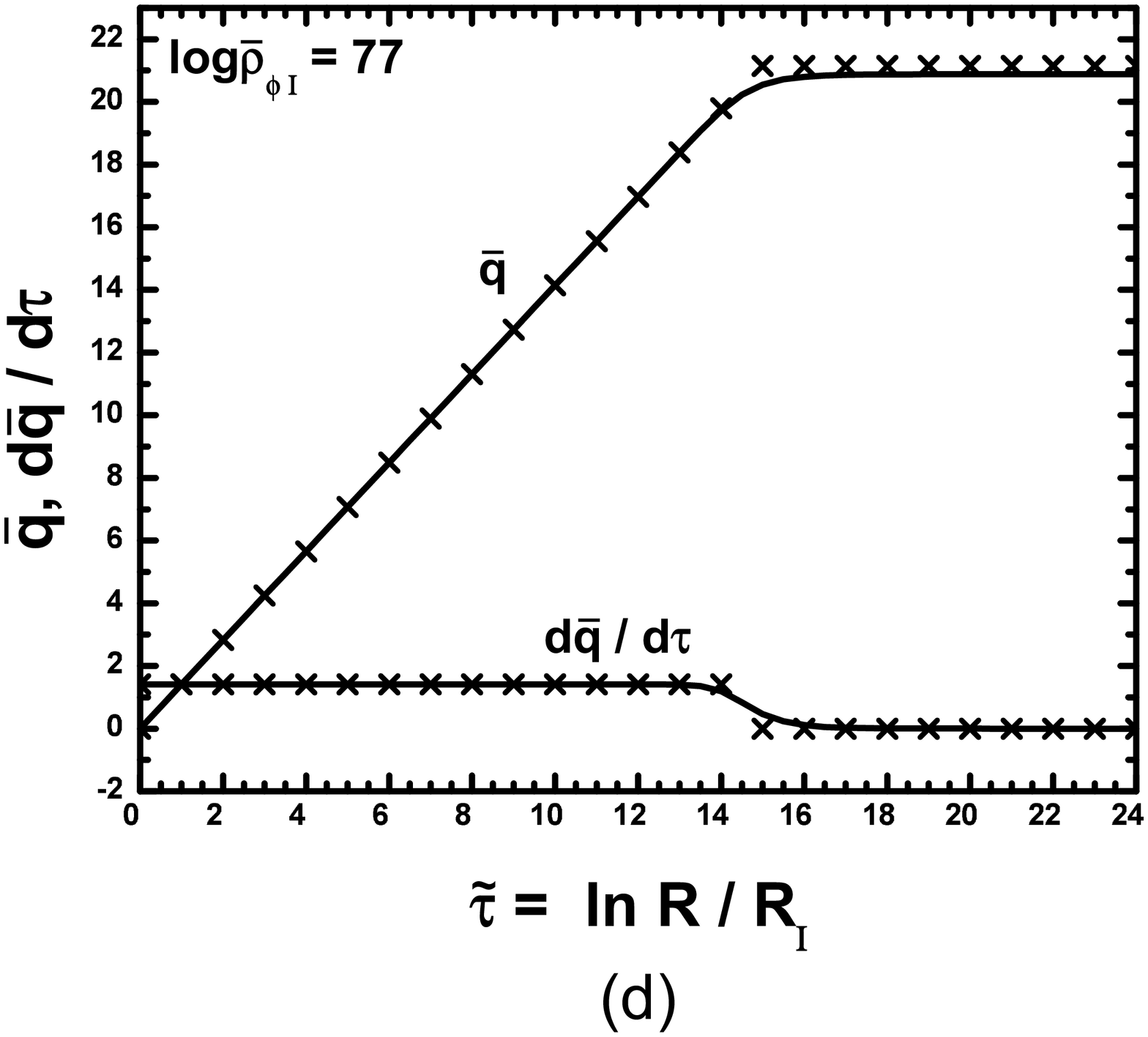,height=3.8in,angle=-90} \hfill
\end{minipage}
\hfill \caption[]{\sl\ftn The same as in fig.~\ref{fig1} but with
$\log\brhofi=77$.} \label{fig2}\vspace*{-.17in}
\end{figure}

\hspace{.562cm} Our numerical investigation depends on the
parameters:
$$\lambda,\ \brhofi,\ m_\phi,\ T_\phi,\ N_{\chi},\
m_{\chi},\ \sv.$$%
Recall that we use $q(0)=0$ and $\brhoqi=(m_\phi/H_0)^2$ (which is
equivalent with $H_{\rm I}=m_\phi$) throughout. Also, \vVo\ is
adjusted so that eq.~(\ref{rhoq0}) is satisfied (we present the
used $\vVo$'s in the explicit examples of figs.~\ref{fig1},
\ref{fig2} and \ref{fig3}).

In order to reduce further the parameter space of our
investigation, we make three extra simplifications. In particular,
since $\lambda$ determines just the value of $w_q$ during the
attractor dominated phase of the $q$-evolution \cite{jcapa} and
has no impact on the $\Omega_\chi h^2$ calculation we fix
$\lambda=0.5$. Note that agreement with eq.~(\ref{wq}{\sf a})
entails $0<\lambda\lesssim1.15$ \cite{jcapa}. Furthermore, since
it is well known that, in any case, $\Omega_\chi h^2$ increases
with $m_{\chi}$ (see, e.g., refs.~\cite{scnr, jcapa}) we decide to
fix it also to a representative value. Keeping in mind that the
most promising CDM particle is the LSP and the allowed by several
experimental constraints and possibly detectable in the future
experiments (see, e.g. ref.~\cite{munoz}) range of its mass is
about $(200-500)~{\rm GeV}$ (see, e.g., fig. 23 of
ref.~\cite{wmapl}), we take $m_{\chi}=350~{\rm GeV}$.

Let us clarify once more that \sv\ can be derived from $m_{\chi}$
and the residual (s)-particle spectrum, once a specific theory has
been adopted. To keep our presentation as general as possible, we
decide to treat $m_{\chi}$ and $\langle\sigma v\rangle$ as
unrelated input parameters, following the strategy of
refs.~\cite{scnr, jcapa}. We focus on the case $\sv=a$ which
emerges in the majority of the particle models (see, e.g.,
refs.~\cite{lkk, su5, wells} and \cite{edjo}-\cite{lah}). We do
not consider the form $\sv=bx$ which is produced in the case of a
bino LSP \cite{Cmssm} without coannihilations \cite{ellis2,
boehm}. However, our numerical and semi-analytical procedure (see
secs.~\ref{Neqs}, \ref{before} and \ref{after}) is totally
applicable in this case also with results rather similar to those
obtained for $\sv=a$, as we showed in ref.~\cite{jcapa}.

The presentation of our results begins with a comparative
description of the two types of $q$-domination (the $q$-TD and
$q$-PD) in sec.~\ref{tocom} and of the various types of
$\chi$-production in sec.~\ref{thnonth}. In sec.~\ref{numan}, we
investigate the behaviour of $\Omega_\chi h^2$ as a function of
the free parameters and in sec.~\ref{scnonsc} we compare the
obtained $\Omega_\chi h^2$ in the KRS with the results of related
scenaria. Finally, in sec.~\ref{NTR} we present areas compatible
with eq.~(\ref{cdmb}).

%%%%%%%%%%%%%%%%%%%%%%%%%%%%%%%%%%%%%%%%%%%%%%%%%%%%%%%%%%%%%%%%%%%%
%%%%%%%%%%%%%%%%

%%%%%%%%%%%%%%%%%%%%%%%%%%%%%%%%%%%%%%%%%%%%%%%%%%%%%%%%%%%%%%%%%%%%
%%%%%%%%%%%%%%%%

\subsection{C{\ssz OMPLETE} V{\ssz ERSUS}
P{\ssz ARTIAL} D{\ssz OMINATION OF} K{\ssz INATION}} \label{tocom}

\hspace{.562cm} The two kinds of $q$-domination, $q$-TD and
$q$-PD, described in sec.~\ref{sec:qdom} are explored in
fig.~\ref{fig1} and fig.~\ref{fig2}, respectively. Namely, in
fig.~\ref{fig1} [fig.~\ref{fig2}], we illustrate the cosmological
evolution of the various quantities as a function of $\btau$ for
$m_\phi=10^6~{\rm GeV}$, $T_\phi=30~{\rm GeV}$ and
$\log\brhofi=69.5$ [$\log\brhofi=77$] (the inputs and some key
outputs of our running are listed in the first [fourth] column of
table~\ref{t1}). We present solid lines [crosses] which are
obtained by our numerical code described in sec.~\ref{Neqs}
[semi-analytical expressions as we explain in secs.~\ref{before}
and \ref{after}] so as we can check the accuracy of the formulas
derived in sec.~\ref{Seqs}. In particular, we design:

$\bullet$ $\log\bar\rho_i$ with $i=q$ (black line and crosses),
$i=\phi$ (gray line and crosses), $i=R$ (light gray line and
crosses) versus $\btau$, in figs.~\ref{fig1}-{\sf (a)} and
\ref{fig2}-{\sf (a)}. In both cases we observe that $\vrho_q$
decreases more steeply than $\vrho_\phi$, and $\vrho_{_{\rm R}}$
remains predominantly constant. On the other hand, in
fig.~\ref{fig1}-{\sf (a)} [\ref{fig2}-{\sf (a)}], we observe that:
{\sf (i)} we obtain 2 [3] intersections of the various lines, {\sf
(ii)} the hierarchy of the various intersection points is
$\btpr<\btkr$ [$\btkp<\btkr<\btpr$],  {\sf (iii)} at the point of
the last intersection ($\btrh=\btkr$ [$\btrh=\btpr$]), we obtain
$(\vrho_q/\vrho_\phi)(\btrh)\gg1$
[$(\vrho_q/\vrho_\phi)(\btrh)\ll1$] as expected from
eq.~(\ref{cond}).

$\bullet$ $T$ versus $\btau$, for
$\brhoqi=(m_\phi/H_0)^2~[\brhoq=0]$ (bold [thin] line and crosses)
in figs.~\ref{fig1}-{\sf (b)} and \ref{fig2}-{\sf (b)} (obviously
the thin lines correspond to a LRS with the same $\brhofi$). In
both cases we observe that $T$ rapidly takes its maximal plateau
value, which is much lower than its maximal value obtained in the
LRS -- see eqs.~(\ref{Tpl}) and (\ref{rhomax}). However, in
fig.~\ref{fig1}-{\sf (b)} the transition from the $\rho_{_{\rm
R}}<\rho_\phi$ to the $\rho_{_{\rm R}}>\rho_\phi$ phase, takes
place at $\btpr\simeq16.08$ [$\btrhp\simeq 12.06$] (where a corner
[kink] is observed on the bold [thin] line), whereas in
fig.~\ref{fig2}-{\sf (b)}, the same transition takes place
practically at a common point $\btpr\simeq 17.47$ for both the LRS
and KRS where a slight kink is observed on both lines. This is
expected since in fig.~\ref{fig2}-{\sf (b)} we obtain $q$-PD and
so, for $\btau>\btkp=14.3$, the KRS and LRS give similar results.

$\bullet$  $\log(sR^3)$ versus $\btau$, for
$\brhoqi=(m_\phi/H_0)^2~[\brhoq=0]$ (bold [thin] line and crosses)
in figs.~\ref{fig1}-{\sf (c)} and \ref{fig2}-{\sf (c)}. In both
cases we observe that the initial entropy $(sR^3)(0)$ is much
lower in the KRS than in the LRS. At the points where we observe a
corner or a kink on the lines of fig.~\ref{fig1}-{\sf (b)}
[figs.~\ref{fig2}-{\sf (b)}], a plateau, which represents the
transition to the isentropic expansion, appears in
fig.~\ref{fig1}-{\sf (c)} [figs.~\ref{fig2}-{\sf (c)}]. The
appearance of the plateau observed on the bold and thin lines for
$q$-TD -- fig.~\ref{fig1}-{\sf (c)} -- occurs at different points
($\btpr\simeq16.08$ and $\btrhp\simeq 12.06$), whereas for $q$-PD
-- fig.~\ref{fig2}-{\sf (c)} -- the same effect is realized at a
common point since $\btpr\simeq 17.5$ and $\btrhp\simeq 17.82$.
This is expected, since for $\btau>\btkp=14.3$ and $q$-PD, the KRS
almost coincides to LRS (with the same $\brhofi$).

$\bullet$ $q$ and $q^\prime$ versus $\btau$, in
figs.~\ref{fig1}-{\sf (d)} and \ref{fig2}-{\sf (d)}. In both cases
we observe the period of the $q$-evolution according to
eq.~(\ref{qk}) with constant inclination ($\sqrt{2}$) and the
onset of the frozen field dominated phase ($q^\prime=0$). However,
we observe that the frozen field phase commences much earlier in
fig.~\ref{fig2}-{\sf (d)} (for $q$-PD) and $q$ takes a value lower
than the one in fig.~\ref{fig1}-{\sf (d)} (for $q$-TD) -- see
eqs.~(\ref{qfqa}) and (\ref{qfff}).

\newpage
\subsection{E{\ssz QUILIBRIUM} V{\ssz ERSUS}
{\ssz NON}-E{\ssz QUILIBRIUM} P{\ssz RODUCTION}} \label{thnonth}

\begin{table}[!h]
\begin{center}
\begin{tabular}{|l||c|c|c|c|} \hline
{\bf \nsz F\ssz IGS.\nsz}&{\bf \nsz 1, 3-${\sf (a_1)}$ \nsz}&{\bf
\nsz 3-${\sf (a_2)}$}& {\bf \nsz 3-${\sf (b_1)}$\nsz } &{\bf \nsz
2, 3-${\sf (b_2)}$ \nsz} \\ \hline\hline
\multicolumn{5}{|c|}{\bf \nsz I\ssz NPUT \nsz P\ssz ARAMETERS \nsz
}\\ \hline
\multicolumn{5}{|c|}{$\lambda=0.5,~m_\phi=10^{6}~{\rm
GeV},~T_\phi=30~{\rm GeV},~m_\chi=350~{\rm GeV}$}\\\hline
$\log\brhofi$&$69.5$&$73.82$&$74.3$&$77$\\
$\sv~({\rm GeV}^{-2})$&$10^{-10}$&$10^{-10}$&$2\times
10^{-9}$&$1.8\times 10^{-9}$\\
$N_\chi$&$10^{-6}$&$0$&$0$&$0$\\
$\vVo$&$7\times10^{11}$&$2.25\times10^9$&$1.18\times10^9$&$5.5\times
10^7$\\ \hline\hline
\multicolumn{5}{|c|}{\bf \nsz O\ssz UTPUT \nsz P\ssz ARAMETERS
\nsz }\\ \hline
$T_{\rm PL}~({\rm GeV})$&1.3&16.05&21.1&100\\\hline
$\btkp$&$-$&$-$&$-$&14.3\\
$\Tkp~({\rm GeV})$&$-$&$-$&$-$&94.7\\\hline
$\btpr$&16.08&16.09&16.1&17.5\\
$\Tpr~({\rm GeV})$&1.3&16.05&21.1&100\\\hline
$\btkr$&22.1&17.1&16.6&15.5\\
$\Tkr~({\rm GeV})$&0.0035&6.31&13.9&70.2\\ \hline
$\brhoqi/\brhofi$&$1.3\times10^{26}$&$6.4\times10^{21}$&$2\times10^{21}$&$4.2\times
10^{18}$\\
$(\vrho_q/\vrho_\phi)(\btrh)$&$2.5\times10^{22}$&$1.2\times10^6$&$11.84$&$2\times
10^{-4}$\\ \hline\hline
$\Omega_q(\vtns)$&0.01&$3\times10^{-9}$&$5.6\times10^{-10}$&$7\times10^{-15}$\\
\hline
$\btast$&15.25&15.7&$15.8$&$15.3$\\
$x^{-1}_{*}$&265&22.7&$17.45$&$4.6$\\
$(f_\chi^{\rm
eq}/f_\chi^{nN})(\btast)$&$0$&12.6&0.0016&$5\times10^{-9}$\\\hline
$\btf$&$-$&$-$&16.5&18.5\\
$x^{-1}_{_{\rm F}}$&$-$&$-$&23.45&23.86\\
$\Omega_\chi h^2$&0.11&0.11&0.11&0.11\\\hline
$\Omega_\chi h^2|_{\rm SC}$&1.87&1.87&0.11&0.12\\
$\Omega_\chi h^2|_{\brhoq=0}$&1.78&1.78&0.1&0.11\\
$\Omega_\chi h^2|_{\brhof=0}$&979&2.86&0.12&0.12\\\hline
\end{tabular}
\end{center}\vspace*{-.155in}
\caption{\sl\ftn  Input and output parameters for the four
examples illustrated in figs.~3-${\sf (a_1)},~{\sf (a_2)},~{\sf
(b_1)}$ and ${\sf (b_2)}$ (see also figs.~1 and 2).}
\label{t1}\vspace*{-.1in}
\end{table}

\begin{figure}[t]\vspace*{-.19in}
\hspace*{-.71in}
\begin{minipage}{8in}
\epsfig{file=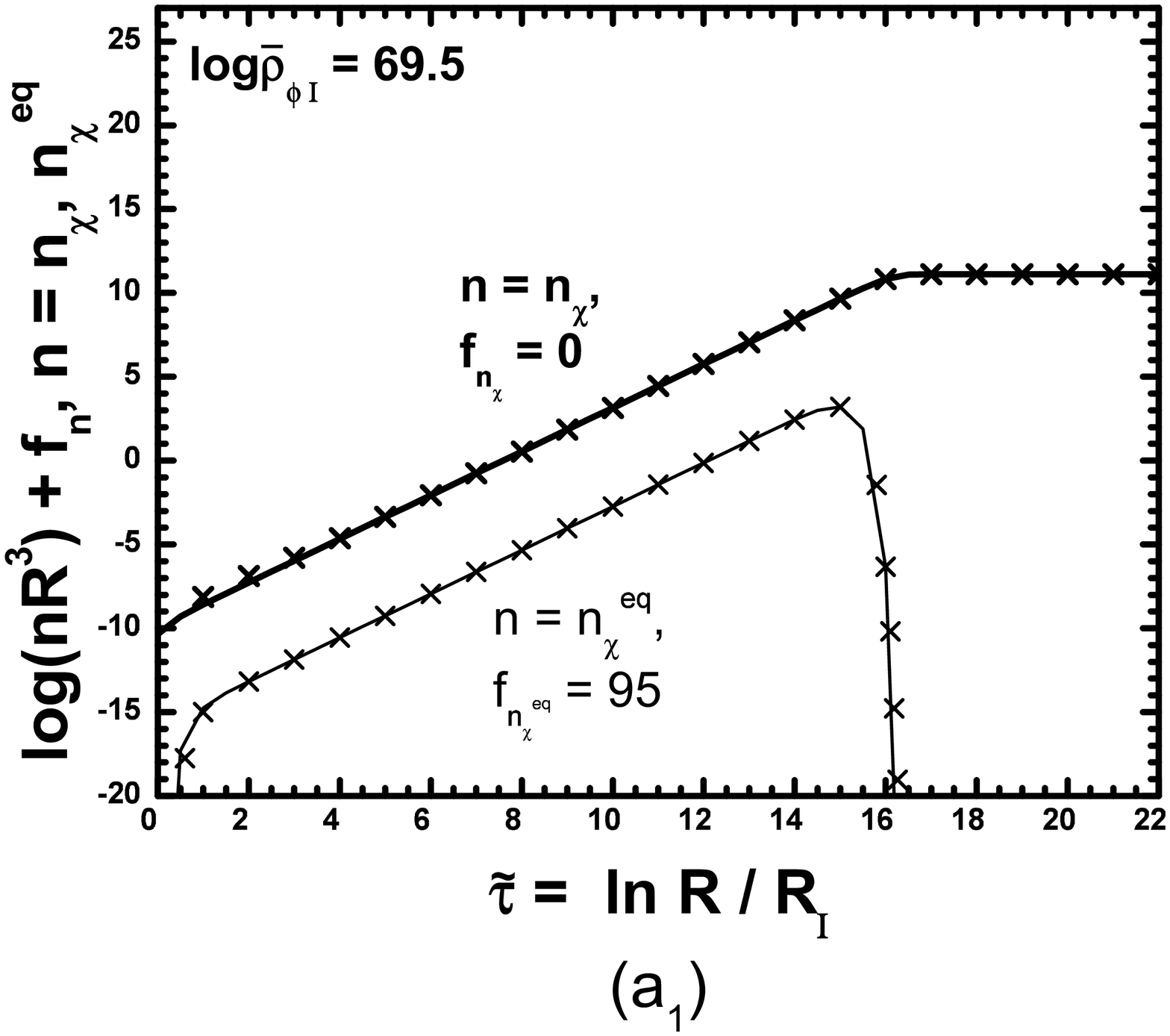,height=3.8in,angle=-90} \hspace*{-1.37 cm}
\epsfig{file=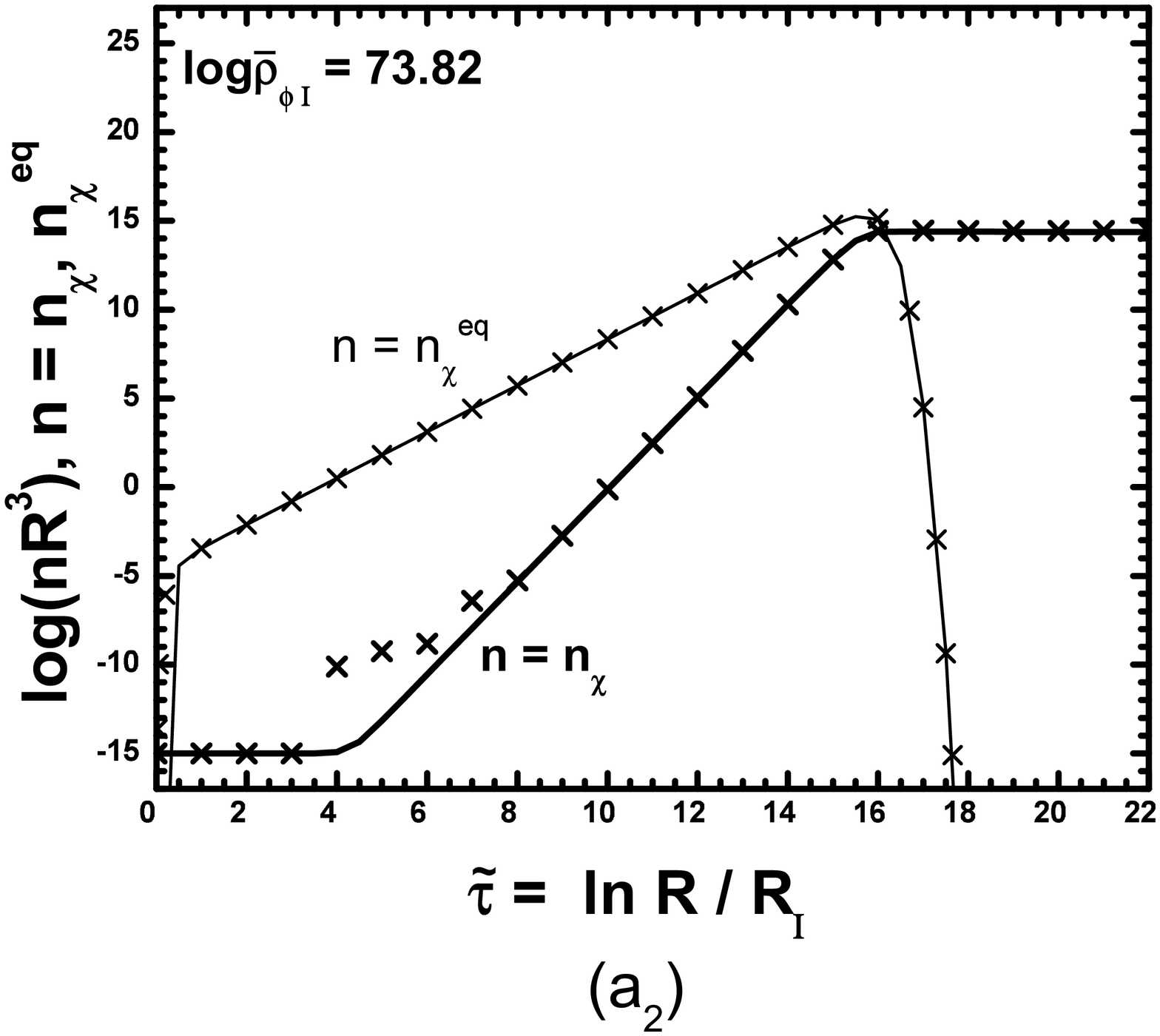,height=3.8in,angle=-90} \hfill
\end{minipage}\vspace*{-.01in}
\hfill\hspace*{-.71in}
\begin{minipage}{8in}
\epsfig{file=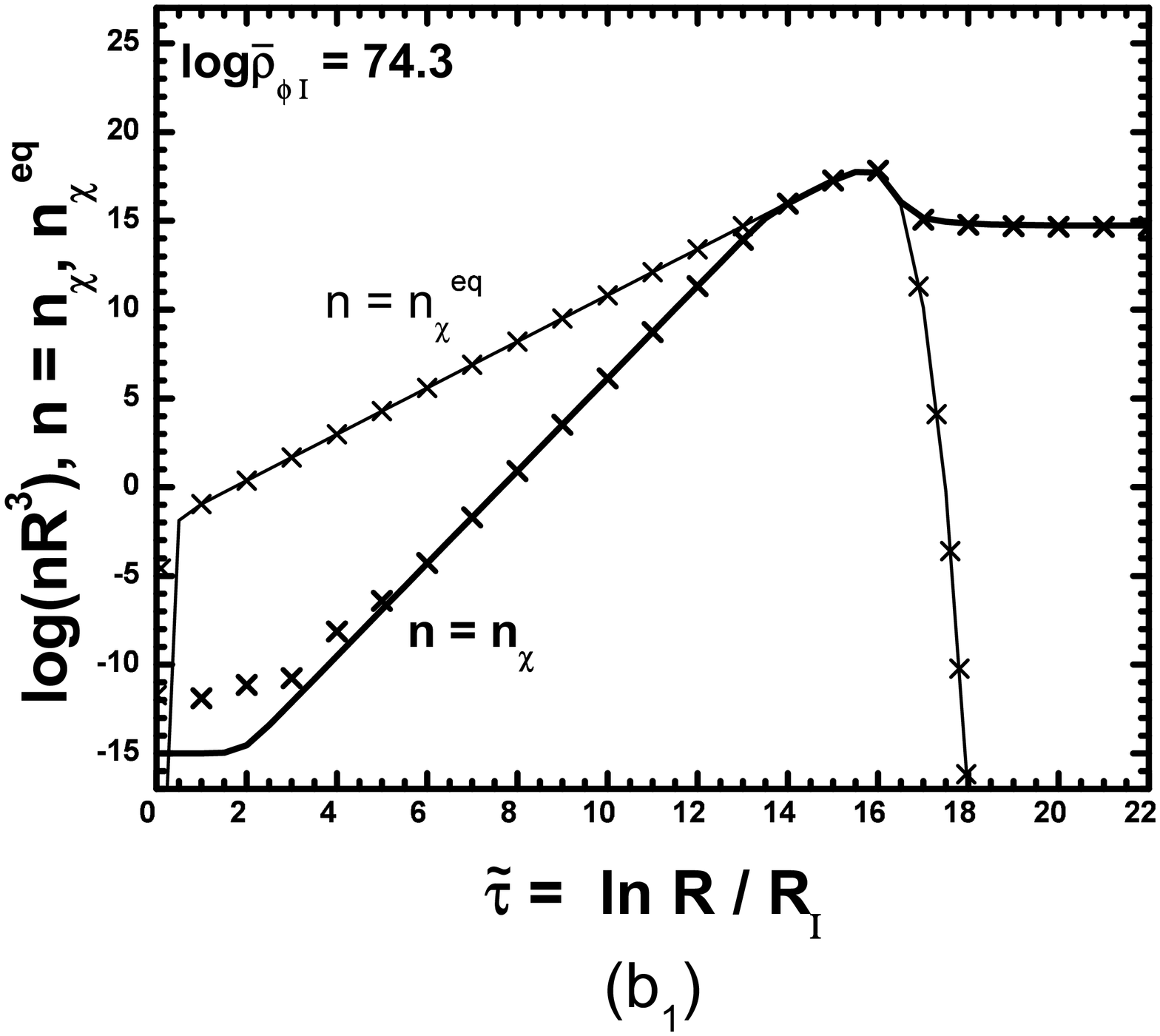,height=3.8in,angle=-90} \hspace*{-1.37 cm}
\epsfig{file=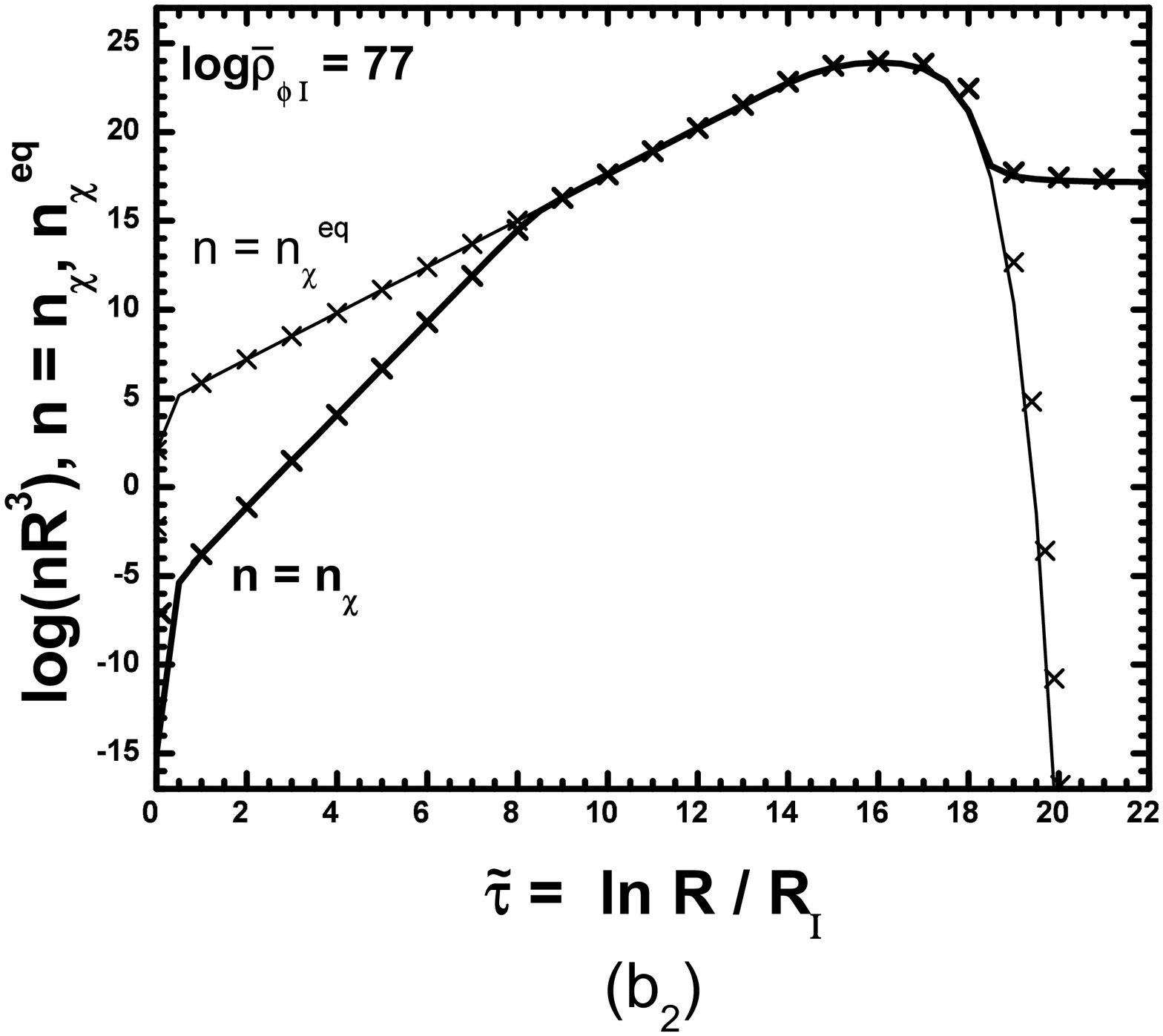,height=3.8in,angle=-90} \hfill
\end{minipage}
\hfill \caption[]{\sl\ftn The evolution as a function of $\btauf$
of the quantities $\log(n_\chi R^3)$ (bold line and crosses) and
$\log(n^{\rm eq}_\chi R^3)+f_{n^{\rm eq}_\chi}$ (thin line and
crosses) for $m_\phi=10^6~{\rm GeV},~T_\phi=30~{\rm GeV}$ and:
${\sf (a_1)}$ $\log\brhofi=69.5,~N_\chi=10^{-6}$ and
$\sv=10^{-10}~{\rm GeV}^{-2}$, ${\sf (a_2)}$
$\log\brhofi=73.82,~N_\chi=0$ and $\sv=10^{-10}~{\rm
GeV}^{-2}$,~${\sf (b_1)}$ $\log\brhofi=74.3,~N_\chi=0$ and
$\sv=2\times 10^{-9}~{\rm GeV}^{-2}$ and ${\sf (b_2)}$
$\log\brhofi=77,~N_\chi=0$ and $\sv=1.8\times 10^{-9}~{\rm
GeV}^{-2}$ ($f_{n^{\rm eq}_\chi}=95$ in the case ${\sf (a_1)}$ and
$f_{n^{\rm eq}_\chi}=0$ elsewhere). The solid lines [crosses] are
obtained by our numerical code [semi-analytical expressions]. In
all cases, we extract $\Omega_{\chi}h^2=0.11$.}
\label{fig3}\vspace*{-.17in}
\end{figure}

The various kinds of $\chi$-production encountered in the KRS, are
explored in fig.~\ref{fig3}. In this, we check also the accuracy
of our semi-analytical expressions (which describe the $f_\chi$
and $f_\chi^{\rm eq}$ evolution), displaying by bold solid lines
[crosses] the results obtained by our numerical code (see
sec.~\ref{Neqs}) [semi-analytical expressions (see
secs.~\ref{after} and \ref{before})]. The thin crosses are
obtained by inserting $T$ (given as we describe in
secs.~\ref{after} and \ref{before}) into eq.~(\ref{neq}).

%%%%%%%%%%%%%%%%%%%%%%%%%%%%%%%%%%%%%%%%%%%%%%%%%%%%%%%%%%%%%%%%%%%%
%%%%%%%%%%%%%%%%

The inputs parameters and some key outputs for the four examples
illustrated in fig.~\ref{fig3} are listed in table~\ref{t1}. In
particular, we present $\btau$'s and the corresponding $T$'s  of
the possible intersections between the various energy-densities.
Comparing the relevant results, we observe that as $\brhofi$
increases, $\brhoqi/\brhofi$ decreases (note that eq.~(\ref{domk})
remains always valid), and so, $(\vrho_q/\vrho_\phi)(\btrh)$ and
$\Omega_q(\vtns)$ decrease too. At the same time, $\btpr$
eventually approaches $\btkr$ and becomes larger than this for
$\log\brhofi=77$, where $q$-PD is achieved (in the other cases we
have $q$-TD). In the same table we provide $\btast$'s, derived
from the maximalization of the intergrand in eq.~(\ref{IIfnsol}),
and we applied the criterion of eq.~(\ref{condf}). In the case of
EP, $\btf$'s and $x_{_{\rm F}}$'s derived from eq.~(\ref{xf}) are
also given. In all cases, we extract $\Omega_{\chi}h^2=0.11$ and
we show the resultant $\Omega_{\chi}h^2$ in several other related
scenaria (see also sec.~\ref{scnonsc}).

We design the evolution of $\log f_\chi$ (bold line and crosses)
and $\log f_\chi^{\rm eq}+f_{n^{\rm eq}_\chi}$ (thin line and
crosses) as a function of $\btau$ for $m_\phi=10^6~{\rm
GeV},~T_\phi=30~{\rm GeV}$ and:

$\bullet$ $\log\brhofi=69.5,~N_\chi=10^{-6}$ ($f_{n^{\rm
eq}_\chi}=95$) and $\sv=10^{-10}~{\rm GeV}^{-2}$ in
fig.~\ref{fig3}-${\sf (a_1)}$. In this case, we obtain $q$-TD (the
evolution of the various energy densities is presented in
fig.~\ref{fig1}-{\sf (a)}). The quantity $n^{\rm eq}_\chi$ turns
out to be strongly suppressed due to very low $T_{\rm PL}\ll
m_\chi/20$ and so, we obtain $f_\chi\gg f^{\rm eq}_\chi$ for any
$\btau<25$. This is a typical example of non-EPI, where the
presence of $N_\chi>0$ is indispensable so as to obtain
interesting $\Omega_\chi h^2$. Fixing $N_\chi=10^{-6}$ and
adjusting $\brhofi$, we achieve $\Omega_{\chi}h^2=0.11$. The bold
crosses are derived by solving numerically eq.~(\ref{Ifn}) for
$\btau<\btrh$ (since \sv\ is rather low, eq.~(\ref{Ifnsol}) is
also valid) and from eq.~(\ref{BEsola}) for $\btrh<\btau<\btfn$.
The constant-$f_\chi$ phase commences at $\btau\simeq16.1$, where
the integrand of $f_\chi^N$ in eq.~(\ref{Ifnsol}) reaches its
maximum.

$\bullet$ $\log\brhofi=73.82,~N_\chi=0$ ($f_{n^{\rm eq}_\chi}=0$)
and $\sv=10^{-10}~{\rm GeV}^{-2}$ in fig.~\ref{fig3}-${\sf
(a_2)}$. In contrast with the previous case, we obtain a less
efficient $q$-TD and so, $T_{\rm PL}$ turns out to be
significantly larger ($T_{\rm PL}\sim m_\chi/20$). This is a
typical example of non-EPII since at $\btast\simeq15.7$, we get
$f_\chi<f^{\rm eq}_\chi$. By adjusting $\brhofi$, we achieve
$\Omega_{\chi}h^2=0.11$. The bold crosses are extracted from
eq.~(\ref{IIfnsol}) for $\btau<\btrh$ and from eq.~(\ref{BEsola})
for $\btrh<\btau<\btfn$. The onset of the constant-$f_\chi$ phase
occurs at $\btast\simeq15.7$.

$\bullet$ $\log\brhofi=74.3,~N_\chi=0$ ($f_{n^{\rm eq}_\chi}=0$)
and $\sv=2\times 10^{-9}~{\rm GeV}^{-2}$ in fig.~\ref{fig3}-${\sf
(b_1)}$. In this case, we obtain a weak (since $\btpr$ is very
close to $\btkr$, as shown in table~\ref{tab1}) $q$-TD with
$T_{\rm PL}>T_{\rm F}$. This is a typical example of EP before the
onset of RD era, since for $\btast\sim15.8$, we get $f_\chi>
f^{\rm eq}_\chi$ and $\btf<\btrh=\btkr$. $\Omega_{\chi}h^2=0.11$
is achieved by adjusting $\sv$. The bold crosses are extracted by
solving numerically eq.~(\ref{fn}) after substituting in it $H$
and $T$ as we describe in secs.~\ref{after} and \ref{before}. They
could be, also, derived from eq.~(\ref{BEsol}) for
$\btf<\btau<\btrh$ and from eq.~(\ref{BEsola}) for
$\btrh<\btau<\btfn$.

\begin{table}[!t]
\begin{center}
\begin{tabular}{|c|c|c|c|c||c|}
\hline {\bf  F\ssz IG. \nsz} & \multicolumn{4}{|c||}{\bf \nsz
R\ssz ANGES OF THE \nsz L\ssz OWER \nsz \boldmath $x$-A\ssz XIS
\nsz P\ssz ARAMETERS \nsz}& {\bf\boldmath $\chi$-P\ssz RO-}\\
\cline{2-5}
&$N_{\chi}=0$&$N_{\chi}=10^{-7}$&$N_{\chi}=10^{-6}$&$N_{\chi}=10^{-5}$
&{\bf\ssz DUCTION}
\\ \hline \hline
\ref{om}-${\sf
(a_1)}$&$-$&$68.56-73.39$&$68.56-73.56$&$68.56-73.74$&non-EPI\\
&$73.75-73.93$&$-$&$-$&$-$&non-EPII\\
&$73.94-74$&$73.4-74$&$73.57-74$&$73.75-74$&EP
\\\hline
\ref{om}-${\sf (a_2)}$
&$-$&$68.56-73.44$&$68.56-73.53$&$68.56-73.72$&
non-EPI\\&$73.5-73.6$&$-$&$-$&$-$& non-EPII\\
&$73.61-74.5$&$73.45-74.5$&$73.54-74.5$&$73.73-74.5$&EP\\ \hline
\ref{om}-${\sf (b_1)}$ &$-$&$1-3.17$&$(-0.9)-3$&$(-0.9)-0.698$&
non-EPI
\\ &$-$&$3.18-3.28$&$-$&$-$& EP
\\ &$3.54-3.69$&$3.29-3.69$&$-$&$-$& non-EPII
\\ \hline
\ref{om}-${\sf (b_2)}$ &$-$&$1-3.17$&$(-0.9)-3$&$(-0.9)-0.698$&
non-EPI
\\ &$-$&$3.18-3.28$&$-$&$-$& EP
\\ &$3.4-3.54$&$3.29-3.54$&$-$&$-$& non-EPII \\ \hline
\ref{om}-${\sf (c_1)}$  &$-$&$5-6.5$&$5.8-6.8$&$6.5-6.8$& non-EPI
\\ &$3.88-4.1$&$-$&$-$&$-$& non-EPII\\ \hline
\ref{om}-${\sf (c_2)}$ &$3-4$&$3-4.2$&$3-4$&3-3.7& EP \\
&$-$&$4.3-6.5$&$5.7-6.8$&$6.5-6.8$& non-EPI\\
&$4-4.3$&$-$&$-$&$-$& non-EPII\\ \hline
\end{tabular}
\end{center}\vspace*{-.155in}
\caption{\sl\ftn The type of $\chi$-production for various ranges
of the lower $x$-axis parameters and $N_{\chi}$'s in fig.
\ref{om}.}\label{t2}\vspace*{-.2in}
\end{table}

$\bullet$ $\log\brhofi=77,~N_\chi=0$ ($f_{n^{\rm eq}_\chi}=0$) and
$\sv=1.8\times 10^{-9}~{\rm GeV}^{-2}$ in fig.~\ref{fig3}-${\sf
(b_2)}$. In this case, we obtain $q$-PD (the evolution of the
various energy densities is presented in fig~\ref{fig2}-{\sf (a)})
and $T_{\rm PL}\gg T_{\rm F}$. This is a typical example of EP
after the onset of the RD era, since for $\btast\simeq15.3$ we get
$f_\chi> f^{\rm eq}_\chi$, and
$\btf\simeq18.5>\btrh=\btpr\simeq17.5$. $\Omega_{\chi}h^2=0.11$ is
achieved by adjusting $\sv$. The bold crosses are extracted
similarly to the previous case. They could be also derived from
eq.~(\ref{BEsola}) for $\btf<\btau<\btfn$.

\addtolength{\textheight}{1.5cm}
\newpage

%%%%%%%%%%%%%%%%%%%%%%%%%%%%%%%%%%%%%%%%%%%%%%%%%%%%%%%%%%%%%%%%%%%%
\begin{figure}[!ht]\vspace*{-.15in}
\hspace*{-.71in}
\begin{minipage}{8in}
\epsfig{file=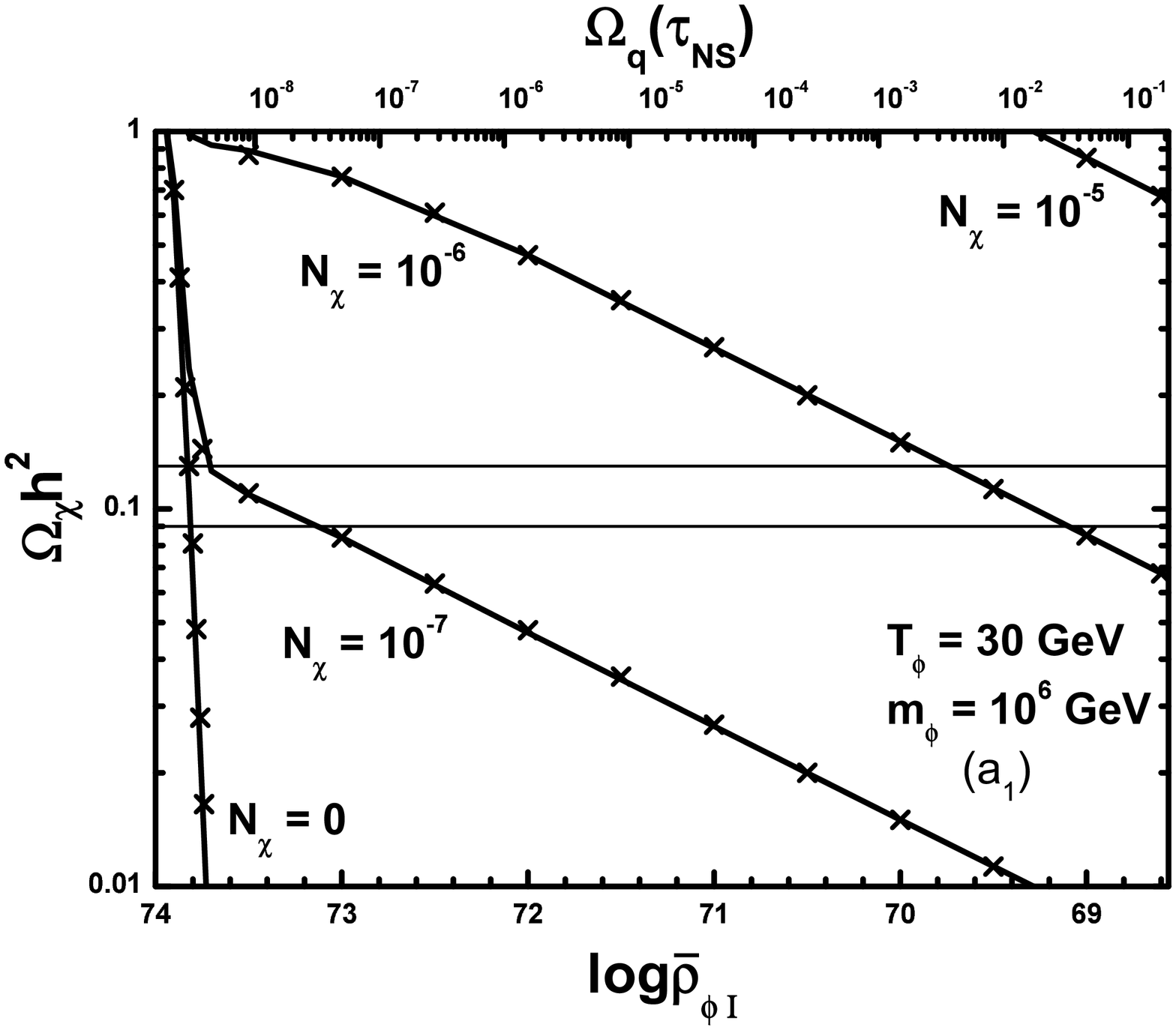,height=3.8in,angle=-90} \hspace*{-1.37 cm}
\epsfig{file=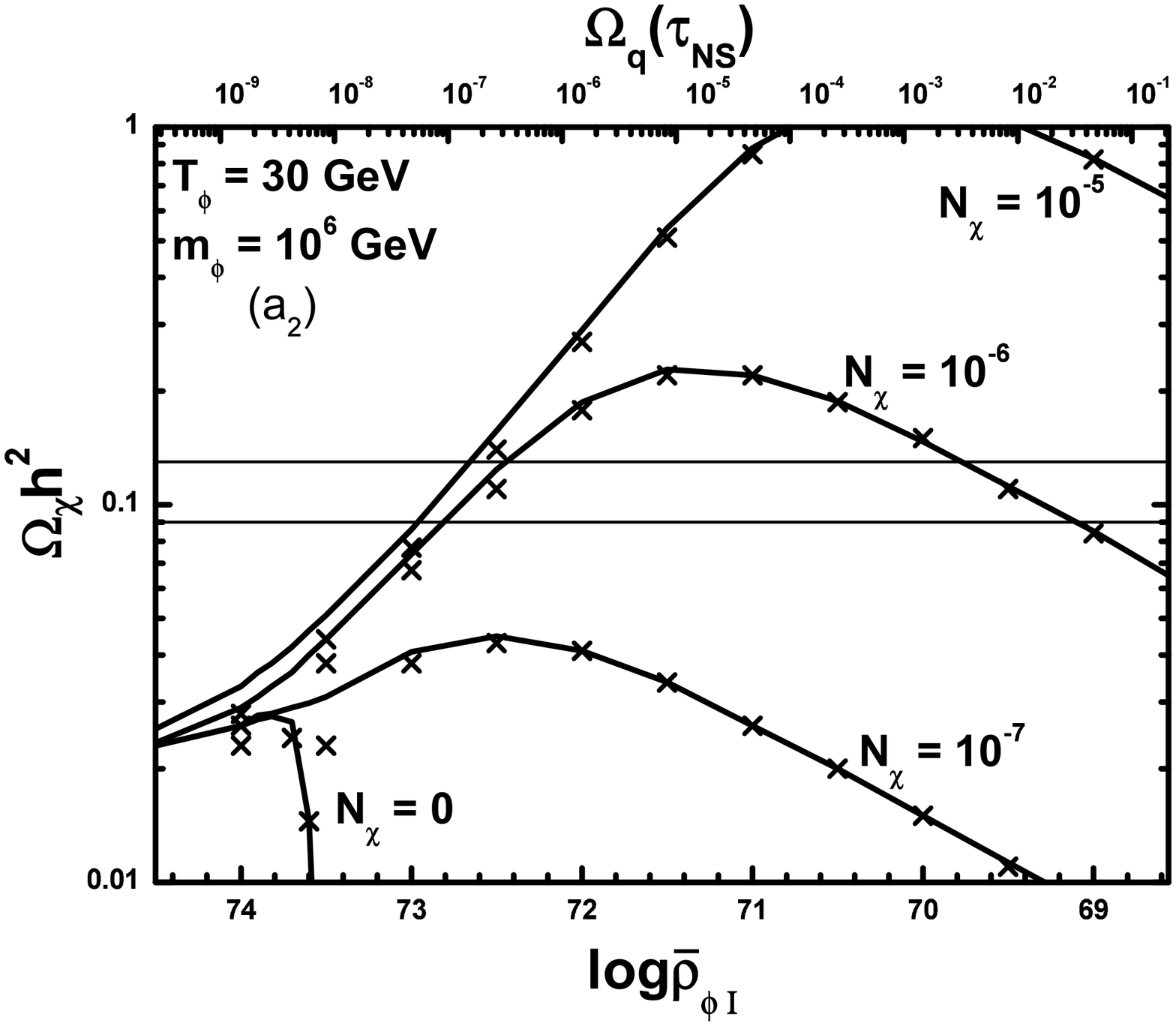,height=3.8in,angle=-90} \hfill
\end{minipage}\vspace*{-.01in}
\hfill\hspace*{-.71in}
\begin{minipage}{8in}
\epsfig{file=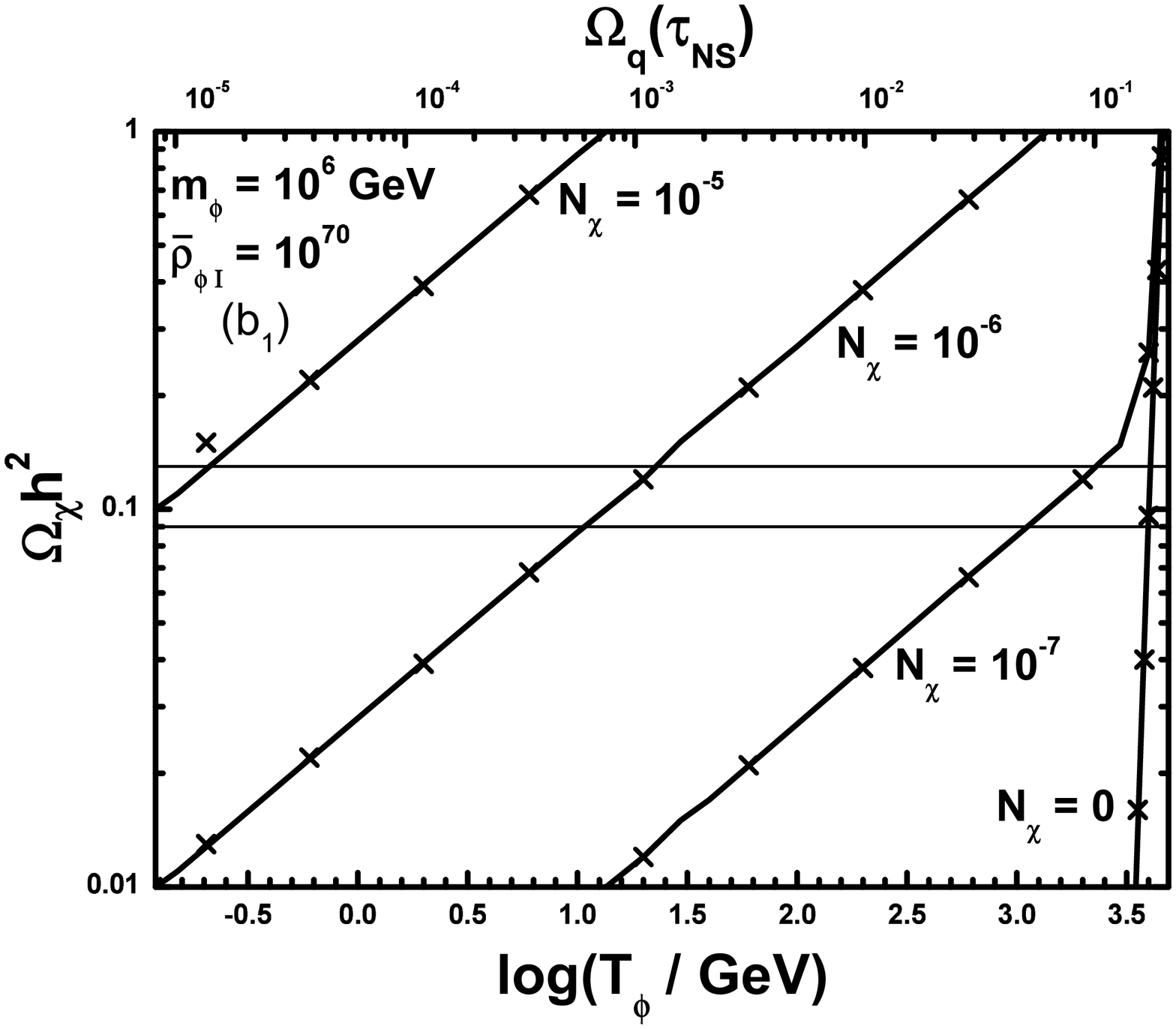,height=3.8in,angle=-90} \hspace*{-1.37 cm}
\epsfig{file=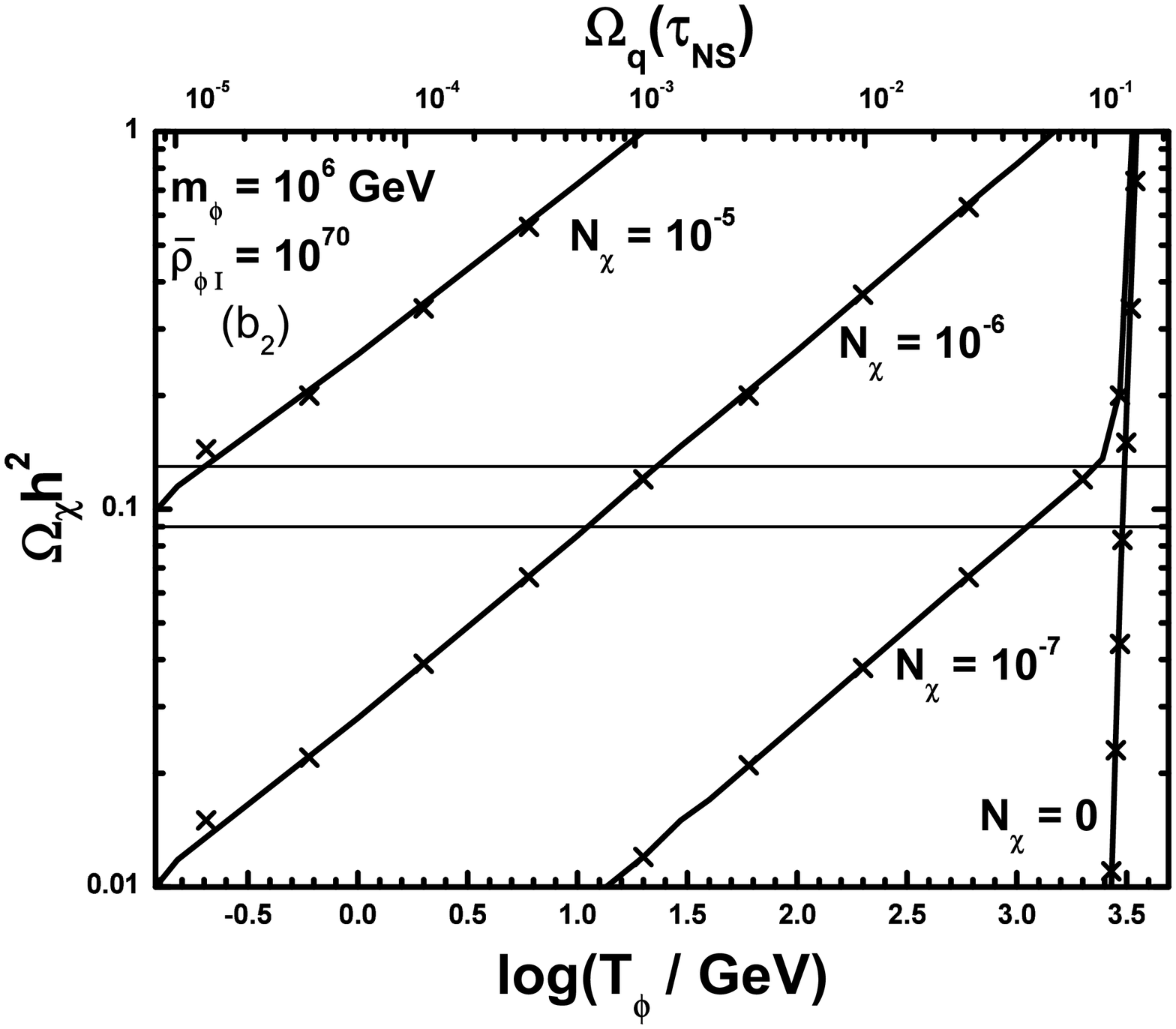,height=3.8in,angle=-90} \hfill
\end{minipage}\vspace*{-.01in}
\hfill\hspace*{-.71in}
\begin{minipage}{8in}
\epsfig{file=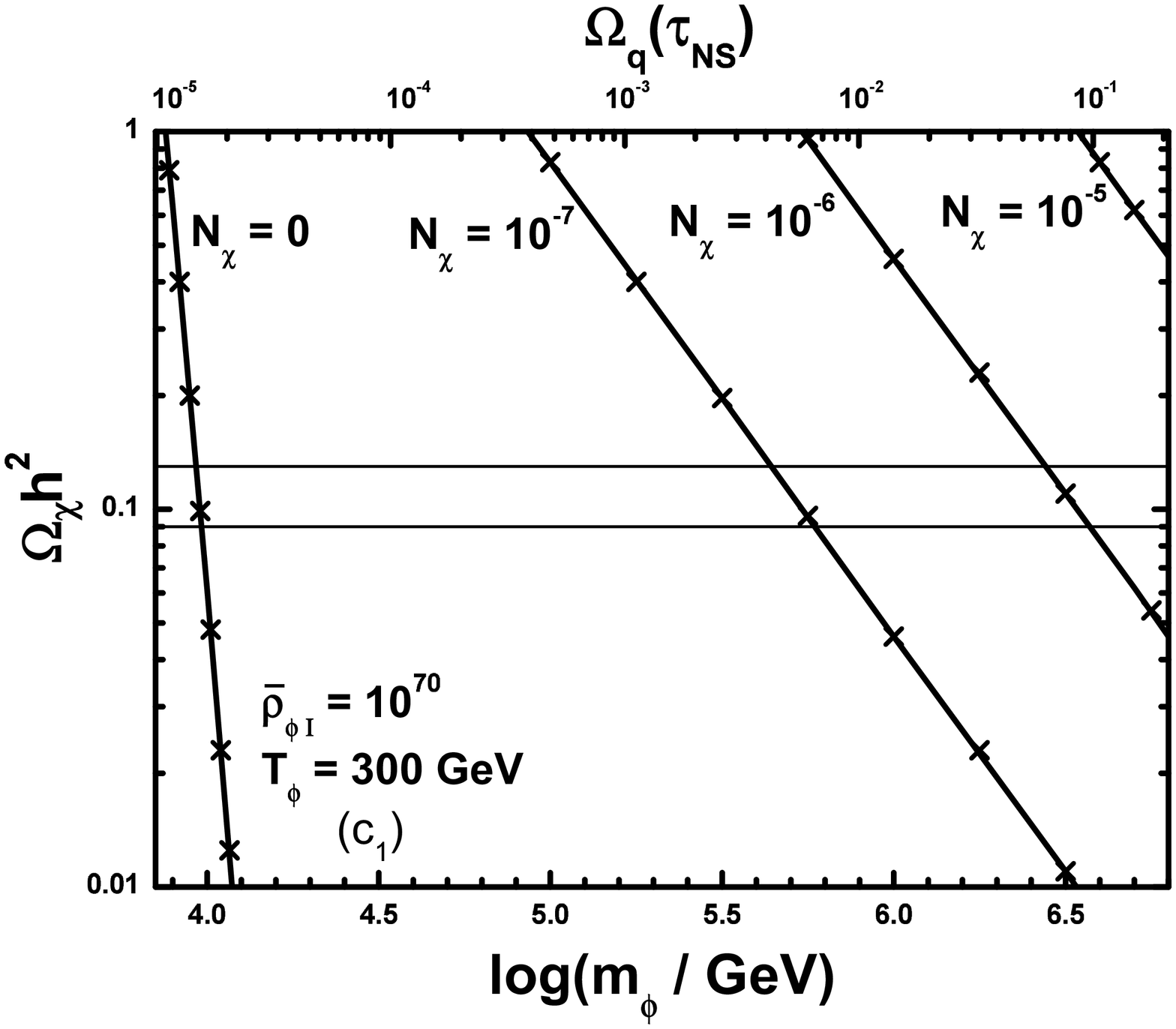,height=3.8in,angle=-90} \hspace*{-1.37 cm}
\epsfig{file=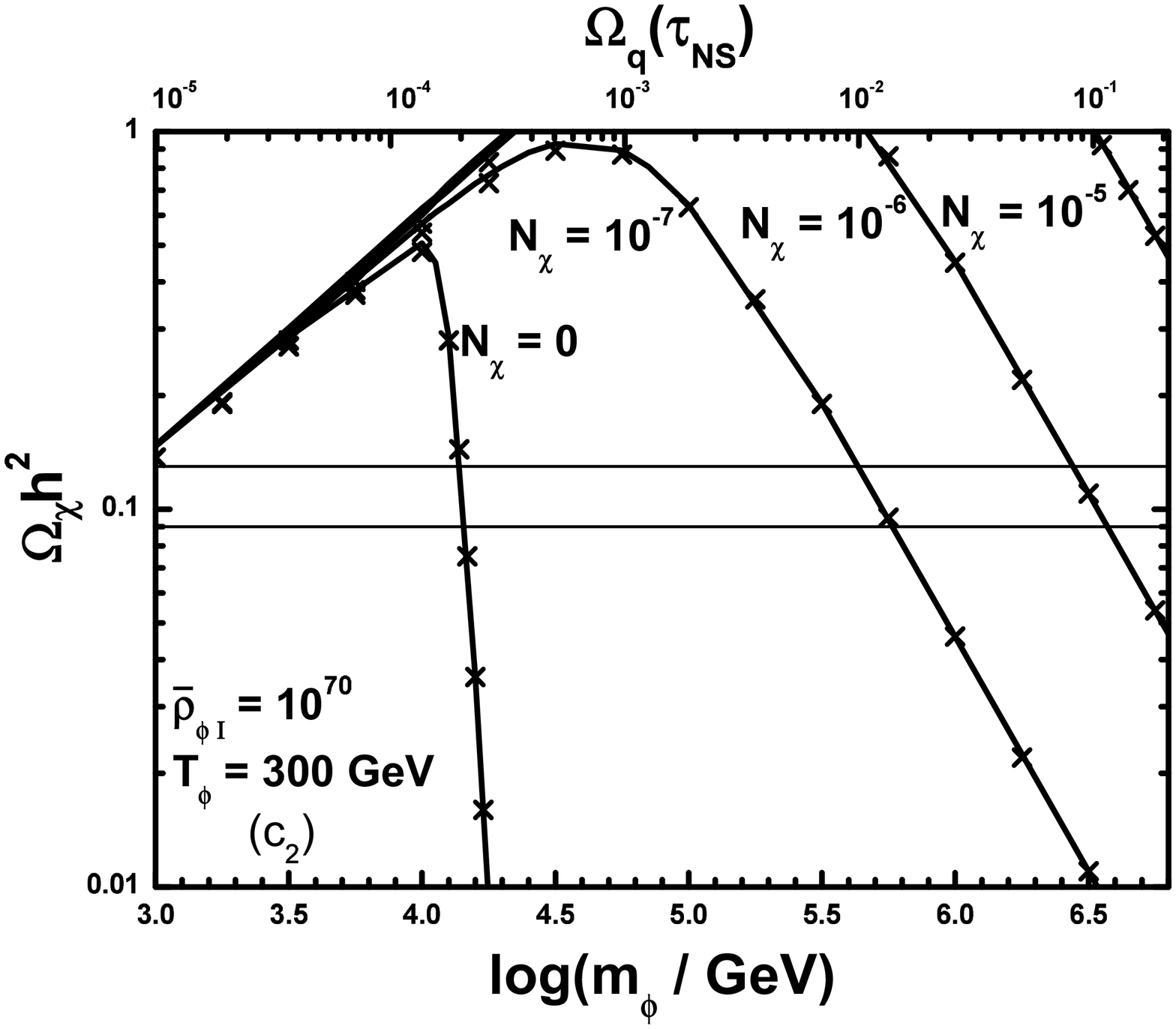,height=3.8in,angle=-90} \hfill
\end{minipage}\vspace*{-.05in}
\hfill \caption[]{\sl\ftn $\Omega_{\chi}h^2$ versus
$\log\brhofi~{\sf (a_1, a_2)}$, $\log T_\phi~{\sf(b_1, b_2)}$ and
$\log m_\phi~{\sf (c_1, c_2)}$ for fixed (indicated in the graphs)
$T_\phi$ and $m_\phi$, $\brhofi$ and $m_\phi$, $\brhofi$ and
$T_\phi$ correspondingly, and various $N_\chi$'s indicated on the
curves. We take $m_{\tilde\chi}=350~{\rm GeV}$ and $\langle\sigma
v\rangle=10^{-10}~{\rm GeV^{-2}}~[\langle\sigma
v\rangle=10^{-8}~{\rm GeV^{-2}}]~{\sf (a_1, b_1, c_1~[a_2, b_2,
c_2])}$. The solid lines [crosses] are obtained by our numerical
code [semi-analytical expressions]. The CDM bounds of
eq.~(\ref{cdmb}) are, also, depicted by the two thin
lines.}\label{om}
\end{figure}
%%%%%%%%%%%%%%%%%%%%%%%%%%%%%%%%%%%%%%%%%

\addtolength{\textheight}{-1.5cm}
\newpage

\subsection{$\Omega_\chi h^2$ {\ssz AS A} F{\ssz UNCTION OF THE}
F{\ssz REE} P{\ssz ARAMETERS}} \label{numan}

\hspace{.562cm} Varying the free parameters, useful conclusions
can be inferred for the behavior of $\Omega_{\chi}h^2$ and the
regions where each ${\chi}$-production mechanism can be activated.
In addition, a final test of our semi-analytical approach can be
presented by comparing its results for $\Omega_{\chi}h^2$ with
those obtained by solving numerically the problem. We focus on
$q$-TD, since the results for $q$-PD are similar to those obtained
in the LRS (see secs.~\ref{fdom} and ref.~\cite{scnr}).

\begin{figure}[t]\vspace*{-.19in}
\hspace*{-.71in}
\begin{minipage}{8in}
\epsfig{file=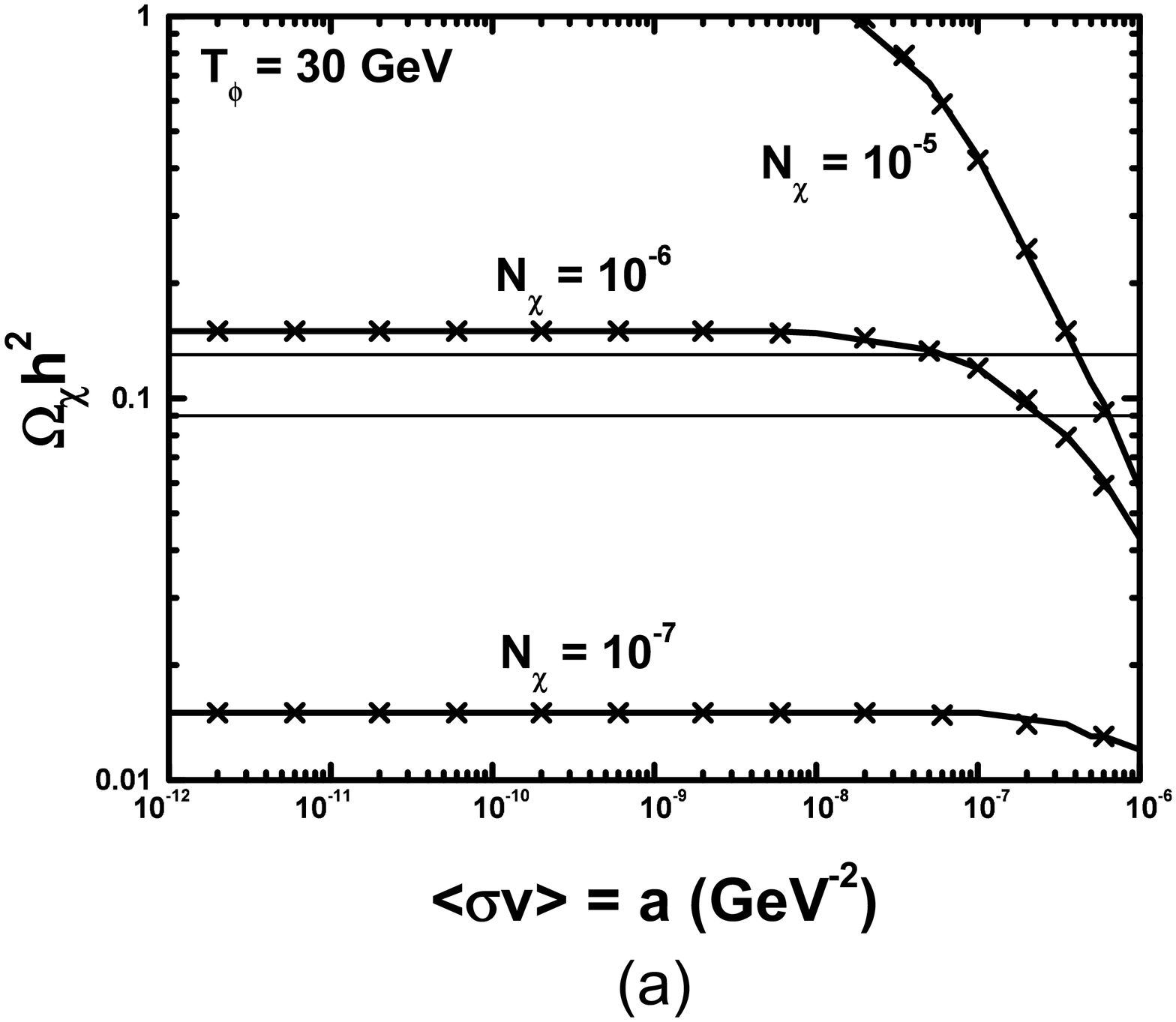,height=3.8in,angle=-90} \hspace*{-1.37 cm}
\epsfig{file=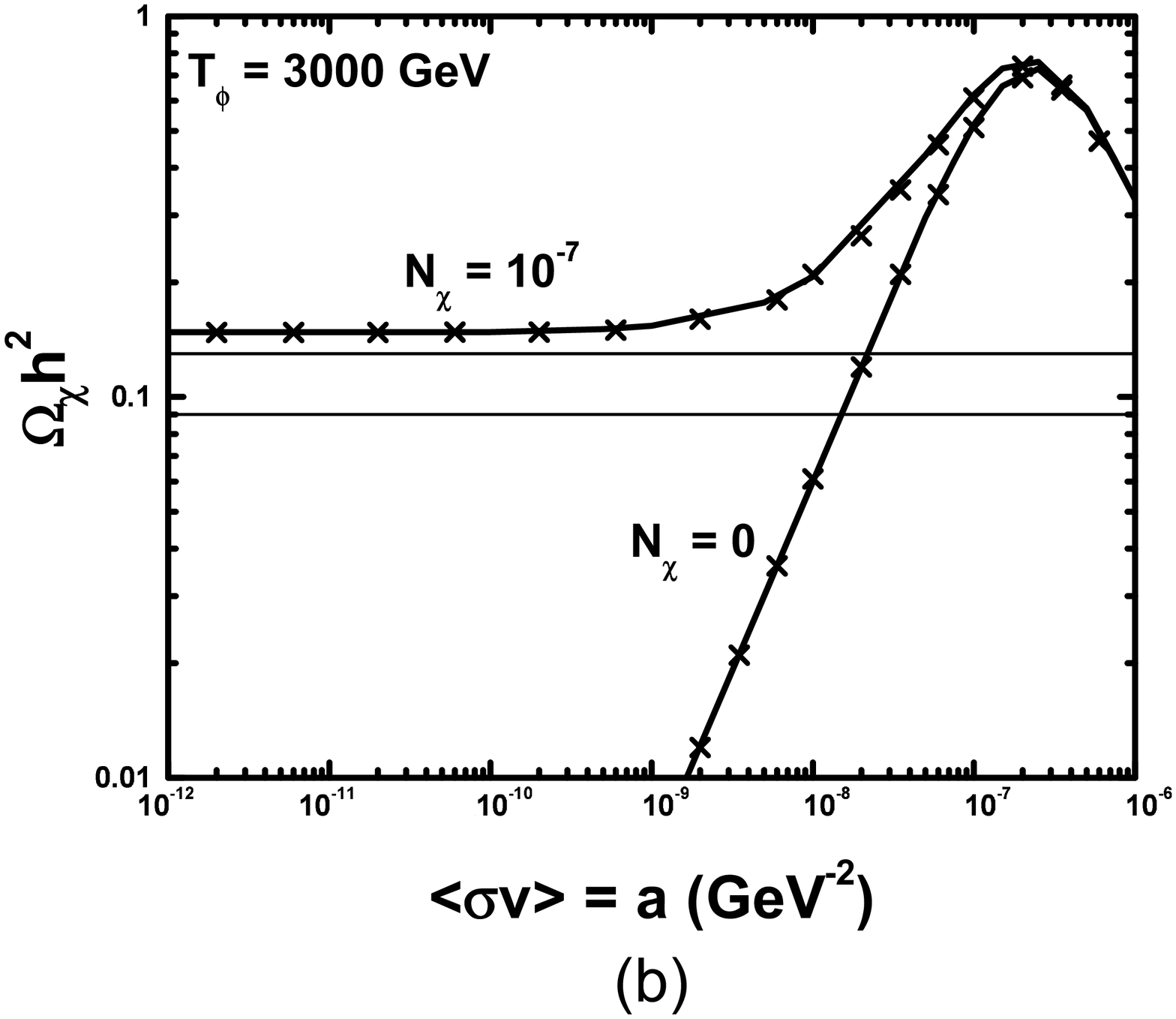,height=3.8in,angle=-90} \hfill
\end{minipage}
\hfill \caption[]{\sl\ftn $\Omega_{\chi}h^2$ as a function of
$\langle\sigma v\rangle=a$ for various $N_\chi$'s, indicated on
the curves, $m_\chi=350~{\rm
GeV},~\brhofi=10^{70},~m_\phi=10^6~{\rm GeV}$ and $T_\phi=30~{\rm
GeV}~[T_\phi=3000~{\rm GeV}]$ ({\sf a [b]}). The solid lines
[crosses] are obtained by our numerical code [semi-analytical
expressions]. The CDM bounds of eq.~(\ref{cdmb}) are, also,
depicted by the two thin lines.} \label{fig4} \vspace*{-.17in}
\end{figure}

Our results are displayed in figs.~\ref{om} and \ref{fig4}. The
solid lines are drawn from the results of the numerical
integration of eqs.~(\ref{fq})-(\ref{fn}), whereas crosses are
obtained by solving numerically eq.~(\ref{fn}) as we describe in
secs.~\ref{before} and \ref{after} (comments on the validity of
eqs.~(\ref{Ifnsol}), (\ref{IIfnsol}) and (\ref{BEsol}) are given,
too). The running of $f_\chi$ after the onset of the RD era -- see
eqs.~(\ref{BEsol}) and (\ref{BEsola}) -- although crucial for the
final result (especially for weak $q$-TD) does not alter the
behaviour of the solution as a function of the free parameters.
The type of ${\chi}$-production for the lower $x$-axis parameters
and the various $N_\chi$'s used in fig.~\ref{om} is presented in
table~\ref{t2}. From this and taking into account the obtained
$T_{\rm PL}$'s in each case (see below), we can induce  that
non-EP (non-EPI [non-EPII] for $N_\chi\neq0$ [$N_\chi\sim0$]) is
dominant for $T_{\rm PL}\lesssim m_\chi/20$, whereas EP is
activated for $T_{\rm PL}>m_\chi/20$.

In figs.~\ref{om}-${\sf (a_1, b_1, c_1)}$ [figs.~\ref{om}-${\sf
(a_2, b_2, c_2)}$], we take $m_{\chi}=350~{\rm GeV}$ and
$\langle\sigma v\rangle=10^{-10}~{\rm GeV^{-2}}$ [$\langle\sigma
v\rangle=10^{-8}~{\rm GeV^{-2}}$]. We design $\Omega_{\chi}h^2$
versus:

$\bullet$  $\log\brhofi$ (or $\Omega_q(\vtns)$) in
fig.~\ref{om}-${\sf (a_1)}$ and ${\sf (a_2)}$ for $T_\phi=30~{\rm
GeV}$, $m_\phi=10^6~{\rm GeV}$ and several $N_\chi$'s indicated on
the curves. We observe that: {\sf (i)} $\Omega_q(\vtns)$ increases
as $\brhofi$ decreases (since $\brhoqi/\brhofi$ increases too) and
so, a lower bound on $\brhofi$ can be derived from
eq.~(\ref{nuc}{\sf a}) -- note that $T_{\rm PL}$ decreases with
$\brhofi$ (see eq.~(\ref{Tpl})) and ranges between (0.8 and
24)~GeV, {\sf (ii)} $\Omega_{\chi}h^2$ decreases with $\brhofi$ in
fig.~\ref{om}-${\sf (a_1)}$  ($\langle\sigma
v\rangle=10^{-10}~{\rm GeV^{-2}}$) whereas it increases as
$\brhofi$ decreases (for large $\brhofi$'s) and decreases with
$\brhofi$ (for low $\brhofi$'s) in fig.~\ref{om}-${\sf (a_2)}$
($\langle\sigma v\rangle=10^{-8}~{\rm GeV^{-2}}$). The last
observation can be explained as follows: $f_\chi$ can be mostly
given by solving numerically eq.~(\ref{Ifn}) -- see
table~\ref{t2}. $f_\chi$ increases with $\brhofi$ and so, when
$\sv$ is large enough ($\sim 10^{-8}~{\rm GeV}^{-2}$) the first
term in the r.h.s of eq.~(\ref{Ifn}) becomes comparable to the
second one and the solution of eq.~(\ref{Ifn}) can be exclusively
realized numerically. For lower $\brhofi$'s, eq.~(\ref{Ifnsol})
can be used and so, $f_\chi$ decreases with $\brhofi$. The latter
behaviour is dominant for low $\sv\sim10^{-10}~{\rm GeV^{-2}}$ as
in fig.~\ref{om}-${\sf (a_1)}$.

$\bullet$  $\log T_\phi$ (or $\Omega_q(\vtns)$) in
fig.~\ref{om}-${\sf (b_1)}$ and ${\sf (b_2)}$ for
$\log\brhofi=70$, $m_\phi=10^6~{\rm GeV}$ and several $N_\chi$'s
indicated on the curves. We observe that: {\sf (i)}
$\Omega_q(\vtns)$ increases with $T_\phi$ (or $\Gamma_\phi$) since
$\phi$ decays more rapidly. Therefore, an upper bound on $T_\phi$
can be derived from eq.~(\ref{nuc}{\sf a}) -- note that $T_{\rm
PL}$ increases with $T_\phi$ (see eq.~(\ref{Tpl})) and ranges
between (0.1 and 23)~GeV, {\sf (ii)} $\Omega_{\chi}h^2$ increases
with $T_\phi$ (or $\Gamma_\phi$) and it turns out almost
$\sv$-independent for $N_\chi\neq0$ -- this is, because $f_\chi$
can be mostly given by eq.~(\ref{Ifnsol}) as shown in
table~\ref{t2}, {\sf (iii)} $\Omega_{\chi}h^2$ is $\sv$-dependent
and increases rapidly for $N_\chi=0$ -- this is because $f_\chi$
can be extracted from eq.~(\ref{IIfnsol}) as shown in
table~\ref{t2}; $f_\chi$ decreases rapidly with $T_\phi$ (and
$T_{\rm PL}$) due to the exponential suppression of $f^{\rm
eq}_\chi$.

$\bullet$ $\log m_\phi$ (or $\Omega_q(\vtns)$) in
fig.~\ref{om}-${\sf (c_1)}$ and ${\sf (c_2)}$ for $\log
m_\phi=70$, $T_\phi=300~{\rm GeV}$ and several $N_\chi$'s
indicated on the curves. We observe that: {\sf (i)}
$\Omega_q(\vtns)$ increases with $m_\phi$ (since
$\brhoqi=(m_\phi/H_0)^2$ increases too) and so, an upper bound on
$m_\phi$ can be derived from eq.~(\ref{nuc}{\sf a}) -- note that
$T_{\rm PL}$ decreases as $m_\phi$ increases (see eq.~(\ref{Tpl}))
and ranges between (3.5 and 31)~GeV, {\sf (ii)} $\Omega_{\chi}h^2$
decreases as $m_\phi$ increases for non-EP (see table~\ref{t2})
and increases with $m_\phi$ for EP  (see fig.~\ref{om}-${\sf
(c_2)}$ and table~\ref{t2}). This is, because $f_\chi$ can be
found from eq.~(\ref{BEsol}{\sf a}) for EP and it increases with
$m_\phi$ or $H$ (given by eq.~(\ref{Hqdom}{\sf a})) since $J_{\rm
F}$ enters the denominator, whereas $f_\chi$ can be derived from
eq.~(\ref{Ifnsol}) [eq.~(\ref{IIfnsol})] for non-EPI [non-EPII]
and it decreases when $H$ increases.

%%%%%%%%%%%%%%%%%%%%%%%%%%%%%%%%%%%%%%%%%%%%%%%%%%%%%%%%%%%%%%%%%%%%
%%%%%%%%%%%%%%%%

The dependence of $\Omega_\chi h^2$ on $\sv$ can be clearly
deduced by fig.~\ref{fig4}. We depict $\Omega_{\chi}h^2$ as a
function of $\langle\sigma v\rangle=a$ for various $N_\chi$'s,
indicated on the curves, $m_\chi=350~{\rm
GeV},~\brhofi=10^{70},~m_\phi=10^6~{\rm GeV}$ and $T_\phi=30~{\rm
GeV}~[T_\phi=3000~{\rm GeV}]$ in fig.~\ref{fig4}-{\sf (a)}
[fig.~\ref{fig4}-{\sf (b)}]. For these parameters we obtain $q$-TD
with $T_{\rm PL}=1.78~{\rm GeV}$ [$T_{\rm PL}=17.5~{\rm GeV}$] and
$\Omega_q(\vtns)=0.0017$ [$\Omega_q(\vtns)=0.14$] in
fig.~\ref{fig4}-{\sf (a)} [fig.~\ref{fig4}-{\sf (b)}].

Obviously, due to very low $T_{\rm PL}$, in fig.~\ref{fig4}-{\sf
(a)} we obtain exclusively non-EPI. When the first term in the
r.h.s of eq.~(\ref{Ifn}) is comparable to the second one (usually
for large $\sv$'s) $\Omega_\chi h^2$ increases as $\sv$ decreases,
whereas when the second term dominates, $\Omega_\chi h^2$ becomes
$\sv$-independent according to eq.~(\ref{Ifnsol}). On the
contrary, in fig.~\ref{fig4}-{\sf (b)}, we obtain EP for
$\sv\gtrsim10^{-7}~{\rm GeV}^{-2}$ and non-EPII for
$\sv\lesssim10^{-7}~{\rm GeV}^{-2}$. We see that $\Omega_\chi h^2$
increases as $\sv$ decreases for EP, in accordance with
eq.~(\ref{BEsol}) -- note that this is the well known behaviour in
the SC. Also, $\Omega_\chi h^2$ decreases with $\sv$ for non-EPII
when the first term in the r.h.s of eq.~(\ref{IIfnsol}{\sf a}) is
dominant (as shown in fig.~\ref{fig4}-{\sf (b)} for $N_\chi=0$ and
for $N_\chi=10^{-7}$ and $\sv$ in the range
$(10^{-9}-10^{-7})~{\rm GeV}^{-2}$) whereas it remains
$\sv$-independent for non-EPII when the second term in the r.h.s
of eq.~(\ref{IIfnsol}{\sf a}) is dominant (as shown in
fig.~\ref{fig4}-{\sf (b)} for $N_\chi=10^{-7}$ and $\sv$ in the
range $(10^{-12}-10^{-9})~{\rm GeV}^{-2}$).

Let us, finally, emphasize that the agreement between numerical
and semi-analytical results is impressive in most of the cases. An
exception is observed in fig.~\ref{om}-${\sf (a_2)}$ for large
$\brhofi$'s  where $(\brhoq/\brhof)(\btrh)<50$ and so, the adopted
approximate formula for $H$ in eq.~(\ref{Hqdom}) is not so
accurate. The need for numerical solution of eq.~(\ref{Ifn}) makes
the discrepancy more evident than in fig.~\ref{om}-${\sf (a_1)}$
where eq.~(\ref{Ifnsol}) is everywhere applicable.

\subsection{C{\ssz OMPARISON} W{\ssz ITH THE}
R{\ssz ESULTS OF} R{\ssz ELATED} S{\ssz CENARIA}} \label{scnonsc}

\hspace{.562cm} It would be interesting to compare $\Omega_\chi
h^2$ calculated in the KRS  with that obtained in the QKS and the
LRS, taking as a reference point the value obtained in the SC
($\brhoq=\brhof=0$), $\Omega_{\chi}h^2|_{_{\rm SC}}$. The relevant
variations can be estimated, by defining the quantities:
\beq\label{dom}\mbox{\sf (a)}~~\Delta\Omega_{\chi}=
\frac{\Omega_{\chi}h^2-\Omega_{\chi}h^2|_{_{\rm
SC}}}{\Omega_{\chi}h^2|_{_{\rm SC}}}~~\mbox{and}~~\mbox{\sf
(b)}~~\Delta\Omega_{\chi}|_{_{\rm CD}}=
\frac{\Omega_{\chi}h^2|_{_{\rm CD}}-\Omega_{\chi}h^2|_{_{\rm
SC}}}{\Omega_{\chi}h^2|_{_{\rm SC}}}\eeq
where CD represents the condition which specifies the scenario
under consideration: $\brhoq=0$ for the LRS or $\brhof=0$ for the
QKS. We restrict our analysis on the parameters used in
fig.~\ref{om} and we present the relevant results in
tables~\ref{t3} and \ref{t4}. In table~\ref{t3} we present
$\Omega_\chi h^2|_{\rm SC}$ and $\Delta\Omega_\chi|_{\brhof=0}$
for several $\Omega_q(\vtns)$'s. For the same $\Omega_q(\vtns)$'s
we arrange $\Delta\Omega_\chi|_{\brhoq=0}$ and $\Delta\Omega_\chi$
in table~\ref{t4}. The corresponding values of $T_{\rm PL}$ and
$T_{\rm RH}$ are also shown. Let us, initially, clarify the basic
features of the $\Omega_{\chi}h^2$ calculation within the other
scenaria. Namely,

\begin{floatingtable}[!h]
\caption{\sl\ftn $\Omega_\chi h^2|_{\rm SC}$ and
$\Delta\Omega_\chi|_{\brhoq=0}$ (for several $\Omega_q(\vtns)$'s)
for the parameters of figs. \ref{om}-${\sf (a_1, b_1, c_1)}$ and
\ref{om}-${\sf (a_2, b_2, c_2)}$.}\label{t3}
\begin{tabular}{|c||c||c|c|} \hline
{\bf  F\ssz IG. \nsz} & {\bf\boldmath $\Omega_\chi h^2|_{\rm
SC}$}&{\bf\boldmath $\Omega_q(\vtns)$}&{\bf\boldmath
$\Delta\Omega_\chi|_{\brhof=0}$}
\\ \hline \hline
\ref{om}-${\sf (a_1)}$,&$1.87$&$0.21$&2242\\
\ref{om}-${\sf (b_1)}$, &&$0.001$&188\\
\ref{om}-${\sf (c_1)}$ &&$10^{-5}$&$25$\\\hline
\ref{om}-${\sf (a_2)}$, &$0.023$&$0.21$&1903\\
\ref{om}-${\sf (b_2)}$, &&$0.001$&160\\
\ref{om}-${\sf (c_2)}$ &&$10^{-5}$&22\\\hline
\end{tabular}
\end{floatingtable}

$\bullet$ $\Omega_{\chi}h^2|_{_{\rm SC}}$ is
$(\brhofi,~m_\phi,~T_\phi,~N_\chi)$ independent and so, it depends
only on $m_\chi$ and $\sv$. Since these variables are fixed in
figs.~\ref{om}-${\sf (a_1, b_1, c_1)}$ and figs.~\ref{om}-${\sf
(a_2, b_2, c_2)}$, we obtain two $\Omega_{\chi}h^2|_{_{\rm SC}}$'s
presented in table~\ref{t3}. $\Omega_{\chi}h^2|_{_{\rm SC}}$
increases as $\sv$ decreases (see also table~\ref{t1}).

$\bullet$ $\Omega_\chi h^2|_{\brhof=0}$ is
($\brhofi,~m_\phi,~T_\phi,~N_\chi$) independent and it exclusively
depends on $\Omega_q(\vtns)$ for fixed $\sv$ and $m_\chi$
\cite{jcapa} (under the assumption that the onset of the KD phase
occurs for $T>m_\chi$). As a consequence, for several fixed
$\Omega_q(\vtns)$'s (see table~\ref{t3}), $\Omega_\chi
h^2|_{\brhof=0}$ takes a certain value for the
figs.~\ref{om}-${\sf (a_1, b_1, c_1)}$ and another for
figs.~\ref{om}-${\sf (a_2, b_2, c_2)}$. It is obvious that we
obtain a sizable enhancement w.r.t $\Omega_{\chi}h^2|_{_{\rm
SC}}$, which increases with $\Omega_q(\vtns)$ and as $\sv$
decreases (see also the last line of table~\ref{t1}).

%A first comparison is displayed in table~\ref{t1}.

$\bullet$ $\Omega_\chi h^2|_{\brhoq=0}$ can be found by solving
numerically \cite{scnr} eq.~(\ref{nf})-(\ref{nx}) where $H$ is
given by eq.~(\ref{Hb}{\sf a}) with $\vrho_q=0$. Note that
$T_\phi$ coincides to $T_{\rm RH}$ in the LRS. As we emphasized in
ref.~\cite{scnr}, the resultant $\Omega_\chi h^2|_{\brhoq=0}$ is
$\brhofi$-independent for $T_{\rm max}>m_\chi$ -- see
table~\ref{t4}, fig.~\ref{om}-${\sf (a_1)}$ and ${\sf (a_2)}$. So,
in principle, $\Omega_\chi h^2|_{\brhoq=0}$ is only
($m_\phi,~T_\phi,~N_\chi$)-dependent for fixed $\sv$ and $m_\chi$.
Since a variation of $m_\phi$ or $T_\phi$ changes
$\Omega_q(\vtns)$ for the KRS, we expect a change to the
corresponding $\Omega_\chi h^2|_{\brhoq=0}$ too. However, due to
the fact that $T_{\rm RH}\gg m_\chi$, $\Omega_\chi
h^2|_{\brhoq=0}$ turns out to be $m_\phi$-independent too -- see
table~\ref{t4}, fig.~\ref{om}-${\sf (c_1)}$ and ${\sf (c_2)}$.
Finally, its $N_\chi$-dependence appears only for $T_{\rm
RH}<m_\chi/20$ -- see table~\ref{t4}, fig.~\ref{om}-${\sf (b_1)}$
and ${\sf (b_2)}$. We observe that $\Omega_\chi h^2|_{\brhoq=0}$
turns out to be very close to $\Omega_{\chi}h^2|_{_{\rm SC}}$,
when $T_{\rm RH}>m_\chi/20$ (see also table~\ref{t1}) and mostly
lower than $\Omega_{\chi}h^2|_{_{\rm SC}}$, for $T_{\rm
RH}<m_\chi/20$ -- see table~\ref{t4}, fig.~\ref{om}-${\sf (b_1)}$
and ${\sf (b_2)}$.

Comparing $\Delta\Omega_\chi$ with $\Delta\Omega_\chi|_{\brhof=0}$
we observe that contrary to the QKS (where $\Omega_\chi
h^2|_{\brhof=0}$ exclusively increases with $\Omega_q(\vtns)$)
$\Delta\Omega_\chi$ depends on the way that the $\Omega_q(\vtns)$
variation is generated. E.g., from table~\ref{t4} we can deduce
that $\Delta\Omega_\chi$ mostly decreases as $\Omega_q(\vtns)$
increases due to a decrease of $\brhofi$ or an increase of
$m_\phi$ and increases with $\Omega_q(\vtns)$, when this is caused
by an increase of $T_\phi$. This can be understood by the fact
that, in the two former cases, $T_{\rm PL}$ decreases whereas in
the latter case, it increases. Obviously, the $\Omega_\chi
h^2$-reduction with $T_{\rm PL}$ is stronger when $N_\chi=0$.

Comparing $\Delta\Omega_\chi$ with $\Delta\Omega_\chi|_{\brhoq=0}$
we observe that: {\sf (i)} $T_{\rm PL}$  turns out to be much
lower than $T_{\rm RH}$, {\sf (ii)} $\Omega_\chi h^2$ increases
with $N_\chi$ much more efficiently than $\Omega_\chi
h^2|_{\brhoq=0}$, {\sf (iii)} $\Delta\Omega_\chi$ can be positive
in many cases (especially for $N_\chi\neq0$ and/or $T_{\rm
PL}\gtrsim m_\chi/20$) in contrast with
$\Delta\Omega_\chi|_{\brhoq=0}$ which is mostly negative, except
for large $\sv$ and $T_{\rm RH}=20~{\rm GeV}$ \cite{fujii, scnr},
{\sf (iv)} $\Omega_\chi h^2$ approaches $\Omega_\chi
h^2|_{\brhoq=0}$ as $\Omega_q(\vtns)$ decreases (for
$\Omega_q(\vtns)\lesssim 10^{-10}$) -- see table~\ref{t1}.

\begin{table}[!t]
\begin{center}
\begin{tabular}{|c|c||c|c|c||c|c|c|} \hline
{\bf  F\ssz IG. \nsz}& {\bf\boldmath $\Omega_q(\vtns)$}
&{\bf\boldmath $T_{\rm PL}$}&\multicolumn{2}{|c||}{\bf\boldmath
$\Delta \Omega_\chi$}&{\bf\boldmath $T_{\rm
RH}$}&\multicolumn{2}{|c|}{\bf\boldmath
$\Delta\Omega_\chi|_{\brhoq=0}$}\\\cline{4-5}\cline{7-8}
&&$({\rm GeV})$&$N_\chi=0$&$N_\chi=10^{-6}$&$({\rm
GeV})$&$N_\chi=0$&$N_\chi=10^{-6}$
\\ \hline \hline
\ref{om}-${\sf
(a_1)}$&$0.21$&0.78&$-1$&$-0.96$&30&\multicolumn{2}{|c|}{$-0.049$}\\
&$0.001$&1.94&$-1$&$-0.91$&30&\multicolumn{2}{|c|}{$-0.049$}\\
&$10^{-5}$&4.15&$-1$&$-0.81$&30&\multicolumn{2}{|c|}{$-0.049$}\\\hline
\ref{om}-${\sf
(a_2)}$&$0.21$&0.78&$-1$&$1.88$&30&\multicolumn{2}{|c|}{$-0.04$}\\
&$0.001$&1.94&$-1$&$5.96$&30&\multicolumn{2}{|c|}{$-0.04$}\\
&$10^{-5}$&4.15&$-1$&9&30&\multicolumn{2}{|c|}{$-0.04$}\\\hline
\ref{om}-${\sf
(b_1)}$&$0.21$&22.6&$0.57$&$1.59$&5000&$-0.1$&$-0.1$\\
&$0.001$&1.5&$-1$&$-0.93$&20&$-0.17$&$-0.15$\\
&$10^{-5}$&0.15&$-1$&$-0.99$&0.15&$-1$&$-1$\\\hline
\ref{om}-${\sf (b_2)}$&$0.21$&22.6&1660&1662&5000&$-0.1$&$-0.1$\\
&$0.001$&1.5&$-1$&4&20&$-0.08$&$0.15$\\
&$10^{-5}$&0.15&$-1$&$-0.51$&0.15&$-1$&$-0.34$\\\hline
\ref{om}-${\sf
(c_1)}$&$0.21$&3.5&$-1$&$-0.97$&300&\multicolumn{2}{|c|}{$-0.1$}\\
&$0.001$&8.8&$-1$&$1.46$&300&\multicolumn{2}{|c|}{$-0.1$}\\
&$10^{-5}$&18.9&$-0.33$&$20$&300&\multicolumn{2}{|c|}{$-0.1$}\\\hline
\ref{om}-${\sf
(c_2)}$&$0.21$&3.5&$-1$&1.04&300&\multicolumn{2}{|c|}{$-0.1$}\\
&$0.001$&8.8&$-1$&90.3&300&\multicolumn{2}{|c|}{$-0.1$}\\
&$10^{-5}$&18.9&19&21&300&\multicolumn{2}{|c|}{$-0.1$}\\\hline
\end{tabular}
\end{center}\vspace*{-.155in}
\caption{\sl\ftn $\Delta\Omega_\chi$ and $T_{\rm PL}$ in the KRS
and $\Delta\Omega_\chi|_{\brhoq=0}$ and $T_{\rm RH}$ in the LRS
for several $\Omega_q(\vtns)$'s and the residual parameters of
figs. \ref{om}-${\sf (a_1, b_1, c_1)}$ and \ref{om}-${\sf (a_2,
b_2, c_2)}$.}\label{t4}\vspace*{-.2in}
\end{table}

\subsection{A{\ssz LLOWED} R{\ssz EGIONS}}
\label{NTR}

\begin{figure}[!ht]\vspace*{-.19in}
\hspace*{-.71in}
\begin{minipage}{8in}
\epsfig{file=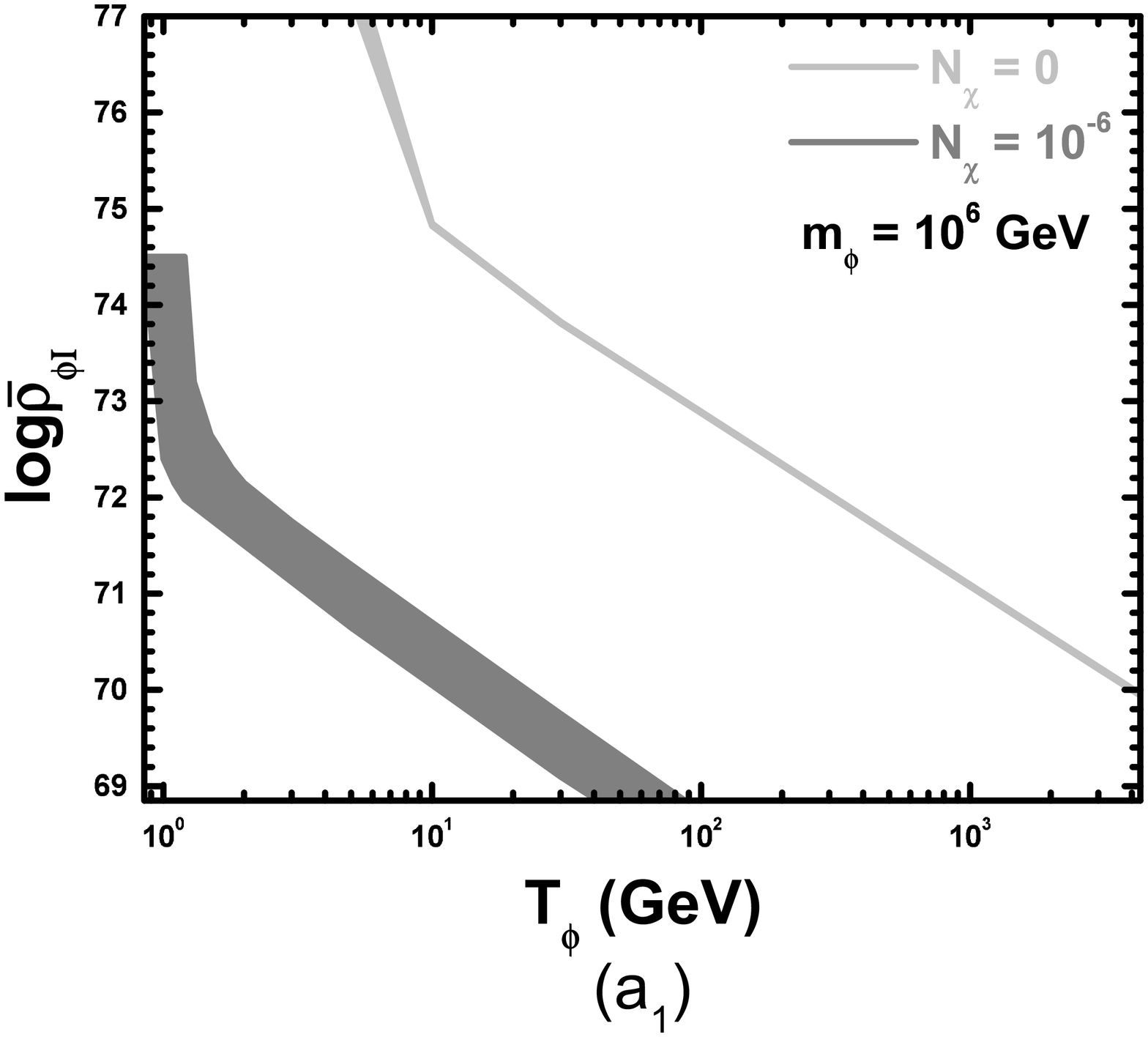,height=3.8in,angle=-90} \hspace*{-1.37 cm}
\epsfig{file=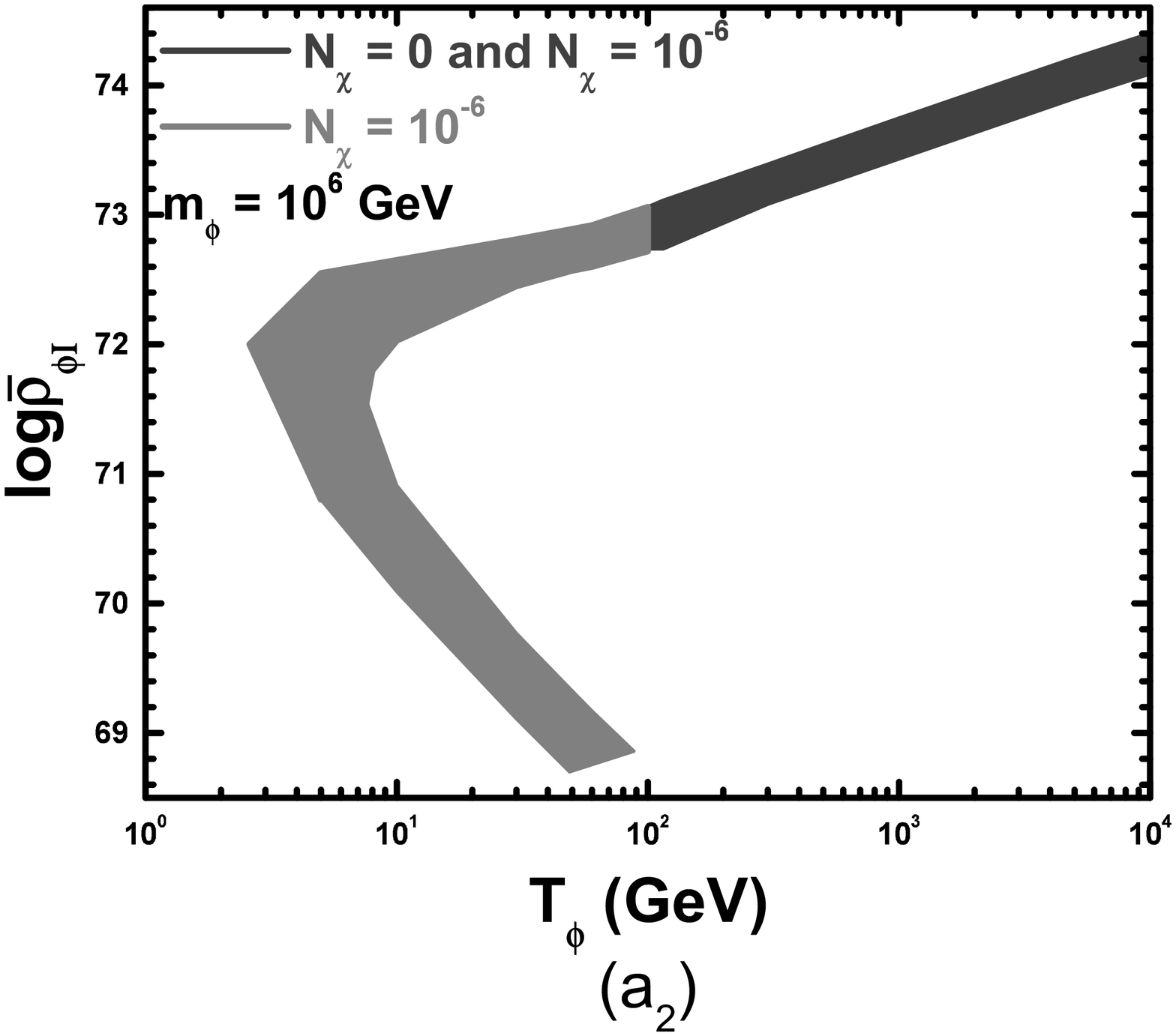,height=3.8in,angle=-90} \hfill
\end{minipage}\vspace*{-.01in}
\hfill\hspace*{-.71in}
\begin{minipage}{8in}
\epsfig{file=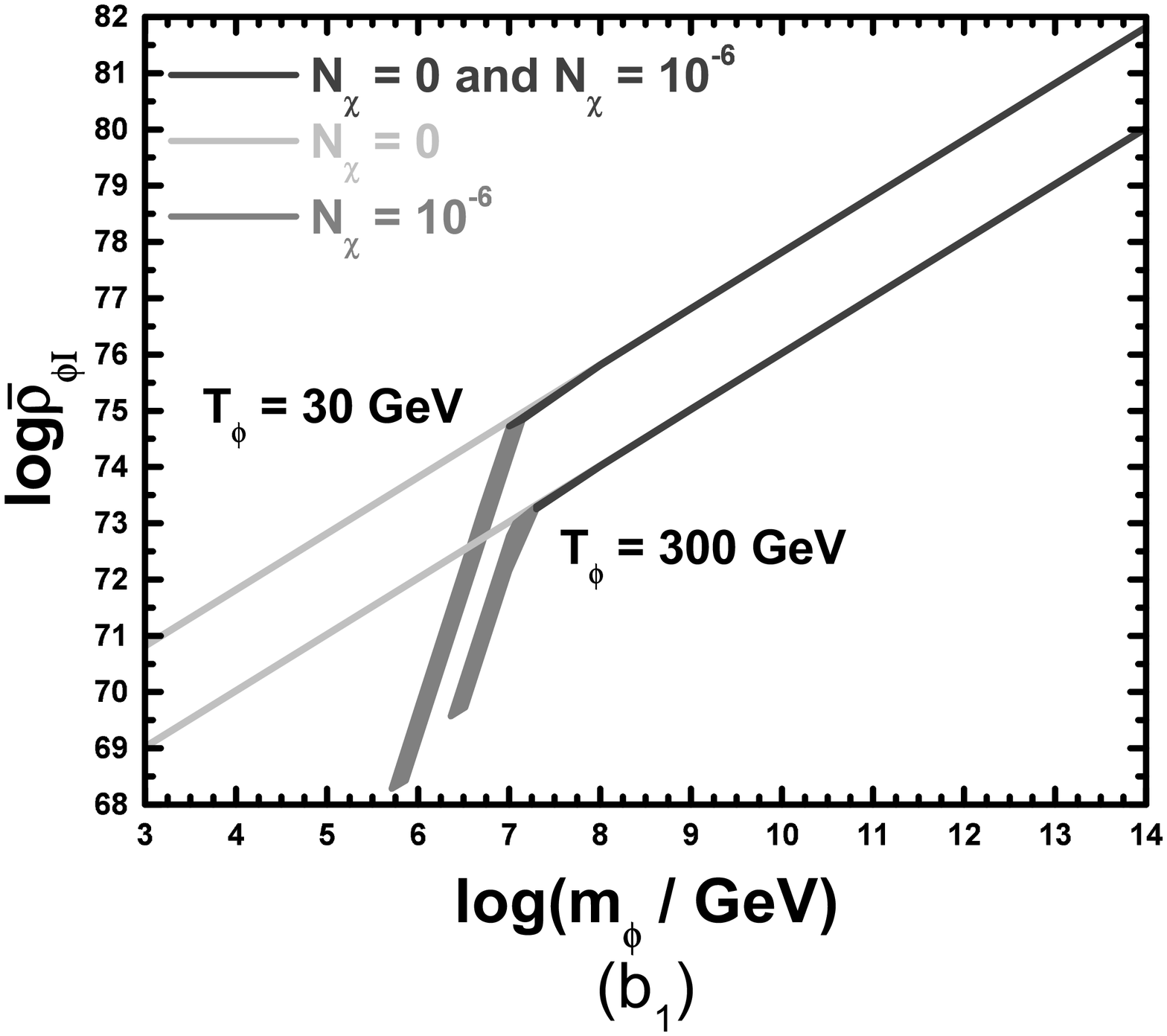,height=3.8in,angle=-90} \hspace*{-1.37 cm}
\epsfig{file=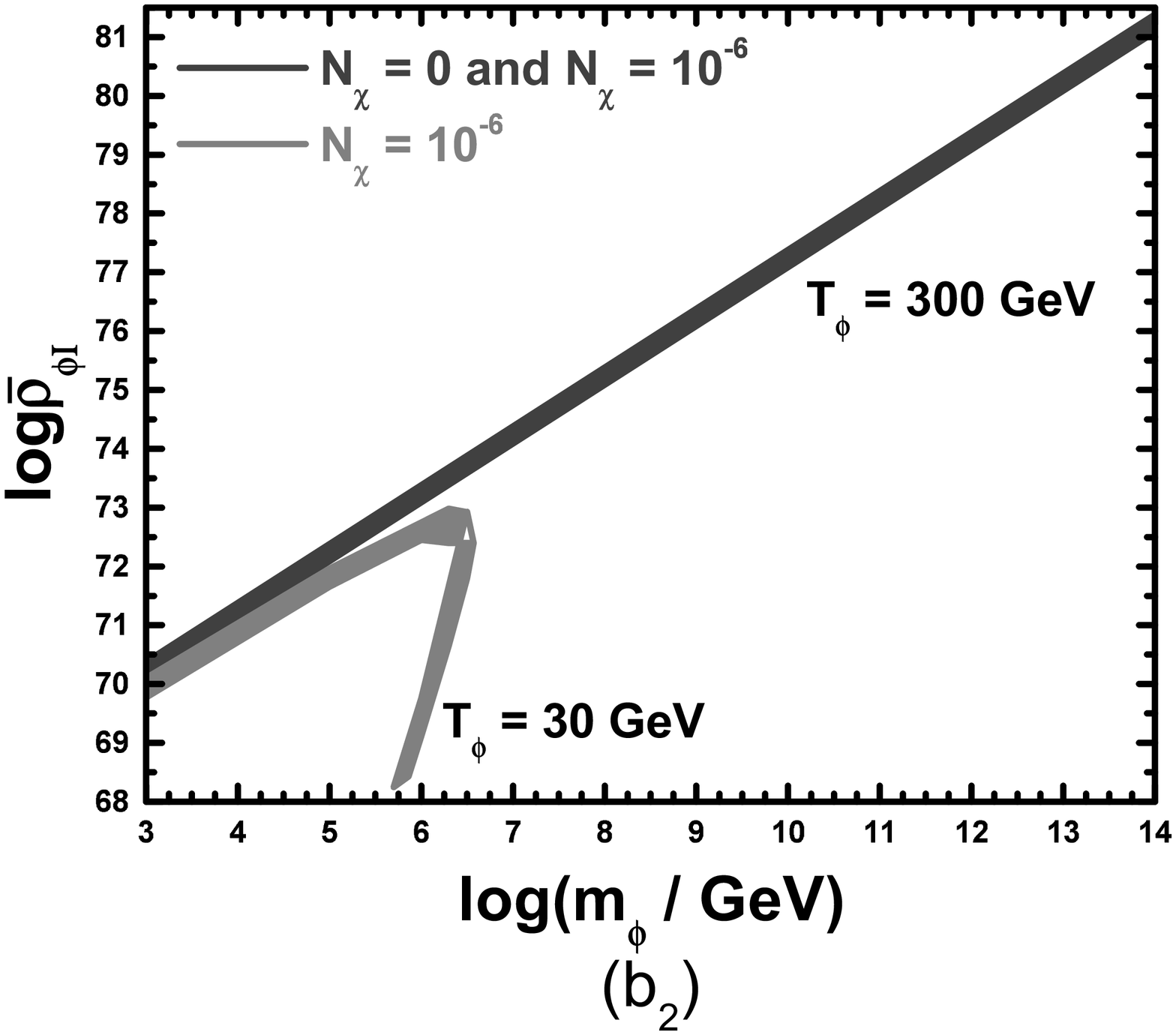,height=3.8in,angle=-90} \hfill
\end{minipage}\vspace*{-.01in}
\hfill\hspace*{-.71in}
\begin{minipage}{8in}
\epsfig{file=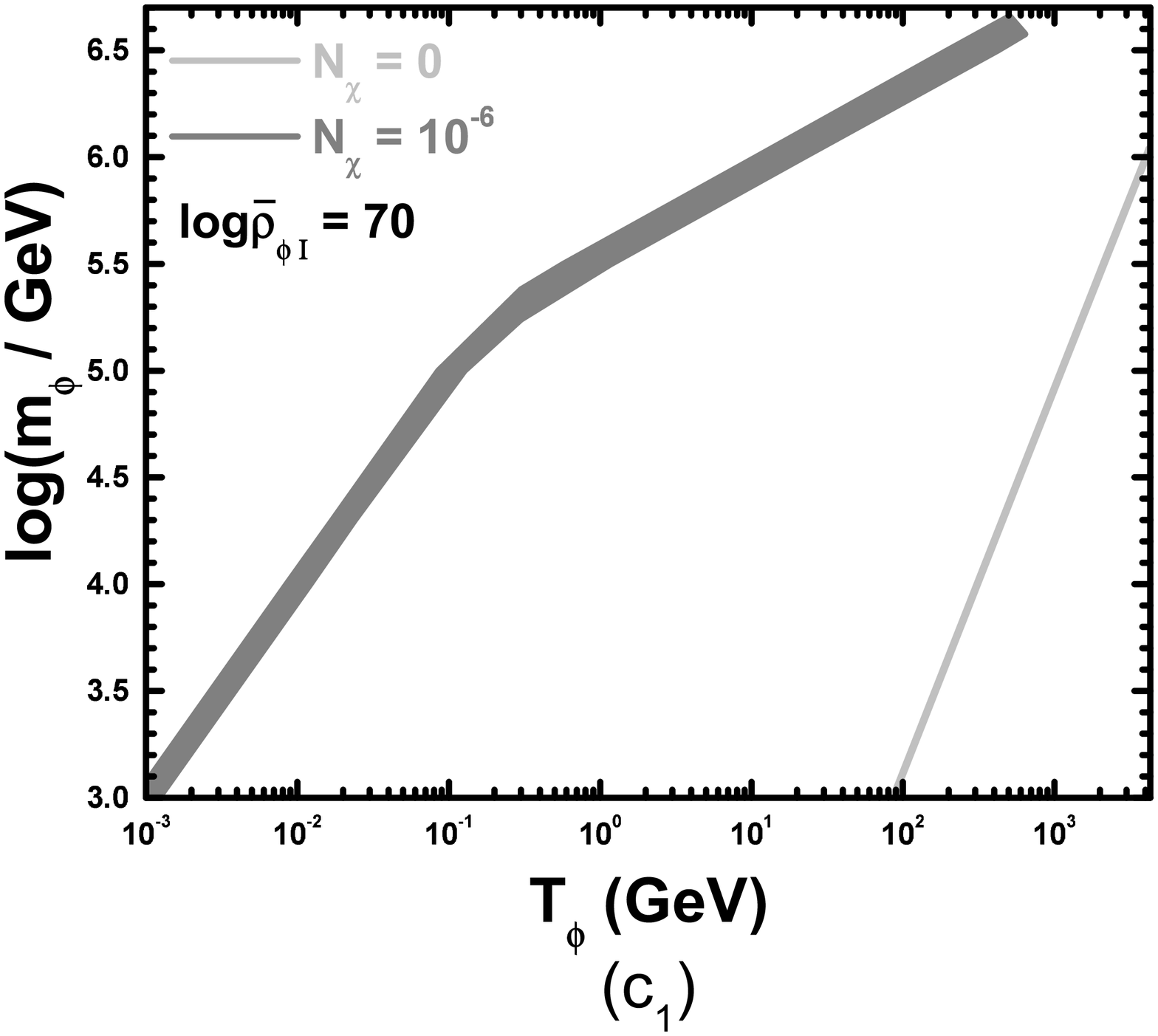,height=3.8in,angle=-90} \hspace*{-1.37 cm}
\epsfig{file=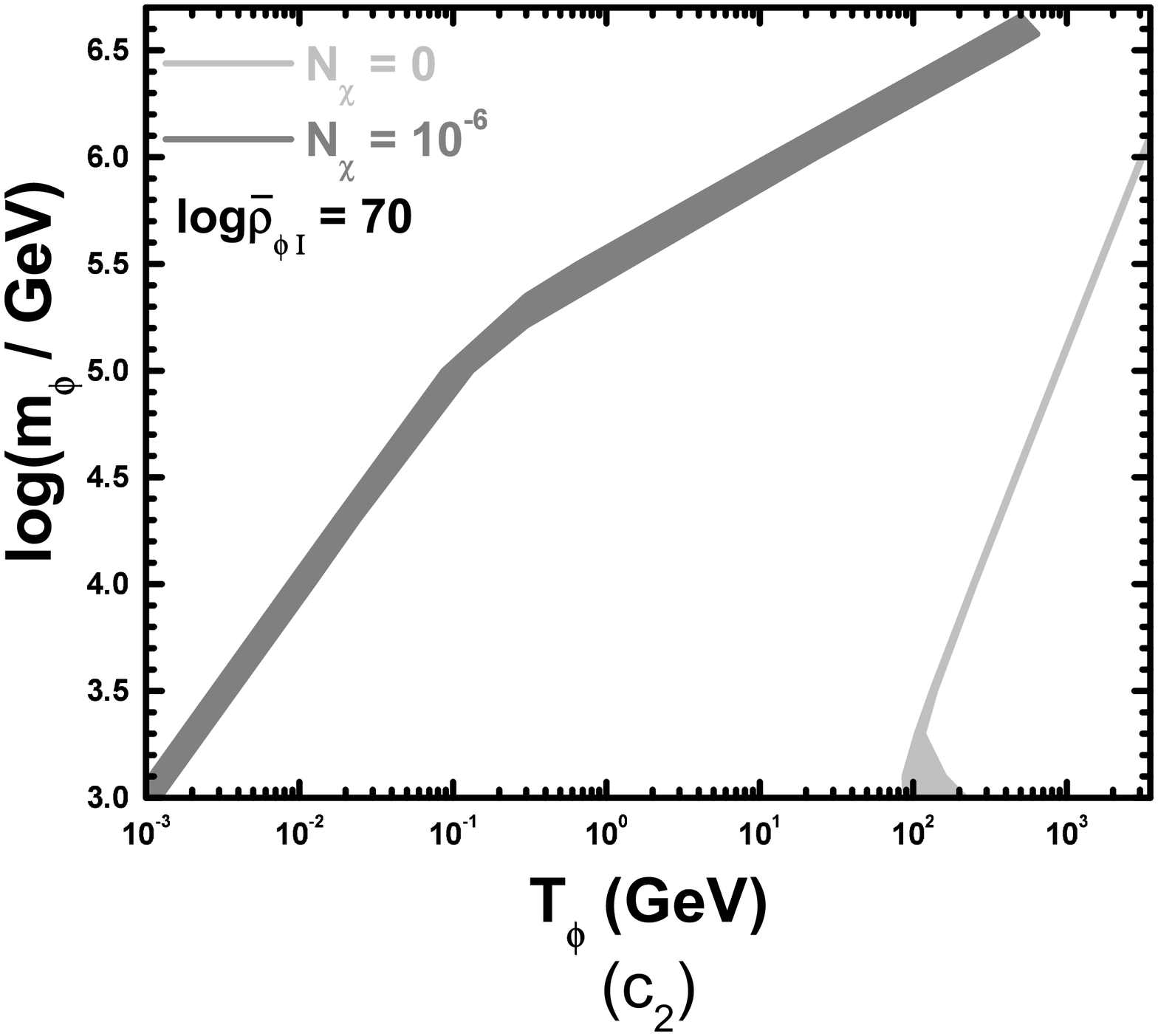,height=3.8in,angle=-90} \hfill
\end{minipage}\vspace*{-.05in}
\hfill \caption[]{\sl\ftn Regions allowed by eq.~(\ref{cdmb}) on
the $T_\phi-\log\brhofi$ plane ${\sf (a_1,~a_2)}$ for
$m_\phi=10^6~{\rm GeV}$, $\log m_\phi-\log\brhofi$ plane ${\sf (
b_1,~b_2)}$ for $T_\phi=30~{\rm GeV}$ and $T_\phi=300~{\rm GeV}$,
and $T_\phi-\log m_\phi$ plane ${\sf (c_1,~c_2)}$ for
$\log\brhofi=70$. We take $N_{\tilde\chi}=0$ or
$N_{\tilde\chi}=10^{-6}$, $m_{\tilde\chi}=350~{\rm GeV}$ and
$\langle\sigma v\rangle=10^{-10}~{\rm GeV^{-2}}~[\langle\sigma
v\rangle=10^{-8}~{\rm GeV^{-2}}]~{\sf (a_1, b_1, c_1~[a_2, b_2,
c_2])}$.} \label{regions}
\end{figure}

\hspace{.562cm} Requiring $\Omega_{\chi}h^2$ to be confined in the
cosmologically allowed range of eq. (\ref{cdmb}), one can restrict
the free parameters. The data is derived exclusively by the
numerical program. Our results are presented in
fig.~\ref{regions}. The allowed regions are constructed for
$N_\chi=0$ and $N_\chi=10^{-6}$. In fig.~\ref{regions}-${\sf (a_1,
b_1, c_1)}$ [fig.~\ref{regions}-${\sf (a_2, b_2, c_2)}$], we fixed
$m_\chi=350~{\rm GeV}$ and $\langle\sigma v\rangle=10^{-10}~{\rm
GeV^{-2}}$ [$\langle\sigma v\rangle=10^{-8}~{\rm GeV^{-2}}$]. We
display the allowed regions on the:

$\bullet$  $T_\phi-\log\brhofi$ plane for $m_\phi=10^6~{\rm GeV}$,
in fig.~\ref{regions}-${\sf (a_1)}$ and ${\sf (a_2)}$. In the
allowed regions of fig.~\ref{regions}-${\sf (a_1)}$ for $N_\chi=0$
[$N_\chi=10^{-6}$] we obtain $q$-PD and EP for $\log\brhofi>74.8$
[$\log\brhofi>72.2$] and $q$-TD with non-EPII [non-EPI] elsewhere,
while $T_{\rm PL}$ ranges between (20 and 40)~GeV [(0.85 and
1.5)~GeV]. Since $f_\chi$ increases with $\rhofi$ -- see
eq.~(\ref{Ifnsol}) [eq.~(\ref{IIfnsol})] for non-EPI [non-EPII]
and eq.~(\ref{BEsol}) for EP -- the upper [lower] boundaries of
the allowed regions come from eq.~(\ref{cdmb}{\sf b})
[eq.~(\ref{cdmb}{\sf a})]. The lower right limit of the allowed
regions comes from eq.~(\ref{nuc}{\sf a}). As $\rhofi$ increases
$\Omega_\chi h^2$ approaches $\Omega_\chi h^2|_{\brhoq=0}$ and it
becomes equal to $0.09$ at

\addtolength{\textheight}{1.5cm}
\newpage
%%%%%%%%%%%%%%%%%%%%%%%%%%%%%%%%%%%%%%%%%%%%%%%%%%%%%%%%%%%%%%%%%%%%
%%%%%%%%%%%%%%%%
\addtolength{\textheight}{-1.5cm}
\newpage

\hspace{-.72cm}the upper left bounds of the allowed regions
(possible further reduction of $T_\phi$ reduces also $\Omega_\chi
h^2$ which turns out to be $\brhof$ independent, any more). In the
dark grey [grey] allowed regions of fig.~\ref{regions}-${\sf
(a_2)}$, we obtain $q$-TD with EP [non-EPI] while $T_{\rm PL}$
ranges between (18 and 400)~GeV [(1 and 15)~GeV]. Due to the
increase of the released entropy which is caused by the increase
of $\brhofi$,  $\Omega_\chi h^2$ decreases as $\rhofi$ increases
and so, the upper [lower] boundary of the dark grey and the almost
horizontal part of the grey allowed region come from
eq.~(\ref{cdmb}{\sf a}) [eq.~(\ref{cdmb}{\sf b})] (note that
eq.~(\ref{Ifnsol}) is not applicable in this part of the grey
allowed region). On the contrary, eq.~(\ref{Ifnsol}) can be
applied for the almost vertical part of the grey allowed region
and so, its left [right] boundary comes from eq.~(\ref{cdmb}{\sf
a}) [eq.~(\ref{cdmb}{\sf b})]. The lower right limit of this
region is found from eq.~(\ref{nuc}{\sf a}) whereas the upper one
is just conventional. \vspace{-.8pt}

%(such a large $T_{\rm PL}$ 400 can be considered as an exception in our scheme)

$\bullet$ $\log m_\phi-\log\brhofi$ plane for $T_\phi=30~{\rm
GeV}$ and $T_\phi=300~{\rm GeV}$, in fig.~\ref{regions}-${\sf
(b_1)}$ and ${\sf (b_2)}$. In the allowed areas we obtain $q$-TD.
In the dark and light grey areas of fig.~\ref{regions}-${\sf
(b_1)}$ [fig.~\ref{regions}-${\sf (b_2)}$] we have non-EPII [EP].
The dark grey regions are $N_\chi$-independent since the first
term in the r.h.s of eq.~(\ref{IIfnsol}{\sf a}) [eq.~(\ref{Ifn})]
dominates over the second one for non-EPII [EP]. In
fig.~\ref{regions}-${\sf (b_1)}$ the upper [lower] boundary of the
allowed regions comes from eq.~(\ref{cdmb}{\sf b})
[eq.~(\ref{cdmb}{\sf a})], whereas in fig.~\ref{regions}-${\sf
(b_2)}$, the origin of the boundaries is the inverse. This is
because higher $\brhofi$'s result to hither $T_{\rm PL}$'s and so,
$T_{\rm RD}$'s. This has as a consequence that the
non-relativistic reduction of $f^{\rm eq}_\chi$ turns out to be
less efficient, thereby increasing $\Omega_\chi h^2$ for non-EPII.
For EP, the same effect causes mainly an increase to the released
entropy and so, a reduction to $\Omega_\chi h^2$. Note, also, that
since $c_{_{q\phi}}\rhofi$ remains constant along the boundaries
of the dark grey areas $T_{\rm PL}$ remains also constant -- see
eq.~(\ref{Tpl}) -- and is equal to $16~{\rm GeV}$ [$18~{\rm GeV}$]
in fig.~\ref{regions}-${\sf (b_1)}$ for $T_\phi=30~{\rm GeV}$
[$T_\phi=300~{\rm GeV}$] and $(32-38)~{\rm GeV}$ in
fig.~\ref{regions}-${\sf (b_2)}$. The upper [lower] limits of the
dark grey or grey areas (in the upper right [lower left] corners)
of these figures correspond to the upper [lower] bound of
eq.~(\ref{para}). In the grey areas of fig.~\ref{regions}-${\sf
(b_1)}$ and ${\sf (b_2)}$, we obtain non-EPI. Eq.~(\ref{Ifnsol})
is applicable in the almost vertical parts of these areas and so,
the left [right] boundary of the allowed regions comes from
eq.~(\ref{cdmb}{\sf b}) [eq.~(\ref{cdmb}{\sf a})] ($T_{\rm PL}$
ranges between (3 and 17-18)~GeV). Eq.~(\ref{Ifnsol}) is not
applicable in the left upper branch of the area in
fig.~\ref{regions}-${\sf (b_2)}$ where the origin of the
boundaries is the inverse ($T_{\rm PL}=(1-7)~{\rm GeV}$). The
lower bound of the almost vertical part of these areas come from
eq.~(\ref{nuc}{\sf a}). Note, finally, that for $T_\phi=30~{\rm
GeV}$ and $\langle\sigma v\rangle=10^{-8}~{\rm GeV^{-2}}$ --
fig.~\ref{regions}-${\sf (b_2)}$ -- we are not able to construct
allowed area for $N_\chi=0$. This is because $T_{\rm PL}\ll
m_\chi/20$ and possible increase of $\brhofi$ (which could
increase $T_{\rm PL}$) leads to $\Omega_\chi h^2$ close to
$\Omega_\chi h^2|_{\brhof=0}$ which is lower than 0.09 due to
large $\langle\sigma v\rangle$. However for $N_\chi=10^{-6}$,
$\Omega_\chi h^2$ increases to an acceptable level.

$\bullet$ $T_\phi-\log m_\phi$ plane for $\log\brhofi=70$, in
fig.~\ref{regions}-${\sf (c_1)}$ and ${\sf (c_2)}$. In the allowed
regions of these figures, we obtain $q$-TD, besides the part of
the grey area for $\log(m_\phi/{\rm GeV})<5.35$ where we get
$q$-PD. For $N_\chi\neq0$ we have non-EPI, whereas for $N_\chi=0$,
we take non-EPII, except for the lower part of the light grey area
(for $\log(m_\phi/ {\rm GeV})<3.3$) where the EP is activated. As
induced from eq.~(\ref{Ifnsol}) [eq.~(\ref{IIfnsol})] for non-EPI
[non-EPII], $f_\chi$ decreases as $m_\phi$ (and so, $\rhoqi$)
decreases. Therefore, the upper [lower] boundary of the allowed
regions comes from eq.~(\ref{cdmb}{\sf a}) [eq.~(\ref{cdmb}{\sf
b})]. The upper bound on the right corners of the allowed regions
is derived from eq.~(\ref{nuc}{\sf a}). The lower limits of the
light grey areas and these of the lower left corner of the grey
areas are extracted from the lower bound of eq.~(\ref{para}{\sf
a}). In the grey [light grey] areas $T_{\rm PL}$ ranges between
(0.1 and 5)~GeV [(16 and 20)~GeV].

\section{C{\ftn ONCLUSIONS}}
\label{con}

\hspace{.562cm} We studied the decoupling of a CDM candidate,
$\chi$, in the context of a novel cosmological scenario termed KRS
(``KD Reheating''). According to this, a scalar field $\phi$
decays, reheating the universe, under the total or partial
domination of the kinetic energy density of another scalar field,
$q$, which rolls down its exponential potential, ensuring an early
KD epoch and acting as quintessence today. We solved the problem
{\sf (i)} numerically, integrating the relevant system of the
differential equations {\sf (ii)} semi-analytically, producing
approximate relations for the cosmological evolution before and
after the onset of the RD era and solving the properly
re-formulated Boltzmann equation which governs the evolution of
the $\chi$-number density. Although we did not succeed to achieve
general analytical solutions in all cases, we consider as a
significant development the derivation of a result for our problem
by solving numerically just one equation, instead of the whole
system above.

%The second way facilitates the understanding of the problem and
%gives, in most cases, accurate results.

The model parameters were confined so as $H_{\rm I}=m_\phi$ and
$\Omega^{\rm I}_q=1$. The current observational data originating
from nucleosynthesis, acceleration of the universe and the DE
density parameter were also taken into account. We considered two
cases depending whether $\phi$ decays before ($q$-TD) or after
($q$-PD) it becomes the dominant component of the universe. We
showed that, in both cases, the temperature remains frozen for a
period at a plateau value $T_{\rm PL}$, which turns out to be much
lower than its maximal value achieved during a pure reheating with
the same initial $\phi$-energy density.

As regards the $\Omega_{\chi}h^2$ computation, we discriminated
two basic types of $\chi$-production depending whether $\chi$'s do
or do not reach chemical equilibrium with plasma. In the latter
case, two subcases were singled out: the type I and type II
non-EP. The type I non-EP is activated for $T_{\rm PL}\ll
m_\chi/20$ and $N_{\chi}\neq0$ is required so as sizable
$\Omega_\chi h^2$ is achieved. The type II non-EP is activated for
$T_{\rm PL}\sim m_\chi/20$. Finally, EP is applicable for $T_{\rm
PL}\gg m_\chi/20$.

Next, we investigated the dependence of $\Omega_{\chi}h^2$ on the
$\Omega_q(\vtns)$ variations, generated by varying the free
parameters $(\brhofi,~m_\phi,~T_{\phi})$. We showed that mostly
$\Omega_\chi h^2$ increases with $\Omega_q(\vtns)$ when $T_{\rm
PL}$ increases and it decreases as $\Omega_q(\vtns)$ increases
when $T_{\rm PL}$ decreases, too. Also, $\Omega_\chi h^2$
decreases as $\sv$ increases for EP and non-EPI for large $\sv$'s,
it decreases with $\sv$ for non-EPII and large $\sv$'s and it
remains constant for non-EPI and non-EPII for low $\sv$'s.
Finally, in any case, $\Omega_\chi h^2$ increases with $N_\chi$
and $m_\chi$.

Comparing the results on $\Omega_{\chi}h^2$ with those in the QKS
and the LRS, we concluded that in the present scenario,
$\Omega_{\chi}h^2$ does not exclusively increases with
$\Omega_q(\vtns)$ (in contrast with the QKS) and it approaches its
value in the LRS as $\Omega_q(\vtns)$ decreases. Finally, regions
consistent with the present CDM bounds were constructed, using
$m_{\chi}$'s and $\sv$'s commonly allowed in several particle
models. In most cases, the required $T_{\rm PL}$ is lower than
about 40~GeV. As a consequence, simple, elegant and restrictive
particle models such as the CMSSM \cite{Cmssm} -- which, due to
the large predicted $\Omega_{\chi}h^2$, is tightly constrained in
the SC \cite{spanos} or almost excluded in the QKS \cite{salati,
prof, jcapa} -- can become perfectly viable in the KRS.

\acknowledgments \hspace{.562cm} The author would like to thank K.
Dimopoulos, G. Lazarides and A. Masiero for enlightening
communications, I.N.R. Peddie for linguistic suggestions and the
Greek State Scholarship Foundation (I. K. Y.) for financial
support.

\end{document}